\documentclass[a4paper,11pt]{article}

\pdfoutput=1 

\usepackage{jheppub} 
\usepackage{rotating}
\usepackage{tikz}
\usepackage{makecell}
\usepackage{float}
\usepackage{environ}
\usetikzlibrary{decorations.pathmorphing,decorations.markings,trees,shapes}
\usepackage[force]{feynmp-auto}
\tikzset{
    photon/.style={decorate, decoration={snake}},
    electron/.style={draw=black, postaction={decorate},decoration={markings,mark=at position .55 with {\arrow[draw=black]{>}}}},
        antielectron/.style={draw=black, postaction={decorate},decoration={markings,mark=at position .55 with {\arrow[draw=black]{<}}}},
    gluon/.style={decorate, draw=magenta,
        decoration={coil,amplitude=4pt, segment length=5pt}} 
}
\usepackage{amsfonts}
\usepackage{amssymb}
\usepackage{amsmath}
\usepackage{amssymb}
\usepackage{wasysym}
\usepackage{bm}
\usepackage{latexsym}
\usepackage{textcomp}
\usepackage{subcaption}
\usepackage{physics}

\usepackage{verbatim}
\usepackage{slashed}
\usepackage{array}
\usepackage{graphicx}
\usepackage{multirow}
\usepackage{graphicx} 
\usepackage{color}
\usepackage{dsfont}

\usepackage{tikz-feynman}
\usepackage{feynmp-auto}

\definecolor{myred}{rgb}{0.7, 0, 0}
\definecolor{myblue}{rgb}{0, 0, 0.7}
\definecolor{mygreen}{rgb}{0.04, 0.7, 0.5}
\definecolor{mygray}{rgb}{0.1, 0.1, 0.1}

\hypersetup{colorlinks,citecolor=myred,linkcolor=myblue,urlcolor=myblue,linktocpage=true}

 \def\be   {\begin{equation}}   \def\ee   {\end{equation}}
 \def\ba   {\begin{array}}      \def\ea   {\end{array}}
 \def\bea  {\begin{eqnarray}}   \def\eea  {\end{eqnarray}}
 \def\bean {\begin{eqnarray*}}  \def\eean {\end{eqnarray*}}
 \def\nn{\nonumber}
 \def\bry{\begin{array}}
 \def\ery{\end{array}}

\def\GeV{\,{\rm GeV}}
\def\MeV{\,{\rm MeV}}

\def\U1B{U(1)_{\rm B}}
\def\eb{\varepsilon_B}
\def\amu{a_\mu}
\def\l{\ell}

\newcommand{\skipnew}[1]{}

\usetikzlibrary{decorations.pathmorphing}
\usetikzlibrary{shapes.geometric}
\newcommand{\beq}{\begin{equation}}
\newcommand{\eeq}{\end{equation}}
\newcommand{\gsim}{\lower.7ex\hbox{$\;\stackrel{\textstyle>}{\sim}\;$}}
\newcommand{\lsim}{\lower.7ex\hbox{$\;\stackrel{\textstyle<}{\sim}\;$}}

\newcommand{\zprime}{$B\,$}

\newcommand{\cA}{\mathcal{A}}
\newcommand{\cB}{\mathcal{B}}
\newcommand{\cO}{\mathcal{O}}
\newcommand{\cL}{\mathcal{L}}
\newcommand{\cF}{\mathcal{F}}

\newcommand{\BR}{\mathcal{B}}

\numberwithin{equation}{section}
\renewcommand{\theequation}{\arabic{section}.\arabic{equation}}

\def\UMD{\small{Maryland Center for Fundamental Physics, University of Maryland, College Park, MD 20742, USA}}

\def\DESY{\small{Deutsches Elektronen-Synchrotron DESY, Notkestr. 85, 22607 Hamburg, Germany}}

\def\TAU{\small{Tel Aviv University, 6195001, Tel Aviv, Israel}}

\def\Weizmann{\small{Department of Particle Physics and Astrophysics, Weizmann Institute of Science, Rehovot 761001, Israel}}

\def\Technion{\small{Physics Department, Technion - Israel Institute of Technology, Haifa 3200003, Israel}}

\usepackage{tikz,hyperref}
\definecolor{lime}{HTML}{A6CE39}
\DeclareRobustCommand{\orcidicon}{%
	\begin{tikzpicture}
	\draw[lime, fill=lime] (0,0) 
	circle [radius=0.16] 
	node[white] {{\fontfamily{qag}\selectfont \tiny ID}};	\draw[white, fill=white] (-0.0625,0.095) 
	circle [radius=0.007];	\end{tikzpicture}
	\hspace{-2mm}}
\foreach \x in {A,...,Z}{%
	\expandafter\xdef\csname orcid\x\endcsname{\noexpand\href{https://orcid.org/\csname orcidauthor\x\endcsname}{\noexpand\orcidicon}}
}

\preprint{
\begin{minipage}{5cm}
\begin{flushright}
UMD-PP-024-11\\
DESY-24-206
 \end{flushright}
\end{minipage}
}

\title{Searching for hadronic scale baryonic and dark forces at $(g-2)_\mu$'s lattice-vs-dispersion front 
}

\author[a]{Kaustubh Agashe\orcidA{},}
\author[a]{Abhishek Banerjee\orcidB{},}
\author[b]{Minyuan Jiang\orcidC{},}
\author[c]{Shmuel Nussinov\orcidD{},}
\author[a]{Kushan Panchal\orcidE{},}
\author[a]{Srijit Paul\orcidF{},}
\author[d]{Gilad Perez\orcidG{},}
\author[e]{Yotam Soreq\orcidH{}}

\affiliation[a]{\UMD}
\affiliation[b]{\DESY}
\affiliation[c]{\TAU}
\affiliation[d]{\Weizmann}
\affiliation[e]{\Technion}

\abstract{ 
The anomalous magnetic moment of the muon ($\amu$) provides a stringent test of the quantum nature of the Standard Model (SM) and its extensions. 
To probe beyond the SM physics, one needs to be able to subtract the SM contributions, which consists of a non-perturbative part, namely, the hadronic vacuum polarization (HVP) of the photon.
The state of the art is to predominantly use two different methods to extract this HVP: lattice computation, and dispersion relation-based, data-driven method.
Thus one can construct different forms of the ``$\amu$  test" which compares the precise measurement of $\amu$ to its theory prediction. 
Additionally, this opens the possibility for another subtle test, where these two ``theory" predictions themselves are compared against each other, which is denoted as the ``HVP-test". 
This test is particularly sensitive to hadronic scale new physics.
Therefore, in this work, we consider an SM extension consisting of a generic, light $\sim(100\MeV-1\GeV)$ vector boson and study its impact on both tests. 
We develop a comprehensive formalism for this
purpose. 
We find that in the case of data-driven HVP being used in the $\amu$ test, the new physics contributions effectively cancel 
for a flavor-universal vector boson. 
As an illustration of these general results, we consider two benchmark models: i)~the dark photon $(A')$ and ii)~a gauge
boson coupled to baryon-number $(B)$. 
Using a combination of these tests, we are able to constrain the
parameter space of $B$ and $A'$, complementarily to the existing limits. 
As a spin-off, our preliminary analysis of the spectrum of the invariant mass of $3\pi$ in events with ISR at the $B-$ factories (BaBar, Belle) manifests the value of such a study in searching for $B\to 3\pi$ decay, thus motivating a dedicated search by experimental collaborations.
}

\begin{document}

\maketitle

\section{Introduction}
\label{sec:intro}

One of the most precise low energy tests of the Standard Model~(SM) of particle physics is through the measurement of the muon's anomalous magnetic moment,  $\amu\equiv(g-2)_\mu/2$. 
The results of the Muon $g-2$ collaboration at the Fermi National Accelerator Laboratory~(FNAL)~\cite{Muong-2:2021ojo,Muong-2:2023cdq,Muong-2:2024hpx} are consistent with the Brookhaven National Laboratory~(BNL) ones~\cite{Muong-2:2006rrc}.
The combined world average is $\amu^{\rm exp}=116592059(22)\times 10^{-11}$~\cite{Muong-2:2023cdq}. 
On the theory side, the most uncertain part is related to the photon's hadronic vacuum polarization~(HVP), which cannot be calculated in perturbation theory~\cite{Jegerlehner:2017gek}. 

There are two competing methods to estimate the HVP.
First, a data-driven (DD) method, obtained by using the measured $\sigma(e^+e^-\to\text{hadrons})$ along with dispersion relations, and second, by using lattice QCD. 
By using either of these methods, the corresponding SM theory predictions, $\amu^{\text{SM (= DD or lat) } }$, can yield independent $\amu$ tests as
\begin{align}
    \label{eq:Deltaamu}
    \Delta\amu^{ \rm SM } \equiv \amu^{\rm exp} - \amu^{ \text{SM (= DD or lat)} } \,. 
\end{align}
Using the data from BaBar~\cite{BaBar:2004ytv}, SND~\cite{Achasov:2002ud,Achasov:2003ir}, and KLOE~\cite{KLOE-2:2016ydq} experiments, the Muon $g-2$ Theory Initiative~(TI) 2020~\cite{Aoyama:2020ynm} prediction of the leading HVP contribution is $(\amu^{\rm HVP})^{\rm TI} =6931(40) \times 10^{-11}$. 
Plugging $(\amu^{\rm HVP})^{\rm TI}$ into the theory predictions for the SM, $\amu^{\rm SM}$, leads to $\sim 5\sigma$ deviation from $a^{\rm exp}_\mu$. 
More recently, based on the measurement of $\sigma(e^+e^-\to 2\pi)$, the CMD-3 collaboration presented an HVP value $3.7\sigma$ higher than that of the TI 2020 result~\cite{CMD-3:2023alj, CMD-3:2023rfe}.  
On the other hand, there have been several lattice calculations of the HVP contribution~\cite{Alexandrou:2022amy,Ce:2022kxy,Kuberski:2024bcj,Djukanovic:2024cmq,RBC:2024fic,RBC:2023pvn,Bazavov:2024xsk}. In this article, we focus on the BMW 2020 result\footnote{There is also the latest 2024 result from BMW \cite{Boccaletti:2024guq}. However, it has data-driven inputs in its computation. Hence, we choose the 2020 number as it is a purely lattice result.},  
$(\amu^{\rm HVP})^{\rm BMW}=7075(55) \times 10^{-11}$~\cite{Borsanyi:2020mff}.
Both CMD-3 and lattice predictions for $\amu$ are consistent with the experimental result. 

Since the TI 2020 data-driven HVP is a measurement, it can be compared to the lattice prediction and define the HVP test~\cite{DiLuzio:2021uty}
\begin{align}
    \label{eq:DeltaamuHVP}
    \Delta \amu^{\rm HVP} 
    \equiv 
     (\amu^{\rm HVP})^{\rm DD} - (\amu^{\rm HVP})^{\rm lat} \, .
\end{align}
The known HVP \textit{puzzle} is when using DD$=$TI and the BMW for the lattice, which leads to $\Delta \amu^{\rm HVP}\times 10^{11}=-144\,(68)$ i.e. a $\sim 2.1\sigma$ discrepancy.
Ref.~\cite{DiLuzio:2024sps} has proposed several tests to shed light on the HVP puzzle in addition to what is discussed above.
Among them are the comparison of the values of $\alpha$ and $\sin^2 \theta_W$ at different scales, and the comparison with data from muonium spectroscopy (see also~\cite{Delaunay:2021uph}).
 We note that a combination of the BaBar~\cite{BaBar:2021cde}, $\tau-$ decay~\cite{Davier:2023fpl}, and CMD-3~\cite{CMD-3:2023alj} data, i.e., {\em exclusion} of the KLOE and SND, ameliorates the HVP tension, while a further combination of the BMW result with them, reduces the $(g-2)_\mu$ anomaly to $2.8\sigma$~\cite{Davier:2023fpl}.

The HVP and the electro-weak precision tests are very sensitive to hadronic scale physics (however, it was shown in~\cite{Balkin:2021rvh} that low energy extensions of the SM exist, in which the $(g-2)_\mu$ sensitivity to the new fields is greatly ameliorated). Thus, it motivated the authors of ~\cite{Passera:2008jk,Keshavarzi:2020bfy,DiLuzio:2021uty,Coyle:2023nmi} to confront models consisting of a new gauge boson, $X$, of mass $m_X\lesssim\cO(\rm GeV)$ having both leptonic and hadronic couplings with these tests. 
In this paper, we consider this class of models, assuming that the new gauge boson couples in a flavor-universal manner, and point out that  there are two 
additional subtle, but important effects. 
First, if the $X$-hadronic couplings ($g_q$) are larger than the $X$-leptonic couplings ($g_\ell$) by at least a loop factor, the $X$-photon mixing diagram can be as important as the 1-loop $X$ contribution to $\amu$, see Fig.~\ref{Fig:feynmann_diagram_HVP} and Fig.~\ref{Fig:feynmann_diagram_1loop_X}, respectively. 
Second, in case that the data-driven method is used to obtain the photon HVP for the $\amu$ test in Eq.~\eqref{eq:Deltaamu}, the $X$ on-shell contribution to the $e^+e^-\to {\rm hadrons}$ cross section (Fig.~\ref{Fig:feynmann_diagram_tree_B}), which needs to be subtracted, cancels the 1-loop $X$ contribution to $(g-2)_\mu$ from Fig.~\ref{Fig:feynmann_diagram_1loop_X},
in the limit where the leptonic branching ratio of $X$ is negligible. 
This effect results in a weaker sensitivity of $a_{ \mu }$ to the 1-loop contribution of $X$, with the correction therein being a factor of 
$(1-\BR(X\to {\rm had}))\sim g_\ell^2/g_q^2$. 
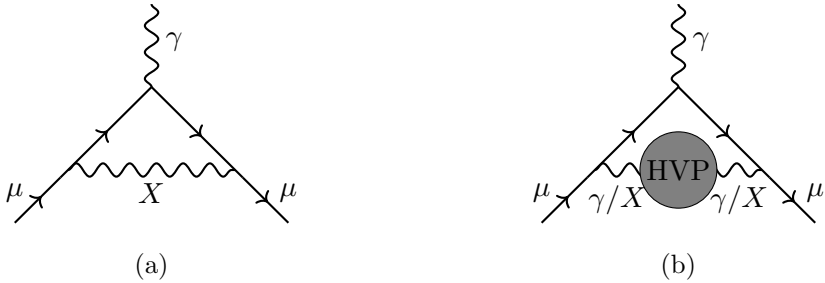
\begin{figure}[t]
\centering
\begin{subfigure}{0.45\textwidth}
    \centering
    \begin{tikzpicture}
    \draw[thick,electron] (0,0) to  [out=-45,in=135] (1.2,-1.2) ;
    \draw[thick,electron] (1.2,-1.2) to  [out=-45,in=135] (1.8,-1.8) node[label=$\mu$] {};
    \draw[thick,antielectron] (0,0) to [out=-135,in=45] (-1.2,-1.2);
    \draw[thick,antielectron] (-1.2,-1.2) to  [out=-135,in=45] (-1.8,-1.8) node[label=$\mu$] {};
    \draw[thick,photon] (0,0) to [out=90,in=-90] (0,1.1); 
    \draw[thick,photon] (-1.1,-1.1) to [out=0,in=-180] (1.1,-1.1) ;
    \node[] at (0,-1.4) {$X$};
    \node[] at (0.3,0.6) {$\gamma$};
    \end{tikzpicture}
    \caption{}
    \label{Fig:feynmann_diagram_1loop_X} 
\end{subfigure}
\vspace{0.5 cm}
\begin{subfigure}{0.45\textwidth}
    \centering
    \begin{tikzpicture}

    \begin{scope}[xshift=6cm]
    \draw[thick,electron] (0,0) to  [out=-45,in=135] (1.2,-1.2) ;
    \draw[thick,electron] (1.2,-1.2) to  [out=-45,in=135] (1.8,-1.8) node[label=$\mu$] {};
    \draw[thick,antielectron] (0,0) to [out=-135,in=45] (-1.2,-1.2);
    \draw[thick,antielectron] (-1.2,-1.2) to  [out=-135,in=45] (-1.8,-1.8) node[label=$\mu$] {};
    \draw[thick,photon] (0,0) to [out=90,in=-90] (0,1.1); 
    \draw[thick,photon] (-1.1,-1.1) to [out=0,in=-180] (1.1,-1.1) ;
    \node[] at (0,-1.4) {$\gamma$};
    \node[] at (0.3,0.6) {$\gamma$};
    \node[draw,circle,fill=gray,inner sep=1.5pt] (amp1) at (0,-1.1) {HVP};
    \node[] at (-0.8,-1.5) {$\gamma/X$};
    \node[] at (0.8,-1.5) {$\gamma/X$};
    \end{scope}
    \end{tikzpicture}
    \caption{} 
    \label{Fig:feynmann_diagram_HVP}
\end{subfigure}
\caption{Relevant Feynman diagrams that contribute to  the $\amu$ test. 
$X$ is the new gauge boson. The diagrams are the $(g-2)_\mu$ contribution from 1-loop $X$ (a), and through the vacuum polarization of photon and $X$ (b).
}
\label{Fig:feynmann_diagrams}
\end{figure}
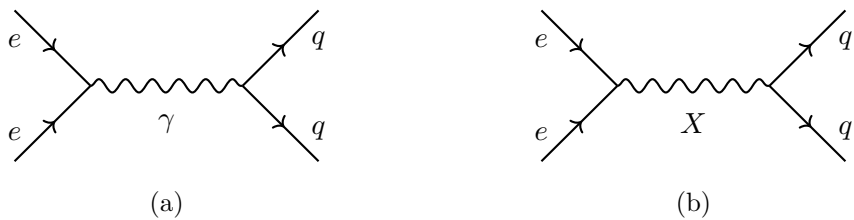
\begin{figure}[t]
\centering
\begin{subfigure}{0.45\textwidth}
    \centering
    \begin{tikzpicture}
    \draw[thick,antielectron] (-1,0) to  [out=135,in=-45] (-2,1);
    \draw[thick,antielectron] (-1,-0) to  [out=-135,in=45] (-2,-1) node[label=$e$] {};
    \draw[thick,electron] (1,0) to  [out=45,in=-135] (2,1)  ;
    \draw[thick,electron] (1,0) to  [out=-45,in=135] (2,-1) node[label=$q$] {};
    \draw[thick,photon] (-1,0) to [out=0,in=-180] (1,0); 
    \node[] at (0,-0.5) {$\gamma$};
    \node[] at (-2,0.6) {$e$};
    \node[] at (2,0.6) {$q$};
    \end{tikzpicture}
    \caption{} 
    \label{Fig:feynmann_diagram_tree_photon}
\end{subfigure}
\begin{subfigure}{0.45\textwidth}
    \centering
    \begin{tikzpicture}
    \draw[thick,antielectron] (-1,0) to  [out=135,in=-45] (-2,1);
    \draw[thick,antielectron] (-1,-0) to  [out=-135,in=45] (-2,-1) node[label=$e$] {};
    \draw[thick,electron] (1,0) to  [out=45,in=-135] (2,1)  ;
    \draw[thick,electron] (1,0) to  [out=-45,in=135] (2,-1) node[label=$q$] {};
    \draw[thick,photon] (-1,0) to [out=0,in=-180] (1,0); 
    \node[] at (0,-0.5) {$X$};
    \node[] at (-2,0.6) {$e$};
    \node[] at (2,0.6) {$q$};
    \end{tikzpicture}
    \caption{} 
    \label{Fig:feynmann_diagram_tree_B}
\end{subfigure}
\vspace{0.5 cm}
\caption{Relevant Feynman diagrams that contribute to  $e^+e^-\to {\rm hadrons}$ process, mediated by the photon (a) and $X$ (b), and thus being relevant for the HVP test as well as the $a_\mu$ test when the data-driven method is used for the theoretical value of photon HVP. 
}
\label{Fig:feynmann_diagrams_xsection}
\end{figure}

To explore the phenomenological aspects of our observations, we consider two models of a new light gauge boson, namely a dark photon extension ($X=A'$)~\cite{Okun:1982xi, Holdom:1985ag, Galison:1983pa, Pospelov:2008zw,Arkani-Hamed:2008hhe,Bjorken:2009mm} and baryon number gauge boson one ($X=B$)~\cite{Carone:1994aa, Aranda:1998fr,Tulin:2014tya}. 
These models offer complementary perspectives with different hierarchies of leptonic and  hadronic charges, and isospin ($I$) representations.   
The dark photon has a lepton coupling similar to that of the quarks and has both $I=1$ and $I=0$ isospin components.  
Baryon number gauge boson on the other hand, couples predominantly to the quarks and has $I=0$, with a small coupling to the leptons being generated through loop-induced kinetic mixing with the photon.

We supplement the above tests with a preliminary resonant search for the baryon number gauge boson in the $3\pi$ decay channel. 
We show that with on-tape BaBar and Belle data, see e.g.~\cite{BaBar:2021cde,Belle-II:2024msd}, such a bump hunt in events with initial state radiation (ISR) can lead to potentially interesting results, and can possibly rule out the parameter space of the model.

The rest of the paper is organized as follows. 
In Section~\ref{section:gen-qft}, we develop the formalism for the $\amu$ and the HVP tests for a generic gauge boson $X$. 
We introduce the quantities relevant to those tests, outline the relevant computational methodologies, and discuss field theoretic insights alongside perturbative estimates in this section. In Section~\ref{sec:constraint}, we analyze existing constraints that are relevant to our study. 
In Section~\ref{sec:models}, we consider two benchmark new physics (NP) models: dark photon (Section~\ref{subsec:dp}) and baryon number gauge boson (Section~\ref{subsec:baryon_number}) and show how they fare against both tests. 
We supplement Section~\ref{subsec:baryon_number} by providing a preliminary analysis of a bump hunt in the $\sigma(e^+e^-\to 3\pi)$ data. 
In Section~\ref{sec:scale_settings}, we discuss possible new physics effects in the lattice scale setting procedure, whereas in Section~\ref{sec:fine_grained_HVP}, we consider the window observables of the lattice simulations and consider the HVP test in various windows. 
We conclude in Section~\ref{sec:conclusion}.

\section{Generalities of the $\amu$ and HVP test}
\label{section:gen-qft}

In this section, we develop the formalism for various quantities related to the HVP and the $a_\mu$ tests for a flavor-universal, but otherwise generic, gauge boson $X$. 
The SM fermions couple to the new gauge boson as $\cL\supset g_X X_\mu J^\mu_X$ with
$J^\mu_X = \sum_{\psi=\ell,u,d} Q_\psi^X\bar{\psi}\gamma^\mu \psi$ where $\ell$, $u$, $d$ denote charged leptons, up and down-type quarks, respectively. 
We define $g_\psi^X=g_X Q_\psi^X$. More explicitly, we define $g_q=g_u^X/Q_u^X = g_d^X/Q_d^X$ and $g_\ell=g_\l^X$, and treat them as independent model parameters.

\subsection{$X$ contributions to the HVP and $a_\mu$ test}
\label{subsec:perturbative_case}

\paragraph{HVP test:} 
This test involves reconciling the HVP contribution to $\amu$ from the data-driven approach and the lattice. 
In the data-driven approach, the HVP contribution to $a_\mu$ is obtained using the data of $e^+e^-\to {\rm hadron}$ cross-section:
\begin{align}
    (a^{\rm HVP}_{\mu})^{\rm DD} 
    &=
    \frac{1}{4\pi^3}\int_{s_{\rm th}}^{\infty}ds ~K(s)\, \sigma(e^+e^-\to {\rm had}) \, ,
    \label{eq:HVP_dd}
\end{align}
where $K(s)$ denotes the kernel function, and $s_{\rm th}$ is the integration threshold.  

Including the new massive vector $X$, the total cross section is given by~\cite{DiLuzio:2021uty}
\bea
    \sigma(e^+e^-\to {\rm had})
    = 
    \int d\Pi \, |\mathcal{A}_{e^+e^- \to \gamma^* \to {\rm had}}+\mathcal{A}_{e^+e^- \to X^* \to {\rm had}}|^2
    = 
    \sigma^{\gamma}+ \sigma^{\gamma-X}+\sigma^{X}\,,
\eea
see Fig.~\ref{Fig:feynmann_diagrams_xsection}. 
Accordingly, the data-driven value of $(a^{\rm HVP}_{\mu})^{\rm DD}$ contains three contributions: 
\begin{align}
    \label{eq:HVPdatadriven}
    (a^{\rm HVP}_{\mu})^{\rm DD} 
    &=\frac{1}{4\pi^3}\int_{0}^{\infty}ds ~K(s)\,(\sigma^{\gamma}
    + \sigma^{\gamma-X}+\sigma^{X})\nonumber\\
    &\equiv 
    (a^{\rm HVP}_{\mu})_{e^+e^-\to {\rm had}}^\gamma
    + (a^{\rm HVP}_{\mu})_{e^+e^-\to {\rm had}}^{\gamma-X}
    +(a^{\rm HVP}_{\mu})_{e^+e^-\to {\rm had}}^X \, .
\end{align}

The lattice method for the HVP, on the other hand, is not contaminated by the presence of $X$, and can be identified the same as $(a^{\rm HVP}_{\mu})_{e^+e^-\to {\rm had}}^\gamma$.
Therefore combing Eq.~\eqref{eq:DeltaamuHVP} and Eq.~\eqref{eq:HVPdatadriven} lead to our master formula for the HVP test
\begin{align}
    \label{eq:hvp_test_def}
    \Delta \amu^{\rm HVP} 
    =(a^{\rm HVP}_{\mu})_{e^+e^-\to {\rm had}}^{\gamma-X}
    +(a^{\rm HVP}_{\mu})_{e^+e^-\to {\rm had}}^X\,. 
\end{align}
\paragraph{$a_{\mu}$ test:} 
This test involves reconciling the total $a_{\mu}$ from the experiment and SM through the relation in Eq.~\eqref{eq:Deltaamu}. 
By breaking it into the different new physics contributions (Fig.~\ref{Fig:feynmann_diagrams}) we can write 
\begin{equation}
    \label{eq:amu-test}
    \Delta \amu^{\rm SM} 
    =  a^{X}_{\mu}+a^{\gamma- X}_{\mu}+a^{XX}_{\mu} \, .
\end{equation}
The direct 1-loop $X$ contribution to $(g-2)_\mu$ can be written as,
\begin{equation}
    \label{eq:1_loop_X}
    \amu^{X}  
    = 
    \int\frac{d^4k}{(2\pi)^4} \frac{1}{k^2-m_X^2}\times g_\l^2\, \cF_\mu(k)
    \equiv
    \frac{g_\l^2}{4\pi^2} \, K(m_X^2)\,,
\end{equation}
where $\cF_{\mu}(k)$ encapsulates the rest propagators and vertices of the diagram, as well as a projection of the tensor structure into $a_\mu$, and is common to all diagrams in Fig.~\ref{Fig:feynmann_diagrams}. 
Note that, the kernel used here is the same as in Eq.~\eqref{eq:HVP_dd}. 
This is because by definition $K(s)$ denotes the direct 1-loop contribution to $\amu$ from a massive gauge boson of mass $\sqrt{s}$ up to the coupling to the muons~\cite{Jegerlehner:2017gek}. A direct evaluation of this kernel function gives~\cite{Aoyama:2020ynm,Pospelov:2008zw},
\bea
    \label{eq:1loop_g_2_mu}
    \amu^X 
    &=& 
    \frac{{g_\l}^2}{4\pi^2} K(m_{X}^2)
    =
    \frac{g_\l^2}{4\pi^2}\int_{0}^{1} dz\, \frac{ z(1-z)^2}{(1-z)^2+ z\, m_X^2/m_\mu^2 }\nn\\ 
    &=& 
    196\times 10^{-11}\left(\frac{g_\l}{3\times 10^{-3}}\right)^2
    \left(\frac{0.6\,\GeV}{m_X}\right)^2\,,
\eea
where the last step holds for $m_X\gg m_\mu$. 

As is well known in the literature, a direct $X$ coupling to the muons reinstates the more than $5\sigma$ tension between the TI 2020 and the experimental result~\cite{Aoyama:2020ynm}. 
However, we notice that including only 1-loop contribution to ameliorate the $g-2$ anomaly may lead to the wrong conclusion, as the same coupling generates the 2-loop $\gamma-X$ mixing term  $\amu^{ \gamma-X}$ as well. 
To see the importance of this effect, let us consider the contribution to $\amu^{\gamma-X}$ from hadronic states in the blob, denoted as $(\amu^{\rm HVP})^{ \gamma-X}$. 
Assuming $Q^X_u\sim Q^X_d\sim \cO(1)$, and including the color factor and the EM charges of the quarks, a perturbative estimates  
gives 
\begin{align}
    \label{eq:amu_estimates_gammaX}
    (a_{ \mu }^{\rm HVP})^{ \gamma-X} 
    &\sim  
    2 \times 3 \times \left( \frac{2}{3} - \frac{1}{3} \right) 
    \frac{ e^2 }{ 16 \pi^2 }\frac{ g_\l \; g_q }{ 16 \pi^2 } 
    \frac{ m^2_{ \mu } }{ m_X^2 } \nn\\
    &\sim 10\times 10^{ -11} \left( \frac{ g_\l }{ 3 \times 10^{ -3 } } \right)
    \left( \frac{ g_q } { 0.15 } \right) \left(\frac{ 0.6\GeV}{m_X} \right)^2\,,
\end{align}
for $m_X \gg m_\mu$. 
These rough estimates show that the mixing term could be as important as the 1-loop $X$ contribution if $g_\l/g_q\lesssim 10^{-3}$. 
Thus one expects that the 2-loop effects will be important if the gauge boson coupling to the hadrons is large compared to that of the leptons.

Considering the non-perturbative nature of the $\gamma-X$ mixing through hadronic states, to calculate $(a_{ \mu }^{\rm HVP})^{ \gamma-X}$ properly, we use a dispersive method similar to the case of SM photon HVP calculation. 
Generally, the hadronic blob, defined as the momentum-space correlator of $J^\nu_{X,q}$ and the EM current $J_{\rm EM}^\mu$, can be written as
$\Pi^{\mu \nu}_{\gamma X}(k^2)=(k^2 g^{\mu\nu}-k^{\mu}k^{\nu})\Pi_{\gamma X}(k^2)+ k^{\mu}k^{\nu}\Pi^L_{\gamma X}(k^2)$ with $\Pi^L_{\gamma X}(k^2)$ being the longitudinal part~\cite{Peskin:1995ev}. 
As shown in Appendix~\ref{app:long-contributions}, the terms proportional to $k^\mu k^\nu$ do not contribute to $(g-2)$, and thus we write 
\begin{equation}
    \label{eqn:gam-X-def}
    (a_{ \mu }^{\rm HVP})^{ \gamma-X} 
    =
    \int\frac{d^4k}{(2\pi)^4}\frac{2e\,g_q\, {\rm Re}(\bar{\Pi}_{\gamma X}(k^2))}
    {(k^2-m_X^2)} \times e g_\ell \mathcal{F}_{\mu} (k)\,,  
\end{equation}
where the bar on $\bar{\Pi}_{\gamma X}$ means it is renormalized. 
Like the case of photon HVP, we can use dispersion relation to obtain $\bar{\Pi}_{\gamma X}$ from its imaginary part ${\rm Im}(\bar{\Pi}_{\gamma X})$, and recast ${\rm Im}(\bar{\Pi}_{\gamma X})$ in terms of the interference contribution in the $e^+e^-\to {\rm hadron}$ cross-section $\sigma^{\gamma-X}$. In the end, we obtain the simple relation:
\begin{equation}
    \label{eq:amu_mix}
    (\amu^{\rm HVP})_{e^+e^-\to {\rm had}}^{\gamma-X}=(\amu^{\rm HVP})^{ \gamma-X}
    \approx \amu^{\gamma-X}\,,
\end{equation}
where the last equality follows from the fact that $\gamma-X$ mixing through leptonic states is always much smaller than hadronic states at least by a factor of $(g_\l/g_q)/4\pi$. 
We note that the above form of the equation is only valid when we renormalize $\Pi_{\gamma X}(k^2)$ at $k^2 = m_X^2$. 
The renormalization scheme independent physical observable is $a^{\gamma-X}_{\mu}+\amu^X$, a result we show in Appendix~\ref{app:renorm-independent}.

Lastly, similar to $\amu^{\gamma-X}$, a perturbative estimate for $\amu^{XX}$ provides
\begin{eqnarray}
    \amu^{XX} 
    & \sim & 
    \frac{ g^2_\ell }{ 16 \pi^2 }\frac{ g^2_q }{ 16 \pi^2 } \frac{ m^2_{ \mu } }{ m_X^2 }
    \sim  
    0.02 \times 10^{-11} \left( \frac{ g_\ell }{ 3 \times 10^{ -3 } } \right)^2 
    \left( \frac{ 0.6\GeV}{ m_X } \right)^2\ll \amu^{\gamma-X}\,,
\end{eqnarray}
for $g_q=0.15$. 
Thus $\amu^{XX}$ is parametrically much smaller than the other two contributions, and beyond the current precision of the measurement. 

At this point, one might naively expect a relation $(\amu^{\text{HVP}})^{X}_{e^+e^- \to \text{had}}\sim a_\mu^{XX}$, similar to Eq.~\eqref{eq:amu_mix}, and conclude that $(\amu^{\text{HVP}})^{X}_{e^+e^- \to \text{had}}$ is also negligible in the HVP test. However, this is not true. To see this, notice that under the narrow width approximation $\Gamma_X \ll m_X$, $\sigma^X$ is dominated by on-shell $X$ contribution:
  $\sigma^X \simeq \sigma(e^+e^- \rightarrow X){\BR}(X\to {\rm had})$. This leads to:
\begin{align}
    (a^{\rm HVP}_{\mu})_{e^+e^-\to {\rm had}}^X &\simeq {\BR}(X\to {\rm had})\frac{1}{4\pi^3}\int_{s_{\rm th}}^{\infty}ds ~K(s)\, \sigma(e^+e^- \rightarrow X)\nn\\
    &=a_\mu^X {\BR}(X\to {\rm had})\,.
    \label{eq:amu-onshell}
\end{align}
With an estimation for both $\amu^X$ and $(\amu^{\rm HVP})^{ \gamma-X}$, and using Eq.~\eqref{eq:amu_mix} we have
\begin{align}
\label{eq:amu_estimates_onshellx2}
\frac{(\amu^{\text{HVP}})^{X}_{e^+e^- \to \text{had}}}{\amu^{\gamma-X}}&\sim \left(\frac{8\pi^2}{e^2} \times \frac{g_\l}{g_q}\right) {\BR}(X\to {\rm had})\sim 17 \left(\frac{g_\l}{3\times 10^{-3}}\right) \left(\frac{0.15}{g_q}\right){\BR}(X\to {\rm had})\,.
\end{align}
Therefore, due to the on-shell enhancement, $(\amu^{\text{HVP}})^{X}_{e^+e^- \to \text{had}}$ can  become larger than $\amu^{\gamma-X}$ and so, contributes significantly to the HVP test. 

For the value of $\amu^{\rm SM}$ in the $\amu$ test, one can use either lattice or data-driven results for the HVP part. 
For the first choice, we have using Eq.~\eqref{eq:Deltaamu}
\begin{equation}
    \label{eqn:amu-lattice-def}
    \Delta\amu^{\rm lat} 
   =
   a^{X}_{\mu}+a^{\gamma- X}_{\mu}+ a^{XX}_{\mu}\simeq
   a^{X}_{\mu}+a^{\gamma- X}_{\mu}\, , 
\end{equation}
which we denote as the ``lattice" $\amu$ test. In the last equality, we ignore the small $a^{XX}_{\mu}$ contribution. 
For the second choice, notice that $a^{\text{DD}}_{\mu}$ is not a pure SM contribution, see Eq.~\eqref{eq:hvp_test_def}, we have
\bea
   \Delta\amu^{\text{DD}}
   =
   \amu^{\text{exp}}-\amu^{\text{DD}}
   &=&a^{X}_{\mu}-(a^{\rm HVP}_{\mu})_{e^+e^-\to {\rm had}}^X+a^{\gamma- X}_{\mu}-(a^{\rm HVP}_{\mu})_{e^+e^-\to {\rm had}}^{\gamma-X}+\amu^{XX}\nn\\
   &\approx& a^{X}_{\mu}-(a^{\rm HVP}_{\mu})_{e^+e^-\to {\rm had}}^X+\amu^{XX}\,.
   \label{eqn:amutest:data}
\eea
where we obtain the last line using Eq.~\eqref{eq:amu_mix}. This is denoted as the ``data driven" $\amu$ test. 
In addition, under the narrow width approximation, we can use Eq.~\eqref{eq:amu-onshell}
and notice significant cancellation in the DD $a_\mu$ test, namely: 
\begin{equation}
\Delta\amu^{\rm DD} 
\simeq a_\mu^X \left[1-{\BR}(X\to {\rm had}) \right]\,,  \label{eqn:amutest:data:sim}
\end{equation}
where we ignore the small $\amu^{XX}$ contribution. 
The above equation shows that a gauge boson which mostly decays to the hadrons i.e. ${\BR}({X\to {\rm had}})\sim 1$, is not sensitive to the data-driven $a_\mu$ test at all.
Even for models such as the dark photon, this consideration is significant, as the branching ratio to hadrons is approximately $\mathcal{O}(80\%-90\%)$ ($\sim\mathcal{O}(50\%)$)
when the dark photon mass is near (away from) the hadronic resonances. 
We expect a similar effect for other observables which use data-driven method to estimate SM non perturbative contributions as in muonium spectroscopy, running of $\alpha$ and $\sin^2\theta_W$ (see~\cite{DiLuzio:2024sps} for a related discussion).

\subsection{Explicit calculations }\label{sec:explicit_calculation}

As we showed above, the crucial input for both the HVP and the $\amu$ tests is the $X$ contribution to the $\sigma(e^+ e^-\to {\rm hadrons})$ cross section, i.e. $\sigma^{\gamma-X}$ and $\sigma^{X}$. Here, we discuss how to determine them from data. This can be done by calculating the relative ratio between the $e^+ e^- \rightarrow X^{*} \rightarrow Y$ and $e^+ e^- \rightarrow \gamma^* \rightarrow Y$  amplitudes with $Y$ an SM final state, which can be obtained by calculating the $X$ mixing with the relevant vector meson, see~\cite{Bjorken:2009mm,Ilten:2018crw,DiLuzio:2021uty} for this approach. 

Depending on the Isospin of $X$, it can interfere with the $\rho$ ($I=1$) meson and contribute to $2\pi$ final state, and/or contribute to $3\pi$, $KK$ etc final states by interfering with the Iso-singlet ($I=0$) vector mesons ($\omega,\phi$).
For instance, for $Y=2\pi$, the ratio is obtained by calculating the $X$ mixing with the $\rho$ meson current, $\rho^\mu \sim (\bar{u} \gamma^\mu u -\bar{d} \gamma^\mu d)$. It can be written as
\begin{equation}
    \label{eq:ee_X_rho}
    \frac{\cA_{e^+ e^- \rightarrow X^{*} \rightarrow 2\pi}}{\cA_{e^+ e^- \rightarrow \gamma^* \rightarrow 2\pi}}
    =
    \frac{\epsilon_\rho\, s}{s-m_{X}^2+im_{X}\Gamma_{X}}\,,
\end{equation}
where, $\epsilon_\rho = g_\l g_q(Q^X_u-Q^X_d)/e^2$.  
Thus the $X$ contribution to $\sigma(e^+ e^- \rightarrow 2\pi)$ can be obtained as,
\bea
   \!\!\!\!\!   \!\!\!\!\!\label{eq:sigma_X_rho}
    \sigma_{2\pi}^{\gamma-X}(s) 
    = \sigma^{\rm SM}_{2\pi} (s) \! \times \!
    \frac{2\epsilon_\rho s (s-m_{X}^2)}{(s-m_{X}^2)^2+m_{X}^2 \Gamma_{X}^2}\,,
    \quad
    \sigma_{2\pi}^{X}(s) 
    = \sigma^{\rm SM}_{2\pi} (s) \! \times \!
    \frac{\epsilon_\rho^2 s^2}{(s-m_{X}^2)^2+m_{X}^2 \Gamma_{X}^2}\,,
\eea
where $\sigma^{\rm SM}_{2\pi}=\sigma(e^+e^-\to\gamma^*\to 2\pi)$ denotes the cross-section through the SM photon, which can be safely approximated using the experimental data. $\sigma_{2\pi}^{\gamma-X}$ and $\sigma_{2\pi}^{X}$  
denotes the SM- and $X$ interference term and pure (on- and off- shell) $X$ contribution respectively.   

Similarly, the $X$ contribution to $\sigma(e^+ e^- \rightarrow 3\pi)$ can be obtained by looking at the mixing with $\omega$ meson ($\omega^\mu \sim \bar{u} \gamma^\mu u +\bar{d} \gamma^\mu d)$, $\phi$ meson ($\phi^\mu \sim \bar{s} \gamma^\mu s$), and the $\omega-\phi$ contribution with the substitution of $\sigma^{\rm SM}_{2\pi}\to \sigma^{\rm SM}_{3\pi}$ in respective $\omega$, $\phi$ and/or $\omega-\phi$ channels along-with 
$\epsilon_\rho\to \epsilon_\omega= 3 g_\l g_q (Q^X_u+Q^X_d)/e^2$, $\epsilon_\rho\to \epsilon_\phi= -3 g_q g_\l Q^X_d/e^2$, and $\epsilon_\rho\to \epsilon_{\omega\phi}=3 g_\l g_q(Q^X_u+Q^X_d) /(2 e^2)$ respectively, as discussed in~\cite{Ilten:2018crw}. 
For $m_X\gtrsim \GeV$, it can also contribute to $e^+ e^- \rightarrow K \bar K$ cross-section. 
The NP contribution to $\sigma(e^+ e^- \rightarrow K \bar K)$ can be obtained by looking at the mixing with $\phi$  meson with the substitution of $\epsilon_\rho\to \epsilon_\phi$ and $\sigma^{\rm SM}_{2\pi}\to \sigma^{\rm SM,\,\phi}_{K\bar K}$. We obtain the SM contribution to $\sigma(e^+e^-\to \rm had)$ through the ratio
\begin{equation}
\label{eq:r_ratio}    
\mathcal{R}_{\mu} = \frac{\sigma(e^+e^- \to \rm had)}{\sigma(e^+e^- \to \rm \mu^+\mu^-)}~.
\end{equation}
In computations of the HVP, we need to use the vacuum polarization (VP) subtracted cross-section, also known as the undressed cross section \cite{Jegerlehner:2017gek}. This is to avoid double counting between the leading and higher order HVP contributions. We use the tree-level value of $\sigma(e^+e^-\to \mu^+\mu^-)$ for the denominator of Eq.~\ref{eq:r_ratio}, which does not need undressing. We incur a small error (at higher order in QED) due to this approximation. However, since the NP contribution to $\sigma(e^+e^-\to \rm had)$ is itself a small correction to the SM contribution, the effect of the error from undressing, to the NP cross section, is doubly suppressed.
One can obtain the decay width of $X$ following the procedure outlined in~\cite{Ilten:2018crw}. 
The partial width of $X$ to the leptons can be calculated perturbatively, while the hadronic for the hadronic width, we use the similar procedure of computing $X$ contribution to  $\sigma(e^+e^-\to {\rm had})$ in various final states. 

\begin{figure}
    \centering
    \begin{subfigure}{0.45\textwidth}
        \includegraphics[width=\textwidth]{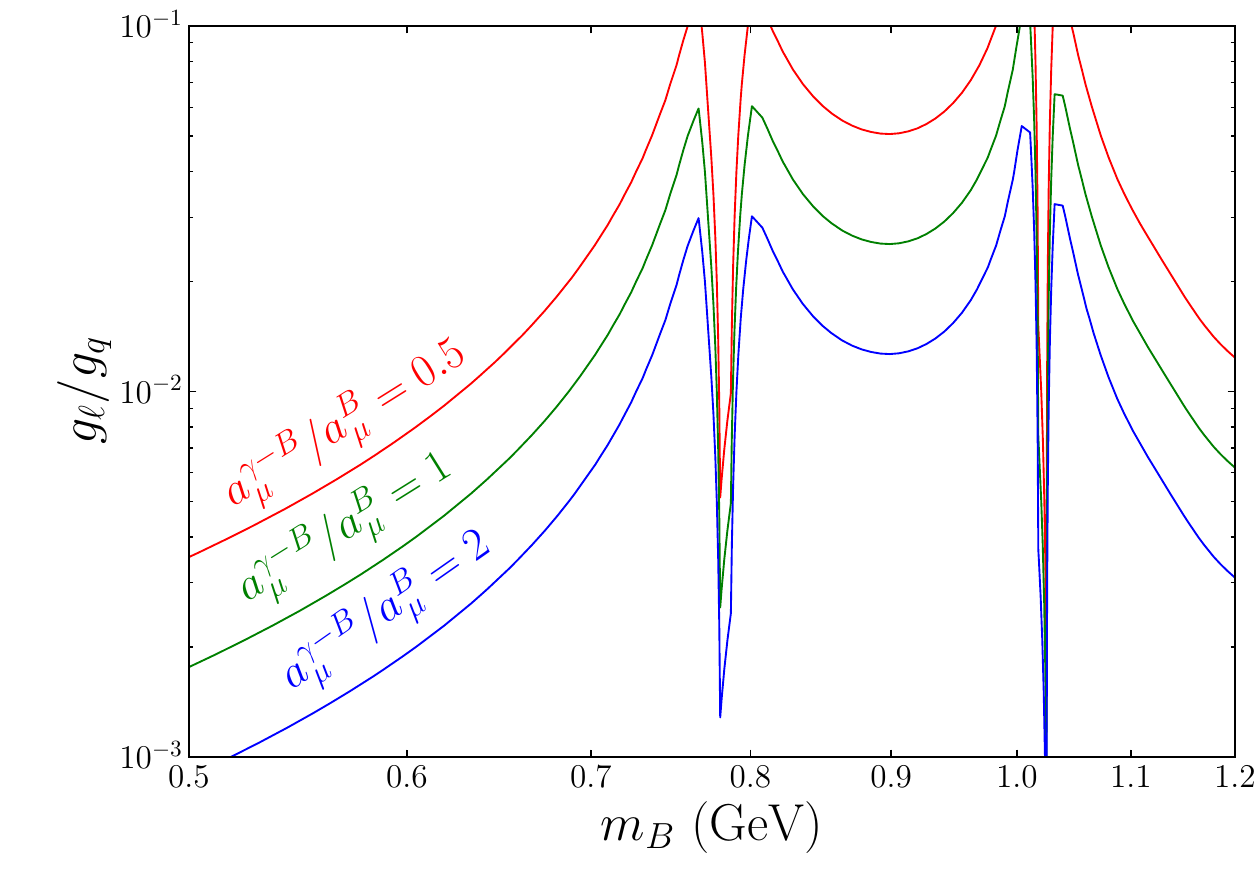}
        \caption{}
        \label{fig:amu_ratio_contour}
    \end{subfigure}
    \begin{subfigure}{0.45\textwidth}
    \includegraphics[width=\textwidth]{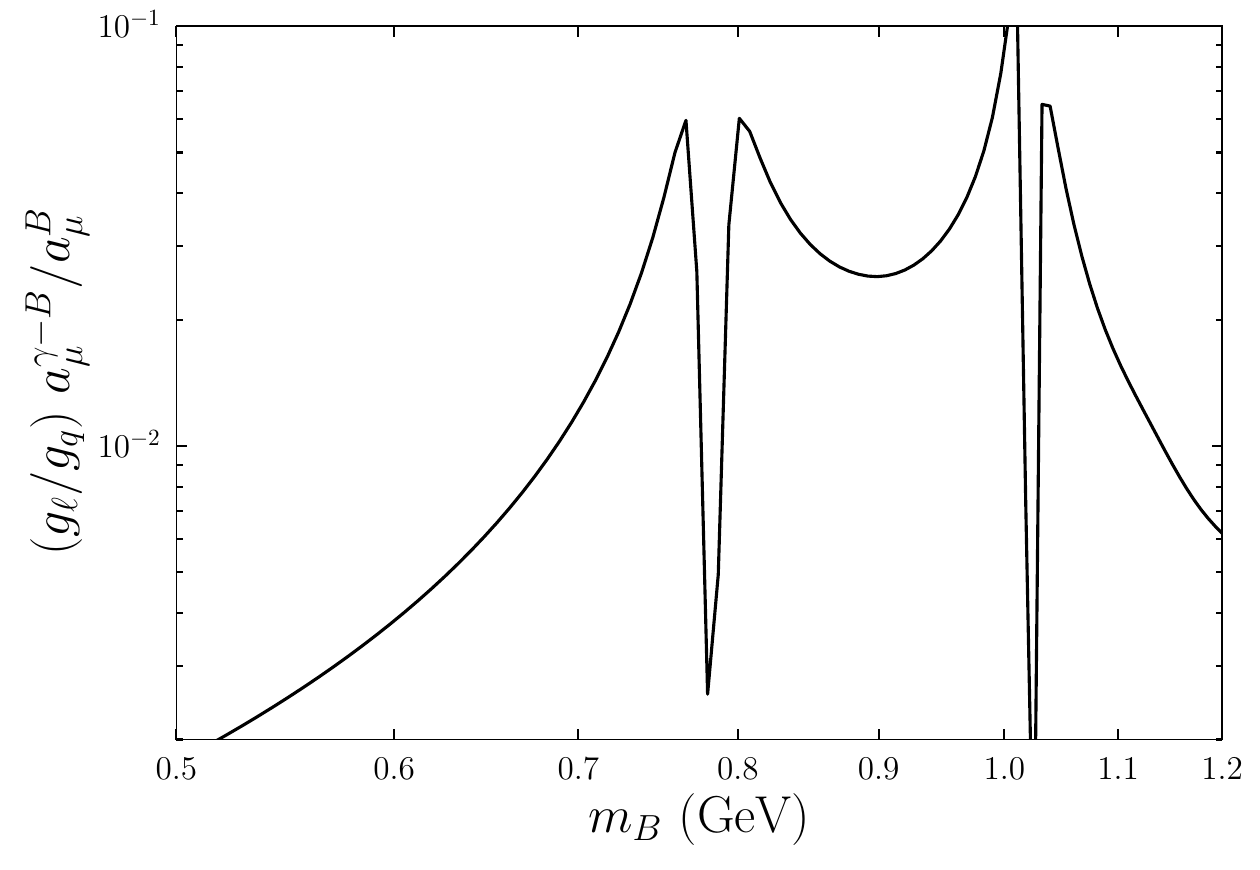}
        \caption{}
        \label{fig:a_mu_ratio_mX}
    \end{subfigure}
    \caption{Plots of the ratio of the $1$-loop and the $\gamma-X$ interference contribution to the $(g-2)_\mu$. In Fig.~(a), we show how does the ratio of the couplings $g_\l/g_q$ changes as a function of $m_X$ ($X=B$ boson) for fixed values of $\amu^{\gamma-X}/\amu^{X}$, whereas Fig.~(b) depicts $\amu^{\gamma-X}/\amu^{X}$ as a function of $m_X$ for fixed values of $g_\l/g_q$. 
    A mild dependence on individual coupling enters through the decay width of $X$ for $m_X$ close to any hadronic resonances. 
    We take $Q^{X}_{u}=Q^{X}_{d}$ and $g_q=0.15$ to obtain these plots. }
   \label{fig:amu_ratio}
\end{figure}

Next, we demonstrate the importance of the $\gamma-X$ mixing contribution to $a_\mu$, which is relevant when $g_\ell \ll g_q$, beyond the above naive estimation.  
For simplicity, we assume that $X$ has universal quark couplings, as in the case of $X=B$. 
We calculate $\amu^{\gamma-X}$ with the data-driven method and we plot the ratio of $\amu^{ \gamma-X}/\amu^X$ in Fig.~\ref{fig:amu_ratio}, for various values of the couplings (i.e. $g_q$ and $g_\l$) and $m_X$. 

Perturbative estimates show that these two term scale as
$\amu^{ \gamma-X}/\amu^X \sim (\alpha/2\pi) (g_q/g_\l)$ which is comparable only for $g_\l/g_q\sim  10^{-3}$. 
However, the hadronic blob enhances the $\amu ^{\gamma-X}$ contribution in a non-trivial $m_X$ dependent way. 
This behavior is depicted in Fig.~\ref{fig:amu_ratio_contour} where we show contours of $\amu^{ \gamma-X}/\amu^X$ as a function of  $g_\l/g_q$ and $m_X$. 
To understand the $m_X$ dependence clearly, in Fig.~\ref{fig:a_mu_ratio_mX}, we plot $(g_\l/g_q)(\amu^{ \gamma-X}/\amu^X)$ as a function of $m_X$. 
From the above plots, we see that the interference term increases as $m_X$ approaches a hadronic resonance, only to decrease significantly once very close to it. 
This behavior can be understood in the following way. 
From Eq.~\eqref{eq:sigma_X_rho} we see that the interference cross section depends on the product of $\sigma_{\rm had}^{\rm SM}(s)$ and $s(s-m_X^2)/((s-m^2_X)^2+m^2_X \Gamma_X^2)$. 
$\sigma_{\rm had}^{\rm SM}(s)$ increases as $s$ approaches a hadronic resonance, and reaches a maximum for  $\sqrt{s}=m_{\rm had}$ where $m_{\rm had}$ is the mass of the resonance. 
On the other hand, $s(s-m_X^2)/((s-m_X^2)^2+\Gamma_X^2m_X^2)$ obtains maximum contribution for  $\sqrt{s}=m_X\pm \Gamma_X$ rather than $m_{\rm had}$. 
Thus the ratio of $\amu^{ \gamma-X}/\amu^X$ increases as $m_X$ nears a resonance, obtaining a maximum value  around $m_X\pm \Gamma_X\simeq m_{\rm had}$, and then drops sharply afterwards 
as $m_X$ approaches $m_{\rm had}$.  
Away from the resonances, $\amu^{ \gamma-X}/\amu^X$ scales inversely with $g_\l/g_q$ for a fixed $m_X$ as expected from perturbative scaling. 

For $g_\l/g_q\gtrsim 10^{-3}$, both the 1-loop and the mixing term would be small, despite their ratios being comparable. 
Hence, a considerable deviation from the ``naive'' $a_{\mu}$ test i.e. keeping only the 1-loop $X$ contribution in $(g-2)_\mu$ in Eq.~\eqref{eq:amu-test}, is only expected when $\amu^{\gamma-X}/\amu^X\sim \cO(1)$ for larger couplings. 
{\color{black} 
We also note that there are higher loop contributions to the $(g-2)$ such as the one shown in Fig.~\ref{fig:gamma-x-gamma} denoted by $\amu^{\gamma-X-\gamma}$. 
The corresponding contribution in $\sigma(e^+e^-\to\rm had)$ would be the interference between the amplitudes given in Fig.~\ref{fig:dressed_cross-sec}. 
Using Eq.~\eqref{eq:1loop_g_2_mu} and Eq.~\eqref{eq:amu_estimates_gammaX}, the typical size of these contributions can be estimated as $\amu^{\gamma-X-\gamma}/\amu^{X}= (\amu^{\gamma-X}/\amu^{X})^2 \sim (\alpha/(2\pi) \, g_q/g_\ell\, \Pi)^2$, where $\Pi$ captures the effects due to the hadronic blob. 
As shown in Appendix \ref{app:2-loop-renorm}, the one loop contribution to the renormalisation of $g_l$, is also proportional to $(\alpha/2\pi)(g_q/g_\ell) \Pi$, which is required to be less than 1 to maintain perturbativity of $g_\ell$. 
Ignoring the effect of the resonant enhancement which is only $\sim\mathcal{O}(10)$ when $m_X$ is very close to hadronic resonances, the above requirement translates to  a bound of $g_\ell/g_q \gtrsim 10^{-3}$. 
Therefore, these and  higher loop order contributions to $(g-2)$ and/or $\sigma(e^+e^-\to\rm had)$ can be safely neglected. 
 }
\begin{figure}
\captionsetup[subfigure]{justification=centering}
\begin{subfigure}{1\textwidth}
\centering
\begin{tikzpicture}
\begin{scope}
\draw[thick,electron] (5,0) to  [out=-45,in=135] (6.8,-1.8) ;
\draw[thick,electron] (6.8,-1.8) to  [out=-45,in=135] (8.8,-3.8) node[label=$\mu$] {};
\draw[thick,antielectron] (5,0) to [out=-135,in=45] (3.2,-1.8);
\draw[thick,antielectron] (3.8,-1.2) to  [out=-135,in=45] (1.2,-3.8) node[label=$\mu$] {};
\draw[thick,photon] (5,0) to [out=90,in=-90] (5,1.1); 
\draw[thick,photon] (2.4,-2.6) to [out=0,in=-180] (7.6,-2.6) ;
\node[draw,circle,fill=gray,inner sep=1.5pt] at (6,-2.6) {Had};
\node[draw,circle,fill=gray,inner sep=1.5pt] at (4,-2.6) {Had};
\node[] at (5.3,0.6) {$\gamma$};
\node[] at (3,-3.2) {$\gamma$};
\node[] at (5,-3.2) {$X$};
\node[] at (7,-3.2) {$\gamma$};
\end{scope}
\end{tikzpicture}
\caption{}
\label{fig:gamma-x-gamma}
\end{subfigure}

\vspace{1em} 

\begin{subfigure}{1\textwidth}
\centering
\begin{minipage}{0.48\textwidth}
\centering
\begin{tikzpicture}
\draw[thick,electron] (-11.2,-1.2) to [out=135,in=-45] (-13.1,0.7);
\draw[thick,antielectron] (-11.2,-1.2) to  [out=-135,in=45] (-13.1,-3.1);
\draw[thick,photon] (-11.2,-1.2) to [out=0,in=-180] (-9.3,-1.2) ;
\draw[thick,electron] (-9.3,-1.2) to [out=45,in=-135] (-7.4,0.7);
\draw[thick,antielectron] (-9.3,-1.2) to  [out=-45,in=135] (-7.4,-3.1);
\node[] at (-10.2,-1.7) {$\gamma$};
\node[] at (-11.7,0.2) {$e^+$};
\node[] at (-11.8,-2.6) {$e^-$};
\node[] at (-8.5,0.2) {$q$};
\node[] at (-8.5,-2.6) {$q$};
\end{tikzpicture}
\end{minipage}
\hfill
\begin{minipage}{0.48\textwidth}
\centering
\begin{tikzpicture}
\draw[thick,electron] (-1.2,-1.2) to [out=135,in=-45] (-3.1,0.7);
\draw[thick,antielectron] (-1.2,-1.2) to  [out=-135,in=45] (-3.1,-3.1);
\draw[thick,photon] (-1.2,-1.2) to [out=0,in=-180] (1.3,-1.2) ;
\draw[thick,electron] (1.3,-1.2) to [out=45,in=-135] (3.2,0.7);
\draw[thick,antielectron] (1.3,-1.2) to  [out=-45,in=135] (3.2,-3.1);
\node[] at (-0.6,-1.7) {$\gamma$};
\node[] at (0.7,-1.7) {$X$};
\node[] at (-1.7,0.2) {$e^+$};
\node[] at (-1.8,-2.6) {$e^-$};
\node[] at (2.1,0.2) {$q$};
\node[] at (2.1,-2.6) {$q$};
\node[draw,circle,fill=gray,inner sep=1.5pt] at (0.05,-1.2) {Had};
\end{tikzpicture}
\end{minipage}
\caption{}
\label{fig:dressed_cross-sec}
\end{subfigure}

\caption{Correspondence between higher-loop $(g-2)$ contributions (above) and the higher loop $\sigma(e^+e^- \to \rm had )$ contributions (below).}
\end{figure}
\section{Existing constraints} 
\label{sec:constraint}

In this section, we discuss the existing relevant constraints on $X$ in addition to the $(g-2)_\mu$.

\paragraph{$(g-2)$ of electron:} As $X$ couples to electrons, it contributes to the anomalous magnetic moment of the electrons, $(g-2)_e$ as well~\cite{Pospelov:2008zw,Endo:2012hp}. 
By requiring the $X$ induced contribution to be less than $\mathcal{O}(10^{-12})$, we obtain a bound on the lepton coupling as 
\bea
    \label{eq:g_2_e_X}
    g_\l\lesssim 1\times 10^{-2}\,\left(\frac{m_X}{0.6\,\GeV}\right)\, ,
\eea   
where we use Eq.~\eqref{eq:1loop_g_2_mu} with $m_\mu\to m_e$. 
Note that compared to $\amu$, the HVP contribution to the theory prediction of $(g-2)_e$ is small by a factor of $(m_e/m_\mu)^2$, and is well beyond the current measurement uncertainties~\cite{DiLuzio:2024sps}.

\paragraph{Existing collider searches:}  
$X$ couples to the charged leptons, and decays visibly to the SM. 
Thus, the visible dark photon $(A')$ searches from various experiments can be recast to our scenario. In the $X$ mass range of our interest, $m_X\gtrsim 0.5\,\GeV$, the relevant constraints come from various $A'\to \mu^+\mu^-$ searches in BaBar~\cite{BaBar:2014zli}, KLOE~\cite{KLOE-2:2014qxg}, CMS~\cite{CMS:2023hwl} and LHCb~\cite{Ilten:2016tkc,LHCb:2017trq,LHCb:2019vmc}, Belle-II \cite{Campajola:2021pdl}, and from $A'\to \pi^+\pi^-$ search in KLOE~\cite{KLOE-2:2016ydq}. 
We use the \href{https://gitlab.com/darkcast/releases}{DarkCast}~\cite{Ilten:2018crw} framework to calculate the exclusion limits from the collider searches for the 
models we consider in the next section. 
Despite the exquisite sensitivity, due to large QCD background, the collider searches have blind spots close to the hadronic resonances (e.g.~\cite{LHCb:2017trq}), leading to unconstrained parameter space.

\paragraph{Rare Meson decay:} 
Various rare meson decays may constrain the $X$ coupling to the SM quarks. 
These constraints are of relevance if $g_q\gg g_\l$ which is the case for baryon number gauge boson i.e. $X=B$.
In particular, in the presence of an $X$ coupling to the bottom quark, through the s channel exchange, $\Upsilon(1S)$ can decay to hadrons. 
The ratio of the branching fractions $\Delta R_\Upsilon = \Gamma(\Upsilon\to Z^{'*},\gamma^*\to jj)/ \Gamma(\Upsilon\to \mu^+\mu^-)$ can be used to constrain $g_q$~\cite{Carone:1994aa,Aranda:1998fr}. 
From the limit on the non-EM dijet decay of $\Upsilon(1S)$, $\Delta R_\Upsilon\lesssim 2.48$ at $95\%$ C.L.~\cite{ARGUS:1986nzm} excludes a coupling of 
\begin{align}
    g_q\lesssim 0.15 \, ,
\end{align}
in the mass range of $m_X\lesssim m_\Upsilon\sim 10\GeV$. 
Similarly, a rare decay of $\eta\to\pi^0\gamma \gamma$, and/or $\eta'\to \pi^+\pi^-\pi^0\gamma$ can be used to constrain $g_q$ as well. 
For some specific $m_X$, these processes yield stringent constraints on $g_q$ than that of the $\Upsilon$ decay, see~\cite{Tulin:2014tya} for a detailed discussion for the case of $X=B$. 

\paragraph{LEP-II bound:} 
The cross section of $e^+e^-\to q\bar{q}$  is measured with $1\%$ accuracy by the LEP and/or LEP-II collaboration 
~\cite{ALEPH:2013dgf} with an appropriate theoretical prediction~\cite{ParticleDataGroup:2024cfk}.  
In the presence of $X$, (mostly) due to the $\gamma-X$ and $Z-X$ interference, the $e^+e^-\to q\bar q$ cross-section receives an additional contribution.  
By requiring the NP contribution is within the accuracy of the measurement, we obtain a bound on the product of the couplings as,
\bea
    \label{eq:LEP_X}
    \left|g_\l\times g_q (11.3 Q^X_u-8.9 Q^X_d)\right|
    \lesssim  10^{-2}\,.
\eea
As the center of mass energy of LEP is much higher than $m_X$, this bound is independent of $m_X$. 
Note that, for most of the parameter space, the LEP bound is weaker than the already discussed constraints. 

\paragraph{Mass difference of pions:} 
For generic $Q^X_u$ and $Q^X_d$, $X$ breaks Isospin, and thus contribute to the Isospin breaking observables such as the mass splitting of the charged and neutral pions. 
Analogous to the case of SM photon, $X$ contribution to pion mass difference is proportional to $g_q(Q^X_u-Q^X_d)$. 
Thus by rescaling the SM prediction from lattice QCD, and comparing it with the experimentally measured mass splitting at 95\% C.L., one obtains a bound of~\cite{DiLuzio:2021uty,Gagliardi:2021vpv} 
\begin{align}
    |g_q(Q^X_u-Q^X_d)|\lesssim 6\times 10^{-2} \, . 
\end{align}
\section{New Physics Models}
\label{sec:models}
In this section, we apply the $\amu$ and HVP tests to phenomenologically interesting models.  
We consider (i)~the dark photon ($A'$)~\cite{Holdom:1985ag,Pospelov:2008zw,Fabbrichesi:2020wbt} and (ii)~the baryon number gauge boson ($B$)~\cite{Carone:1994aa,Aranda:1998fr,Tulin:2014tya}. 
In all the exclusion plots, the bounds from our proposed HVP and $\amu$ tests are depicted by the dark red and green lines respectively, whereas all the relevant existing bounds (discussed in Sec.~\ref{sec:constraint}) are shown by gray lines and shaded regions. 
The naive $\amu$ test i.e. 1-loop contribution to $\amu$ is depicted by blue dashed lines. 

\subsection{Dark Photon}
\label{subsec:dp}

We consider the dark photon interaction with the SM as $\cL\supset e\varepsilon\, J_{\rm EM}^{\mu} A'_\mu\,$ with $\varepsilon$ being the kinetic mixing~(KM) parameter.
Thus, from the generic case of $X$ boson, we obtain the dark photon phenomenology by substituting $Q^X_u=2/3$, $Q^X_d = -1/3$, $Q^X_l=1$ and $g_\l=g_q=e\varepsilon$. 
The dark photon morphs the properties of the SM photon, and like the SM photon it contributes to all the hadronic final states in the $\sigma(e^+ e^-\to {\rm had})$.   
\begin{figure}[t]
\centering

    \includegraphics[width=0.75\linewidth]{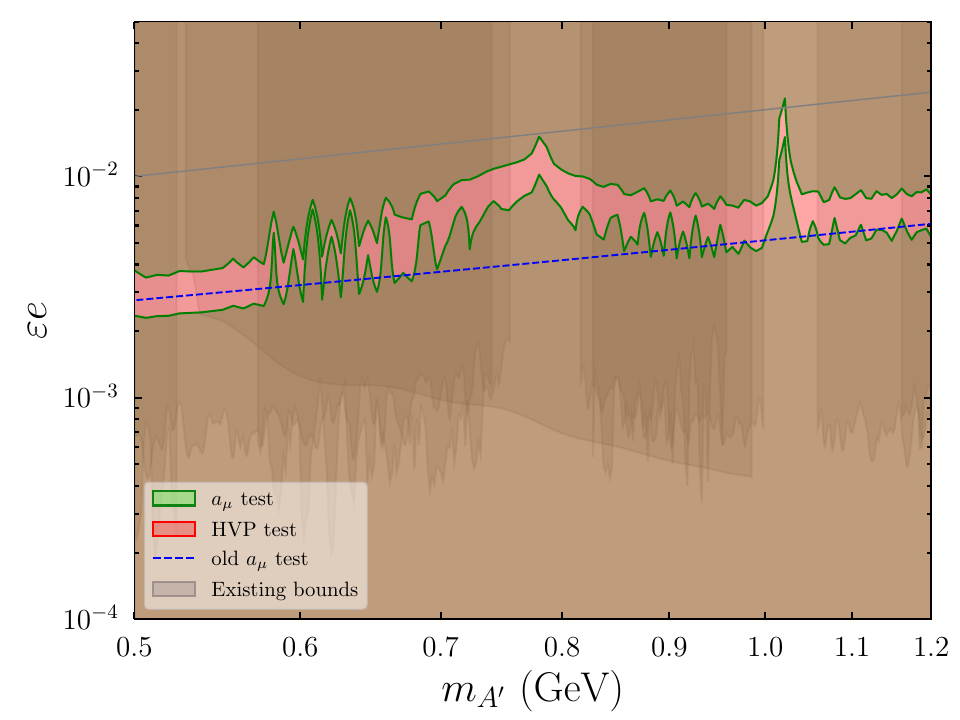}
    \caption{
    Parameter space of the dark photon coupling $\varepsilon e$ as a function of its mass $m_{A'}$ for the data driven $\amu$ test (c.f. Eq.~\eqref{eqn:amutest:data:sim}). We use TI 2020 as the DD value for both the $\amu$ and HVP tests.  The red and green shaded regions are excluded by the $\pm2\sigma$ HVP and $a_{\mu}$ tests, respectively, whereas the gray lines depict the existing bounds see Section~\ref{sec:constraint} for details). {\color{black}The brown colour in the plot is due to the overlap of the green and red shades. Therefore, brown regions are excluded by both the HVP and $\amu$ tests.} The blue dashed line shows the contribution from $\amu^{A'}$ (naive $\amu$ test).
    }
    \label{fig:dark_photon_dd}
\end{figure}
\begin{figure}[t]
    \centering
    \includegraphics[width=0.75\textwidth]{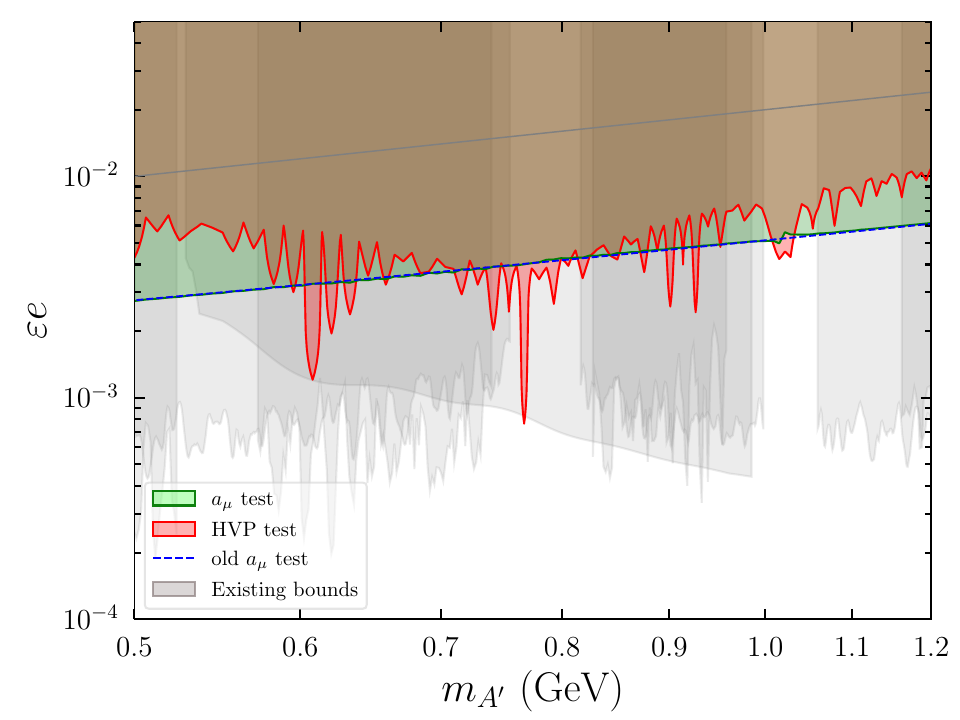}

    \caption{Parameter space of the dark photon coupling $\varepsilon e$ as a function of its mass $m_{A'}$ for the lattice $\amu$ test (c.f. Eq.~\eqref{eqn:amu-lattice-def}). We use BMW result for the $\amu$ test~\cite{Boccaletti:2024guq}. 
    For the HVP test, we take a conservative approach. We consider a possibility that in the future, the central values of the data driven HVP and the lattice result match  i.e. $(a^{\rm HVP}_{\mu})^{\rm DD}=(\amu^{\rm HVP})^{\rm lat}$. 
    The rest of the figure matches the description of  Fig.~\ref{fig:dark_photon_dd} }
    \label{fig:dark_photon_lat}
    \end{figure}
Next, we discuss the result of subjecting the dark photon to the $a_{\mu}$ and HVP tests. In Fig.~\ref{fig:dark_photon_dd} and~\ref{fig:dark_photon_lat}, we show the constraints from the HVP and $\amu$ test in dark red and green respectively, on the parameter space of $\varepsilon e$ and $m_{A'}$. 
In Fig.~\ref{fig:dark_photon_dd}, we consider the data driven  $\amu$ test (c.f. Eq.~\eqref{eqn:amutest:data:sim}).  
We use TI 2020 as the DD value for both the $\amu$ and HVP tests. 
Using them, we exclude hitherto uncharted territory of the dark photon parameter space; even for arbitrarily small coupling. 
This can be understood as follows: to satisfy the HVP test, we need (c.f. Eq.~\ref{eq:hvp_test_def},~\ref{eq:amu_mix},~\ref{eq:amu-onshell}) {$\Delta \amu^{\rm HVP} 
\approx a_{\mu}^{\gamma-A'}
+a_\mu^{A'} {\BR}(A'\to {\rm had})=(-144\pm68) \times 10^{-11}$. 
Since $a_\mu^{A'}$ is always positive, and the interference is small, the HVP test is never satisfied. On the other hand, to satisfy the DD $\amu$ test, a large positive contribution from $a_\mu^{A'}$ is required to resolve the $5\sigma$ tension as (c.f. Eq.~\eqref{eqn:amutest:data:sim}) $\Delta\amu^{\rm DD} = \amu^{A'}[1-\cB(A'\to {\rm had})]=(249\pm 48)\times 10^{-11}$ where we use the DD $\amu$ value from~\cite{Aoyama:2020ynm}. 
Since $\cB({A'\to {\rm had}})\sim 1(\sim 1/2)$ when $m_{A'}$ is near (away)
a hadronic resonance, the cancellation in Eq.~\eqref{eqn:amutest:data:sim} is significant, and thus making it virtually impossible to satisfy the $\amu$ test as well.

In Fig.~\ref{fig:dark_photon_lat},
we consider the lattice version of the $\amu$ test (c.f. Eq.~\eqref{eqn:amu-lattice-def}) as $\Delta\amu^{\rm lat}\approx \amu^{\gamma-A'}+\amu^{A'}=(105\pm 61)\times 10^{-11}$ where we take the BMW 2020 result as lattice value~\cite{Borsanyi:2020mff}.
We find that $\amu$ test improves the bound on the dark photon parameter space by $\cO(10)$ for $m_{A'}$ close to the $\phi$ resonance, where the collider searches have blind spots. 
The blue dashed line depicts the bound from direct 1-loop contribution $\amu^{A'}$, which is almost identical to the full $a_{\mu}$ test. 
This is expected because as $g_q= g_\l=e\varepsilon\ll 1$, the mixing term $\amu^{\gamma-A'}$ is small compared to the 1-loop contribution 
even with the hadronic enhancement factor. 
For the HVP test, we take a conservative approach. 
We consider a possibility that in future the central values of the data driven HVP and the lattice result match with their respective error bars i.e. $(a^{\rm HVP}_{\mu})^{\rm DD}=(\amu^{\rm HVP})^{\rm lat}$. We take error bars to be the current combined value of $68 \times 10^{-11}$. 
We notice that the exclusion from the HVP test overlaps significantly with that of the $\amu$ test. 
We finish the section by noting that, if one considers the possibility of matching the DD HVP to the lattice result within $\pm 1\sigma$ of the current error combined bar i.e. $(a^{\rm HVP}_{\mu})^{\rm DD}=(\amu^{\rm HVP})^{\rm lat}\pm 1\sigma$, the bound from the HVP test improves upon the $\amu$ test by a factor of $\cO(3-4)$. We defer this discussion to Appendix~\ref{app:additional_figures_dd_lat}.

\subsection{Baryon Number Gauge Boson}
\label{subsec:baryon_number}

Next, we discuss the second model -- gauging the global baryon number $\U1B$ i.e. $X=B$. 
At the tree level $B$ couples to the quarks, whereas a small coupling to the leptons is generated through the kinetic mixing with the hypercharge gauge boson. 
Thus, unlike the case of dark photon, the $\gamma-B$ mixing contribution would be relevant for both the $\amu$ and the HVP test. 
Below the weak scale, we consider the following low-energy Lagrangian as~\cite{Tulin:2014tya} 
\bea
    \label{eq:lag_model}
    \cL
    \supset 
    - \frac{\varepsilon_B}{2}B_{\mu \nu}F^{\mu \nu} 
    + g_q\, B_{\mu} \bar q \gamma^\mu q \, \longrightarrow \, 
    B_{\mu} \left[(g_q+Q^{\rm EM}_q g_\l)\bar q \gamma^\mu q-g_\l \bar \l \gamma^\mu \l\right]   \,,
\eea 
where, $q$ denotes the quarks with $g_q$ being the flavor universal coupling of $B$,  $\varepsilon_B$ is the kinetic mixing between $B$ and photon and the arrow denotes a basis transformation of $A\to A+\varepsilon_B B$ and $B\to B$ where the gauge kinetic terms are diagonalized. 
Upon diagonalizing, we obtain the $B$ coupling to the SM quarks and charged leptons as $g_{u,d}=g_q+ Q^{\rm EM}_{u,d}g_\l\simeq g_q$, and $g_\l$, respectively. Note that there is a minus sign in front of $g_\ell$ in the Lagrangian above because $Q^{\rm EM}_{\ell}=-1$.

To estimate the size of $g_\l$, one can consider a UV completion such that either $U(1)_Y$ and/or $U(1)_B$ is embedded in a non-abelian gauge group at some scale $\Lambda$ making  $\eb(\Lambda)=0$. 
In such cases, the natural size of the lepton coupling can be estimated to be $g_\l\gtrsim \alpha\,g_q/(4\pi)[c_1\log(\Lambda/v_{\rm EW})+c_2\log(v_{\rm EW}/m_B)]$ where $c_i$'s depend on the specific UV model~\cite{Aranda:1998fr}. 
In our case, we are interested in large couplings, and thus, we do not resort to any particular UV completion.
Thus, hierarchical coupling strengths between various SM sectors are obtained naturally in the case of $\U1B$~\cite{Carone:1994aa}. 
We note that, in the basis where $B$ couplings to the quarks are flavor universal, there is no lepton coupling to $B$ but $\varepsilon_B\neq 0$. 
On the other hand, a flavor non universal quark coupling elucidates non zero $g_\l$. 
Thus, only the sum of $\eb$ and $g_\l$ represents a basis independent physical quantity, similar to the case of choosing oblique/non-oblique basis as discussed in~\cite{Agashe:2003zs}. 
In the first case, free theory parameters are $g_q$ and $\varepsilon_B$, whereas in the second case, they are $g_q$ and $g_\l$. 

Another remark is in order. 
There are triangle anomalies involving the mixed $U(1)_B$ and SM gauge currents~\cite{Gross:1972pv}. 
These necessitate the need of adding new physics (anomalons) to cancel the anomalies. 
Introducing anomalons leads to generation of Wess-Zumino terms involving $B$ with SM $W$ and $Z$ bosons, and may result in a much stronger bound on $g_q$ than the previously considered ones for $m_B\sim\cO(\GeV)$~\cite{Dror:2017ehi, Dror:2017nsg}. 
However, these can be evaded by UV model building as discussed in~\cite{DiLuzio:2022ziu, Bonnefoy:2020gyh, Kamenik:2017tnu,Bizot:2015zaa}. 
\begin{figure}[t]
    \centering
     \begin{subfigure}{0.45\textwidth}
    \captionsetup{justification=centering}
    \includegraphics[width=\textwidth]{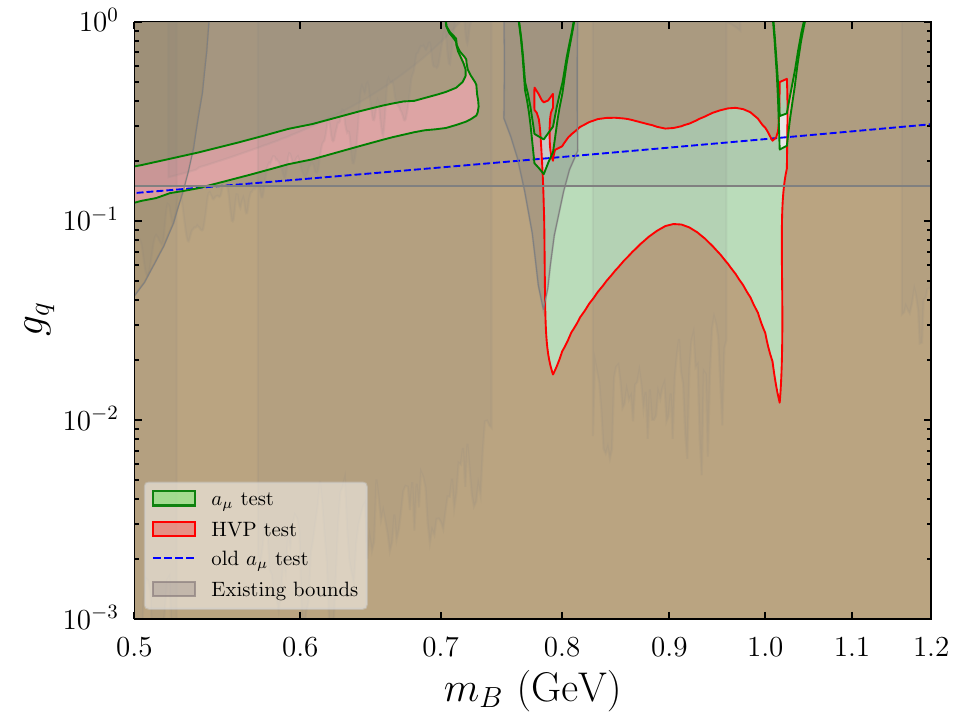}
    \caption{$g_\l/g_q=0.02$}
    \label{fig:gq_mb_SM_dd_02_pos}
    \end{subfigure}~
    \begin{subfigure}{0.45\textwidth}
    \captionsetup{justification=centering}
    \includegraphics[width=\textwidth]{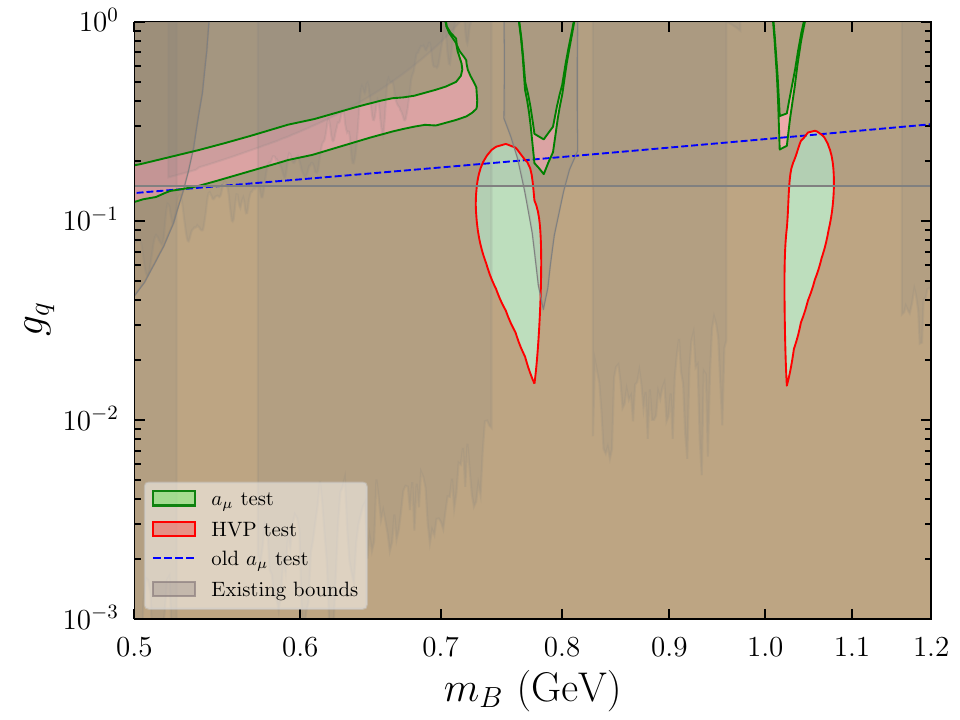}
    \caption{$g_\l/g_q=-0.02$}
    \label{fig:gq_mb_SM_dd_02_neg}
    \end{subfigure}
\caption{
    Parameter space of the quark coupling $g_q$ as the function of the mass of \zprime mass ($m_B$) for $g_\l/g_q=0.02$ (Fig.~\ref{fig:gq_mb_SM_dd_02_pos}) and $g_\l/g_q=-0.02$ (Fig.~\ref{fig:gq_mb_SM_dd_02_neg}). 
    In this plot, we consider the data-driven $a_{\mu}$ test (c.f. Eq.~\eqref{eqn:amutest:data}). Both for the HVP test (c.f. Eq.~\eqref{eq:hvp_test_def}) and the $a_{\mu}$, we take TI 2020 as the data driven value. 
    The green and red regions show the $\pm 2\sigma$ exclusion by the $\amu$ and HVP tests respectively. The blue dashed line shows 1-loop B contribution $\amu^B$ (c.f. Eq.~\eqref{eq:1_loop_X}), whereas the gray bands depict the existing constraints (see Sec.~\ref{sec:constraint} for details).
}
\label{fig:gq_mb_SM_dd}
\end{figure}
\begin{figure}[t]
\centering
    \begin{subfigure}{0.45\textwidth}
    \includegraphics[width=\textwidth]{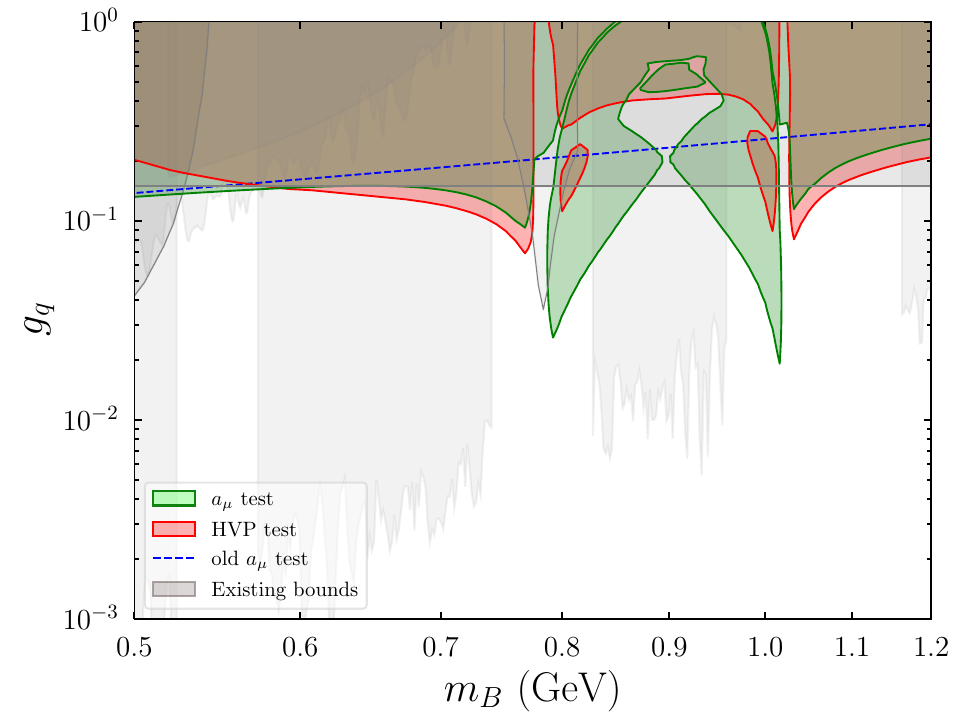}
    \caption{$g_\l/g_q=0.02$}
    \label{fig:first_0_sigma}
    \end{subfigure}~
    \begin{subfigure}{0.45\textwidth}
    \captionsetup{justification=centering}
    \includegraphics[width=\textwidth]{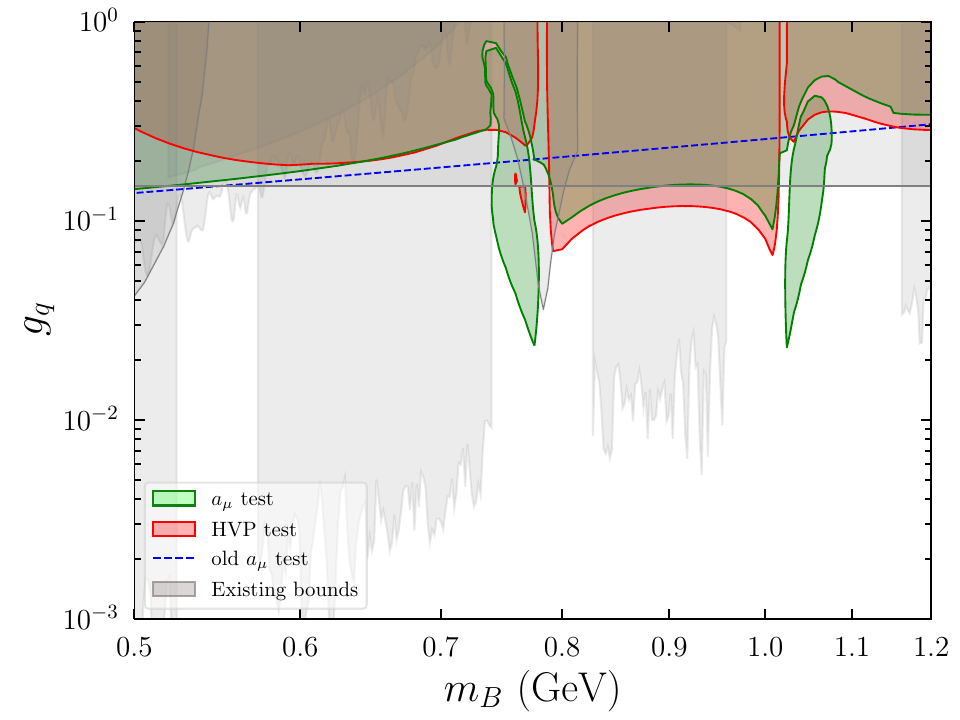}
    \caption{$g_\l/g_q=-0.02$}
    \label{fig:zero_02_neg}
    \end{subfigure}
\caption{ 
    Parameter space of the quark coupling $g_q$ as the function of the mass of \zprime mass ($m_B$) for  $g_\l/g_q=0.02$ (Fig.~\ref{fig:first_0_sigma}) and $g_\l/g_q=-0.02$ (Fig.~\ref{fig:zero_02_neg}). 
    In these plots, we perform the lattice $a_{\mu}$ test (c.f. Eq.~\eqref{eqn:amu-lattice-def}) by taking the BMW result as the SM contribution. 
    For the HVP test (c.f. Eq.~\eqref{eq:hvp_test_def}), we consider a futuristic conservative possibility of the central values of the data driven HVP and the lattice result to match with each other within their respective
    error bars. 
    The rest of the figure matches the description of  Fig.~\ref{fig:gq_mb_SM_dd} 
    }
\label{fig:mb_gq_r_02}
\end{figure}
\begin{figure}[t]
\centering
    \begin{subfigure}{0.45\textwidth}
    \captionsetup{justification=centering}
    \includegraphics[width=\textwidth]{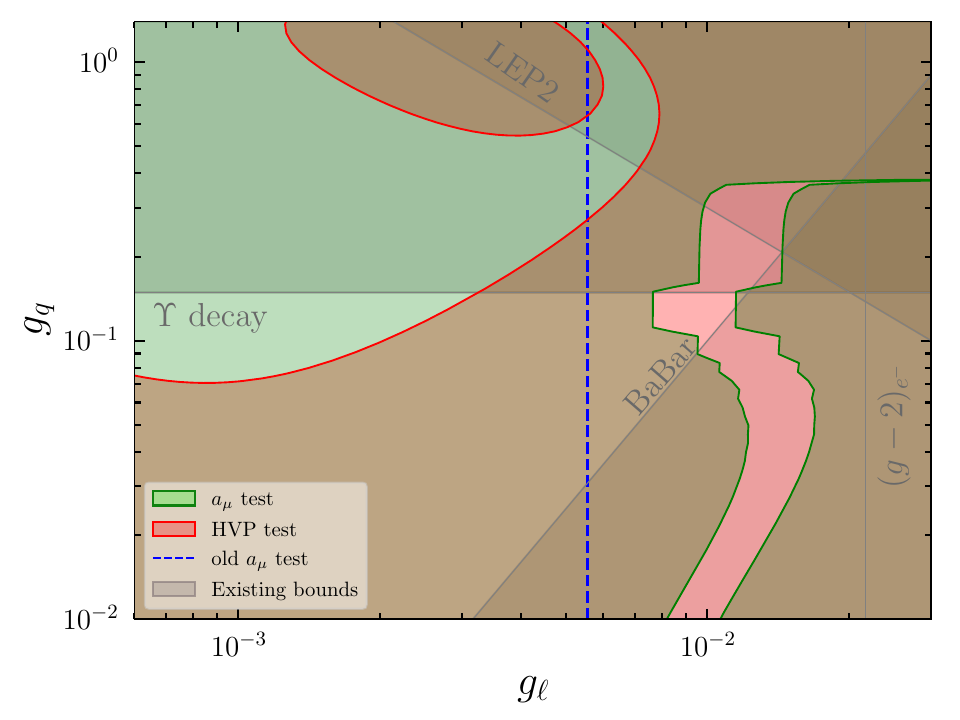}
    \caption{$(a^{ \rm  HVP }_{ \mu })^{\text{DD}}  =  (a^{ \rm  HVP }_{ \mu })^{\text{TI}}$, $g_\l\times g_q>0$}
    \label{fig:gl_gq_dd_sm}
    \end{subfigure}~
    \begin{subfigure}{0.45\textwidth}
    \captionsetup{justification=centering}\includegraphics[width=\textwidth]{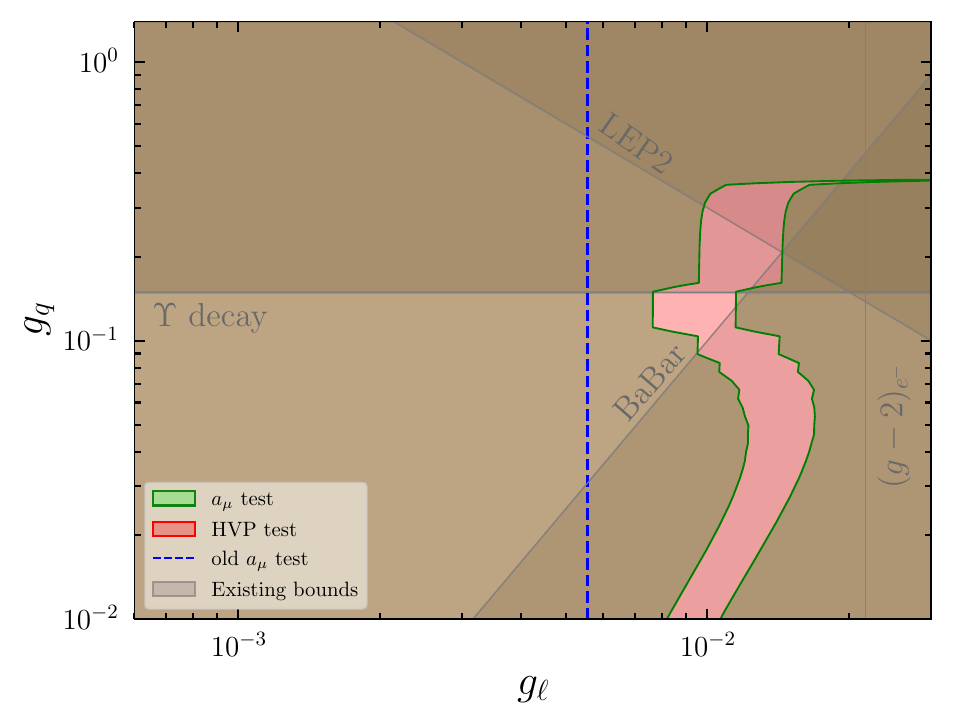}
    \caption{$(a^{ \rm  HVP }_{ \mu })^{\text{DD}} = (a^{ \rm  HVP }_{ \mu })^{\text{TI}}$, $g_\l\times g_q<0$}
    \label{fig:gl_gq_dd_sm_neg}
    \end{subfigure}
    \begin{subfigure}{0.45\textwidth}
    \captionsetup{justification=centering}\includegraphics[width=\textwidth]{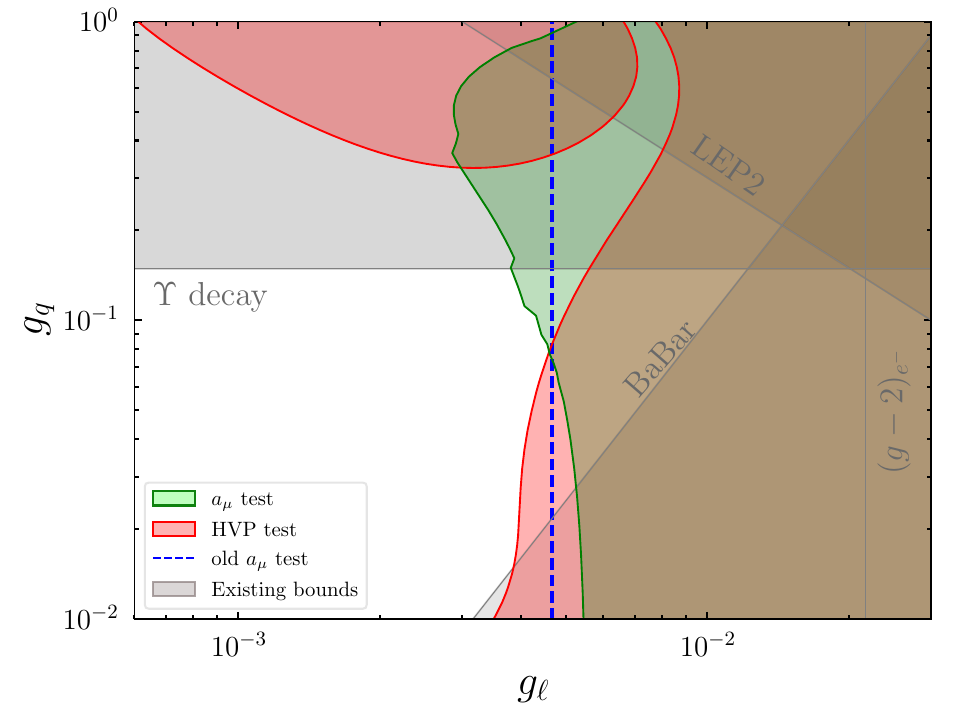}
    \caption{$(a^{ \rm  HVP }_{ \mu })^{\text{DD}}= (a^{ \rm  HVP }_{ \mu })^{\rm{lat}}$, $g_\l\times g_q>0$}
    \label{fig:gl_gq_dd_lat}
    \end{subfigure}~
    \begin{subfigure}{0.45\textwidth}
    \captionsetup{justification=centering}\includegraphics[width=\textwidth]{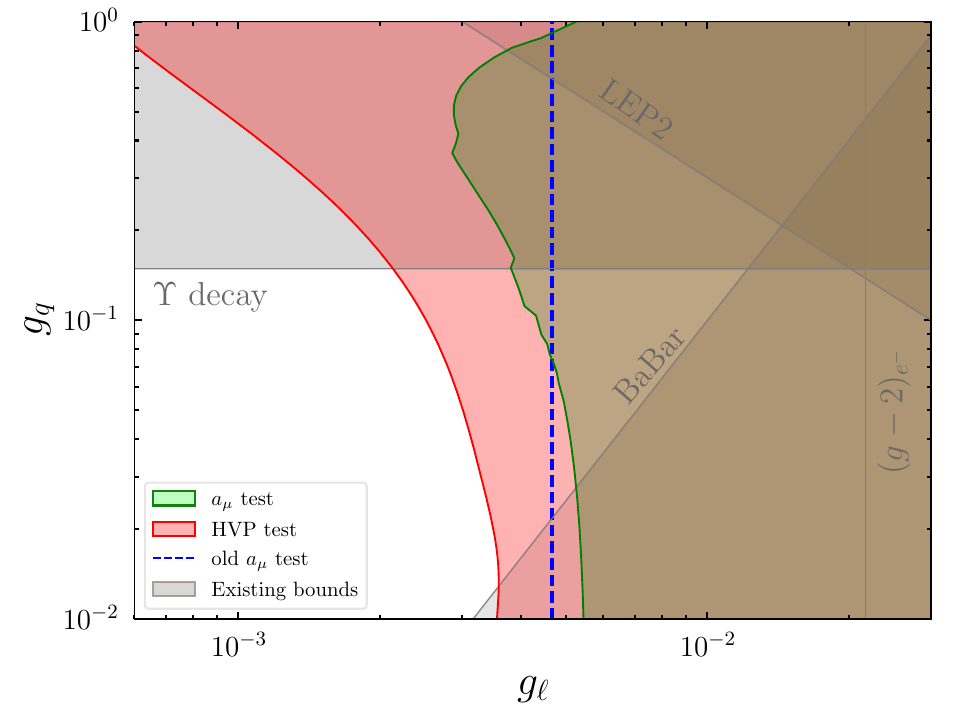}
    \caption{$(a^{ \rm  HVP }_{ \mu })^{\text{DD}}= (a^{ \rm  HVP }_{ \mu })^{\rm{lat}}$, $g_\l\times g_q<0$}
    \label{fig:gl_gq_dd_lat_neg}
    \end{subfigure}
    
    \caption{Plot of $g_q$ vs $g_\l$ for a fixed of $m_B=0.9$ GeV both for positive and negative ratio. 
    For the plots with $g_\l\times g_q<0$, we always consider $g_q>0$ and plot $|g_\l|$. 
    In Fig.~\ref{fig:gl_gq_dd_sm} and~\ref{fig:gl_gq_dd_sm_neg}, we perform the data-driven $a_{\mu}$ test (c.f. \eqref{eqn:amutest:data}) and take the TI 2020 value for the HVP test, whereas in Fig.~\ref{fig:gl_gq_dd_lat} and ~\ref{fig:gl_gq_dd_lat_neg}, we consider the lattice $\amu$ test (c.f. \eqref{eq:amu-test}) and take the data-driven HVP to be equal to the lattice result for the HVP test. 
    The gray lines depict various existing constraints (see Sec.~\ref{sec:constraint} for details).
    The colour scheme is the same as Fig.~\ref{fig:gq_mb_SM_dd}.  
}
\label{fig:gl_gq}
\end{figure}
We obtain the $B$ contribution to $\sigma(e^+e^-\to {\rm had})$ by following the discussion around Eq.~\eqref{eq:ee_X_rho} and~\eqref{eq:sigma_X_rho} and using $Q^X_{u,d}=1$.  
As already discussed, $\U1B$ is mostly Isospin singlet, with a small Isospin breaking contribution coming from the mixing with the SM photon. 
Thus it contributes mostly to the Iso-singlet final states such as $3\pi$, $K\bar K$ etc. 
As can be seen from Eq.~\eqref{eq:ee_X_rho}, $B$ contribution to the $2\pi$ final states is proportional to $\epsilon_\rho\propto g_\l^2 \ll \epsilon_{\omega,\phi}$.

In Fig.~\ref{fig:gq_mb_SM_dd}, we perform the data-driven $\amu$ test (c.f. Eq.~\eqref{eqn:amutest:data}). 
Both for the HVP test (c.f. Eq.~\eqref{eq:hvp_test_def}) and the $\amu$ tests, we take TI 2020 as the data driven value. 
Using them, we exclude hitherto uncharted territory of the $B$ boson parameter space shown by the dark red and green shaded region respectively; even for arbitrarily small couplings. 
To understand this, note that NP is necessary to satisfy either of the tests at the $\pm 2\sigma$ level.  
To satisfy the HVP test, we need (c.f. Eq.~\ref{eq:hvp_test_def},~\ref{eq:amu_mix},~\ref{eq:amu-onshell}) $\Delta \amu^{\rm HVP} 
\approx a_{\mu}^{\gamma-B}
+a_\mu^{B} {\BR}(B\to {\rm had})=(-144\pm 68)\times 10^{-11}$. 
Since ${\BR}(B\to {\rm had})\sim 1$ for $m_B\gtrsim 0.6\GeV$ and $a_\mu^{B}$ is always positive, satisfying the HVP test requires large negative $\amu^{\gamma-B}$. 
As shown in Fig.~\ref{fig:amu_ratio}, the mixing term is only relevant for $g_q/g_\l\gg 1$ with sizable $g_\l$.
Thus, a large negative mixing with not so large $g_\l$ can be obtained close to the hadronic resonances, and we see two small closed shaped regions where the HVP test is satisfied. 
In the rest of the parameter space, we rule out arbitrary small couplings using the HVP test. 
At the same time, satisfying DD $\amu$ test requires (c.f. Eq.~\eqref{eqn:amutest:data:sim})  $\Delta\amu = \amu^{B}[1-\cB(B\to {\rm had})]= (249 \pm 48)\times 10^{-11}$. 
As ${\BR}(B\to {\rm had})\sim 1$ for large $m_B$ and the cancellation is exact in the narrow width approximation, a significant deviation from this is necessary to satisfy the DD $\amu$ test. 
This can be obtained either in the small mass limit, or for large $g_\l$ and $g_q$, or both.
This observation can be validated by the observation that the region of the parameter space that is not shaded in green, requires large and thus previously excluded $g_q$.  
Both the tests exclude arbitrarily small couplings irrespective of the sign of the interference term as can be seen by the overlapping green and red regions in Fig.~\ref{fig:gq_mb_SM_dd_02_pos} and Fig.~\ref{fig:gq_mb_SM_dd_02_neg}.
Exclusion plots for various other values of $g_\l/g_q$ (positive and negative) can be found in Appendix~\ref{app:additional_figues_gq_mb}.   

In  Fig.~\ref{fig:mb_gq_r_02}, 
we consider the lattice $a_{\mu}$ test (c.f. Eq.~\eqref{eqn:amu-lattice-def}). 
Unlike the case of dark photon, we see a considerable deviation from the naive $\amu$ test for $m_B\gtrsim 0.6\GeV$. This is because the mixing term becomes important for $B$ boson as $g_q\gg g_\l$ as shown in Fig.~\ref{fig:amu_ratio}. 
As before (c.f. Fig.~\ref{fig:dark_photon_lat}), we consider the futuristic conservative possibility for the HVP test: $(a^{\rm HVP}_{\mu})^{\rm DD}=(\amu^{\rm HVP})^{\rm lat}$.  
Using both the tests, we improve the sensitivity on the $B$ parameter space by a factor of $4$ where the existing collider searches have limited to zero sensitivities. 
As before red and green shaded regions depict the exclusion limits from the HVP and the $\amu$ test respectively.  
 To understand the exclusion limits, notice that to satisfy the HVP test with $(a^{\rm HVP}_{\mu})^{\rm DD}=(\amu^{\rm HVP})^{\rm lat}$ and the lattice $\amu$ test within the $\pm 2\sigma$ value, no new physics is required, making $g_\l,g_q\to 0$ consistent unlike the case discussed in Fig.~\ref{fig:gq_mb_SM_dd}. 
However, a positive $\amu^{\gamma-B}$ along with a positive $\amu^B$, can provide a NP contribution above $+2\sigma$ value, and excludes the parameter space. 
On the other hand, a large mixing term, can provide a significant negative contribution and exclude the parameter space at $-2\sigma$ value as well. 
This happens for $m_B\sim m_{\rm had}$ and depends on the sign of $g_\l/g_q$ as can be see from both Fig.~\ref{fig:first_0_sigma} and Fig.~\ref{fig:zero_02_neg}. 

From the above plots, we see the complementarity of the HVP test and the $\amu$ test i.e. there are regions of parameter space that satisfy either of the tests but excluded by the other one. To gain more insight, in Fig.~\ref{fig:gl_gq}, we plot $g_q$ as a function of $g_\l$ for a fixed mass. As an example we set $m_B=0.9\GeV$. 
We show similar plots for a different $m_B$ in Appendix~\ref{app:additional_figures_gq_gl}. 
Similar to  Fig.~\ref{fig:gq_mb_SM_dd}, in Fig.~\ref{fig:gl_gq_dd_sm} and~\ref{fig:gl_gq_dd_sm_neg}, we consider the DD $\amu$ test for positive and negative $g_\l/g_q$ respectively with TI 2020 values. 
We see that the $\amu$ test is satisfied only in a small region where $g_\l/g_q$ is large, effectively making the cancellation in Eq.~\eqref{eqn:amutest:data:sim} inexact. 
At the same time, a large $g_\l$ increases the $\amu^X$ contribution (c.f. Eq.~\eqref{eq:1_loop_X}), leading to fulfillment of the test. 
However, as $m_B\gtrsim m_\phi$ and $\epsilon_\phi <0$ for positive couplings, $\sigma^{\gamma-B}$ (c.f. Eq.~\eqref{eq:sigma_X_rho}) and thus, $\amu^{\gamma-B}$ are positive, and HVP test is never satisfied in this case. 
Despite $g_\l\times g_q<0$, we see similar behaviour in Fig.~\ref{fig:gl_gq_dd_sm_neg} as well.  
Similar to Fig.~\ref{fig:mb_gq_r_02}, in Fig.~\ref{fig:gl_gq_dd_lat} and~\ref{fig:gl_gq_dd_lat_neg}, we consider the lattice $\amu$ test to show the parameter space of $g_\l$ vs $g_q$ for $m_B=0.9\GeV$. 
We see that, for regions not already excluded by existing constraints, the HVP test fares slightly better than the $\amu$ test in this case, improving the bound on both $g_\l$ and $g_q$ by more than a factor of $4$. 
Again, as expected, the interference term is relevant when $g_q\gtrsim g_\l$, and for small $g_q$, both the tests approach the bound from the 1-loop contribution. 
This behavior can be seen in the plots by converging red, green and blue dashed lines. 
We finish the section by noting that, if one considers the possibility of matching the DD HVP to the lattice result within $\pm 1\sigma$ of the current combined error bar
i.e. $(\amu^{\rm HVP})^{\rm DD} = (\amu^{\rm HVP})^{\rm lat} \pm 1\sigma$, the complementarity of the HVP test and the $\amu$ test becomes much more manifested and the bounds improve by a factor of $\mathcal{O}(2-4)$. 
We defer this discussion to Appendix~\ref{app:additional_figures_dd_lat}. 

\paragraph{Resonant $B$ search:} 

We end this section by estimating the reach of a bump hunting in the $I=0$ hadronic spectrum~\cite{BaBar:2021cde, Belle-II:2024msd,Fanelli:2016utb}. 
Among other channels~\cite{delRio:2021xag, Fanelli:2016utb}, a bump hunt in $\sigma(e^+e^-\to 3\pi)$ channel can be used to find/rule out the $B$ boson. 

To perform a preliminary analysis, we consider a bump hunt with Gaussian
signal shape and data driven background estimation. 
We obtain the upper-limit (ul) of NP cross section as $\sigma_{\rm NP}^{\rm ul}\lesssim 2 \sqrt{N_{\rm eve}}/(\mathcal{L}_{\rm eff}\times \epsilon_{\rm eff}\times R)\,$, where $N_{\rm eve}$ of a given bin is obtained by summing the number of events in 4 neighboring bins, 
$\mathcal{L}_{\rm eff} $ is obtained by multiplying ISR differential luminosity with the effective energy resolution. $\epsilon_{\rm eff}$ and $R$ are the detection efficiency as a function of mass, and a radiative correction that accounts for the distortion of the mass
spectrum due to emission of several photons respectively. The choice of only taking the 4 neighboring bins is well suited for a narrow resonance, where the signal contribution diminishes rapidly outside the central
bin, provided that the bin width exceeds the intrinsic width of the resonance \cite{Ilten:2016tkc}. In contrast, for broader resonances or in cases where the new physics signal extends beyond a single bin, a statistically significant
excess of events would be observable across multiple adjacent bins relative to the expected background.

First, we consider the BaBar $e^+e^{-}\to 3\pi+\gamma$ measurement~\cite{BaBar:2021cde}. 
In this case, there are correlated errors for $\sqrt{s}\lesssim 1.1\GeV$. 
However, to simplify our analysis, we ignore this effect. 
We consider 3 sample resonance points as $m_B$: a) $0.75\GeV$, b) $0.9\GeV$, and c) $1.1\GeV$. 
By using, $\mathcal{L}_{\rm int}=469 \,{\rm fb}^{-1}$ and $R=1.0077$ as given and taking mass independent detection efficiency to be $\epsilon_{\rm eff}\simeq 11\%$, we obtain an upper limit on $g^{\rm ul}_\l\times \sqrt{\cB(B\to 3\pi)}\lesssim \{ 3\times 10^{-4},\,2\times 10^{-4},\,2\times 10^{-4}\}$ for the above data points respectively. 
Next, we do a similar analysis using Belle-II $e^+e^-\to 3\pi+\gamma$ measurement~\cite{Belle-II:2024msd}. 
We use $\mathcal{L}_{\rm int}=191 \,{\rm fb}^{-1}$, $R=1.0080$ and $\epsilon_{\rm eff}\simeq 10\%$, we obtain an upper limit on $g^{\rm ul}_\l\times \sqrt{\cB(B\to 3\pi)}\lesssim \{3\times 10^{-4},\, 2 \times 10^{-4},\, 10^{-3}\}$ for the same sample points respectively. 
With an improved luminosity of $\mathcal{L}_{\rm int}=50 \,{\rm ab}^{-1}$, the reach of Belle-II would be improved by a factor of $4$. In addition to the bump hunt in the $3\pi$ channel, another way to look for $B$ is in the decays of Charmonium and/or Bottomonium. 
In particular, we can look for final states where the photon is replaced by $B$ in the radiative decays of such states.
A similar analysis for $\phi\to\eta B\,, B\to \pi^0\gamma$ has been performed by KLOE~\cite{delRio:2021xag, Tulin:2014tya}. 
However, this is only relevant for $0.3\GeV\lesssim m_B\lesssim 0.5 \GeV$, as $\cB(B\to \pi^0\gamma)$ becomes negligible otherwise.

\section{Implications for and from lattice computation}\label{sec:lattice}

In this section we discuss the implication of the new physics models on the scale setting on the lattice~\cite{FlavourLatticeAveragingGroupFLAG:2024oxs}. 
In particular, we estimate the NP contribution to the most commonly used $\Omega^-$ baryon physical scale, which may impact all lattice predictions.
Next, we emphasize the impact of new physics on the window quantities proposed in Ref.~\cite{RBC:2018dos} by re-evaluating the bounds from the HVP test.

\subsection{Lattice scale setting error due to New Physics} \label{sec:scale_settings}

Lattice QCD predictions of physical quantities are intrinsically dependent on a energy scale $\mathcal{S}$, which is set by matching an independently computed precise lattice observable $O_{\rm{lat}}$ to its corresponding experimental value $O_{\rm{exp}}$. The predictions are sensitive to the choice of this observable and the uncertainty associated with it. 
The experimental value of $O_{\rm{exp}}$ always includes electromagnetic and strong isospin-breaking corrections. 
The presence of a new gauge boson $X$ alters this decomposition, and introduces an additional, irreducible contribution to $O_{\rm exp}$. Therefore, one can expand $O_{\rm{exp}}$ as
\begin{equation}
    O_{\rm{exp}} = \bar{O} + O_{\rm IB} + O_{\gamma} +  O_{X}~,
\end{equation}
where $\bar{O}$, $O_{\rm IB}$, $O_{\gamma}$ denote the iso-symmetric, isospin breaking, and electromagnetic contributions respectively, whereas $O_{X}$ denotes the possible NP contribution. The matching procedure is,
 $O_{\rm{exp}} = (\mathcal{S})^{[O]}\langle O\rangle_{\rm{lat}}$ with $[O]$ being the energy dimension of $O$. 
In BMW20~\cite{Borsanyi:2020mff}, the mass of  $\Omega^-$ baryon ($M_{\Omega^{-}}=1672.5\MeV$), 
is used as the physical scale setting observable. 

To estimate the correction due to $X$, we rescale the EM contribution of $M_{\Omega^{-}}$ as $\delta M_{X}^{\Omega^{-}}\sim (g^X_d/e)^2 M_{\gamma}^{\Omega^{-}}$
where $M_{\gamma}^{\Omega^{-}}$ denotes the EM self energy correction of omega baryon mass which is estimated to be $<(0.4\%)M_{\Omega^-}$~\cite{Borsanyi:2020mff}.  
As the self-energy correction to the fermion masses due to gauge bosons is logarithmic in nature, by adding a possible log correction, we obtain 
\bea
\delta M_{X}^{\Omega^{-}}\sim  \frac{(g_d^X)^2}{4\pi \alpha} M_{\gamma}^{\Omega^{-}} \,\ln(\frac{m_X^2}{\Lambda^2})\sim 52\MeV \times (g_d^X)^2 \,,
\eea
where $\Lambda\sim \GeV$ is some characteristic scale and we use $m_X=600\MeV$. For $X=B$, by using $g_d^X=g_q\lesssim 0.15$, we obtain the NP correction of $\delta M^{\Omega^{-}}_{X}\sim 1\MeV$, and for $X=A'$, by using $\varepsilon=0.1$, we get the mass correction to be of $0.05\MeV$. 
These corrections are within $0.4\%$ scale-setting uncertainty quoted in the BMW20~\cite{Borsanyi:2020mff} estimate. 
Note that the other possible NP corrections to $M^{\Omega^{-}}$ are either of higher power of NP couplings and/or loop factor suppressed, and hence are not considered here. 
Also, a similar analysis can be performed for other scale setting observables, such as the pion decay constant (see~\cite{Boccaletti:2024guq}) as well. 
However, as long as the EM contribution to those scales is well within the quoted uncertainty, the NP contributions will not lead to any new errors, and thus can be safely neglected.

\subsection{Fine grained HVP test} 
\label{sec:fine_grained_HVP}
\begin{figure}[t]
    \centering
   
    \begin{subfigure}{0.45\textwidth}
    \centering
    \includegraphics[width=\textwidth]
    {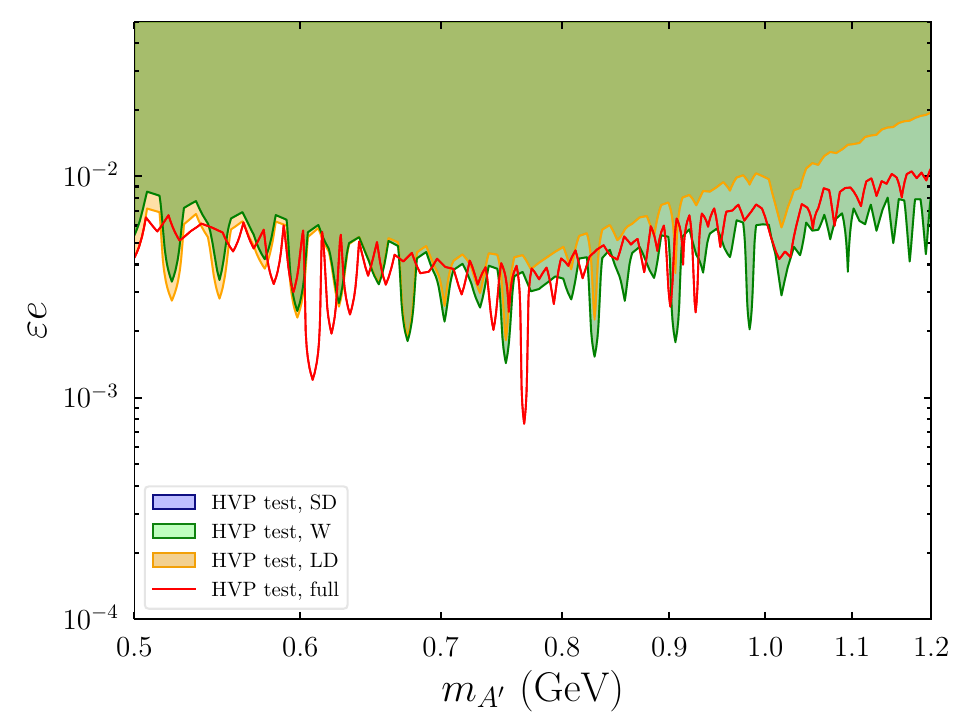}
    \caption{Dark photon}
    \label{fig:finegrained_HVP_dp}
    \end{subfigure}~
    \begin{subfigure}{0.45\textwidth}
    \centering
    \includegraphics[width=\textwidth]
    {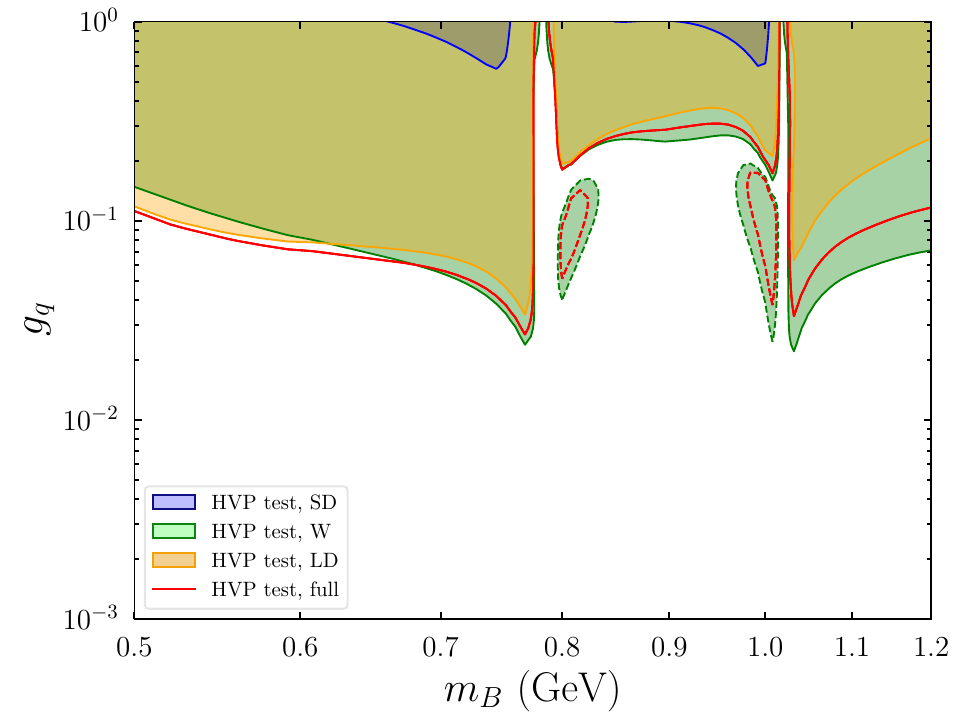}
    \caption{Baryon number}
    \label{fig:finegrained_HVP_b}
    \end{subfigure}
\caption{Bounds from fine grained HVP test for $A'$ (Fig.~\ref{fig:finegrained_HVP_dp}) and $B$ (Fig.~\ref{fig:finegrained_HVP_b}) considering $(a_\mu^{\rm HVP})^{\rm DD}=(a_\mu^{\rm HVP})^{\rm lat}$. 
Different colors depict the HVP tests in different windows as: blue: $a_\mu^{\rm SD}$ using ETMC~\cite{Alexandrou:2022amy}; green: $a_\mu^{\rm W}$ using BMW 24~\cite{Boccaletti:2024guq}, orange: $a_\mu^{\rm LD}$ using Mainz 24~\cite{Kuberski:2024bcj}; 
red=full HVP test (c.f. Fig.~\ref{fig:dark_photon_lat} and~\ref{fig:gq_mb_SM_dd} respectively), whereas the solid and dashed contours show $+2\sigma$, and $-2\sigma$ exclusions respectively. 
}
\label{Fig:}
\end{figure}

In lattice QCD, the HVP contribution to $\amu$ is calculated based on the integral over Euclidean time, $t$, of the two-point functions of quark EM current  convoluted with a kernel $\tilde{K}(t)$ as~\cite{Bernecker:2011gh}, 
\bea
\label{eq:latticeHVP}
(a_\mu^{\rm HVP})^{\rm lat}=-\frac{\alpha^2 }{3\pi^2}\int_0^\infty dt\, \tilde{K}(t) \sum_{\mu=1,2,3}\int d^3 x\,  \langle J^\mu_{\rm EM}(\vec{x},t)J^\mu_{\rm EM}(0)\rangle\,.
\eea
The above integral is often separated into short distance ($a_\mu^{\rm SD}$), intermediate distance ($ \amu^{\rm W}$), and long distance ($a_\mu^{\rm LD}$) contributions~\cite{RBC:2018dos}. 
These are called the ``window quantities".

These window quantities can be compared with the DD HVP; hence providing us with additional information of the energy dependence of the deviations between them.
In the data-driven methods, these window quantities can be obtained by adding the weight functions $\tilde{\Theta}_{\rm win}(s)$ centered at different center of mass energy to the integral Eq.~\eqref{eq:HVP_dd}, 
\begin{align}
(a_\mu^{\mathrm{HVP}})^{\rm{DD}}_{\rm win}&=\frac{1}{4\pi^3}\int^\infty_{s_\text{exp}} ds\,  K(s)\, \tilde{\Theta}_{\rm win}(s) \sigma(e^+e^-\to{\rm had})(s).
\end{align}
The explicit form of the three weight functions $\tilde{\Theta}_{\rm win}(s)$ can be found in~\cite{Colangelo:2022vok} and in Appendix~\ref{app:window}. 
We also summarize the current results of the window observables from different lattice groups in Table~\ref{tab:win} in Appendix~\ref{app:window}. 
We find that the large deviation shows up for the intermediate and long distance regime when BaBar/KLOE results are used in the DD method compare to the lattice results, while results from CMD-3 agrees within precision with the lattice result. 
However, in the short-distance window, these two methods are consistent with each other.

We utilize these window observables to provide additional fine grained HVP test of our benchmark models. The fine grained HVP test is a more futuristic test. However, as many current lattice collaborations report their HVP results in different windows, this test could become more relevant. The goal of using the HVP test in various windows, is to get a more precise estimation of how new physics enters the HVP as a function of the energy, $\sqrt{s}$. For example, if the new gauge boson $X$ is light ($m_X \sim 1$ GeV), it is expected to primarily affect just the long distance or the intermediate distance windows.
In the presence of $X$, the data-driven results of these window observables modify as,
\begin{equation}
(\Delta \amu^{\rm HVP})_{\rm win}=\frac{1}{4\pi^3}\int^\infty_{s_{\rm th}} ds \,K(s)\, \tilde{\Theta}_{\text{win}}(s)  (\sigma^{\gamma-X}(s)+\sigma^{X}(s))\,,
\end{equation}
using the discussion in Section~\ref{sec:explicit_calculation} (c.f. Eq.~\eqref{eq:sigma_X_rho}). 
As an illustration, in Fig.~\ref{fig:finegrained_HVP_dp} and~\ref{fig:finegrained_HVP_b}, we show the bounds obtained from the fine grained HVP test for the dark photon and $B$ model respectively, assuming the discrepancies between lattice and data-driven eventually disappear in the future. 
We see that for the mass range we are considering, $a_\mu^{\rm SD}$ is the least sensitive to the new physics and yields the weakest bound. For the dark photon, the $+ 2\sigma$ HVP bounds from $a_\mu^{\rm W}$ and $a_\mu^{\rm LD}$ are comparable at $m_{A'} \lsim 0.8$ GeV, while $a_\mu^{\rm W}$ gives the strongest bound when $m_{A'} \gtrsim 0.8$ GeV. For the largest masses, $a_\mu^{\rm W}$ gives a $+2\sigma$ bound on $\varepsilon e$ that is stronger by a factor of $\mathcal{O}(2)$ compared to the bound from $a_\mu^{\rm LD}$. None of the HVP tests give a $-2\sigma$ constraint for this model. Similarly, for $B$, the $+ 2\sigma$ HVP bounds from $a_\mu^{\rm W}$ and $a_\mu^{\rm LD}$ are comparable at $m_B \lsim 1$ GeV, while $a_\mu^{\rm W}$ gives strongest bound when $m_B \gtrsim 1$ GeV. For this model, only $a_\mu^{\rm W}$ and the full HVP test give $-2\sigma$ constraints, and they are both very similar.

\section{Conclusions}
\label{sec:conclusion}
\begin{figure}[t]
    \centering
    \includegraphics[width=0.75\linewidth]{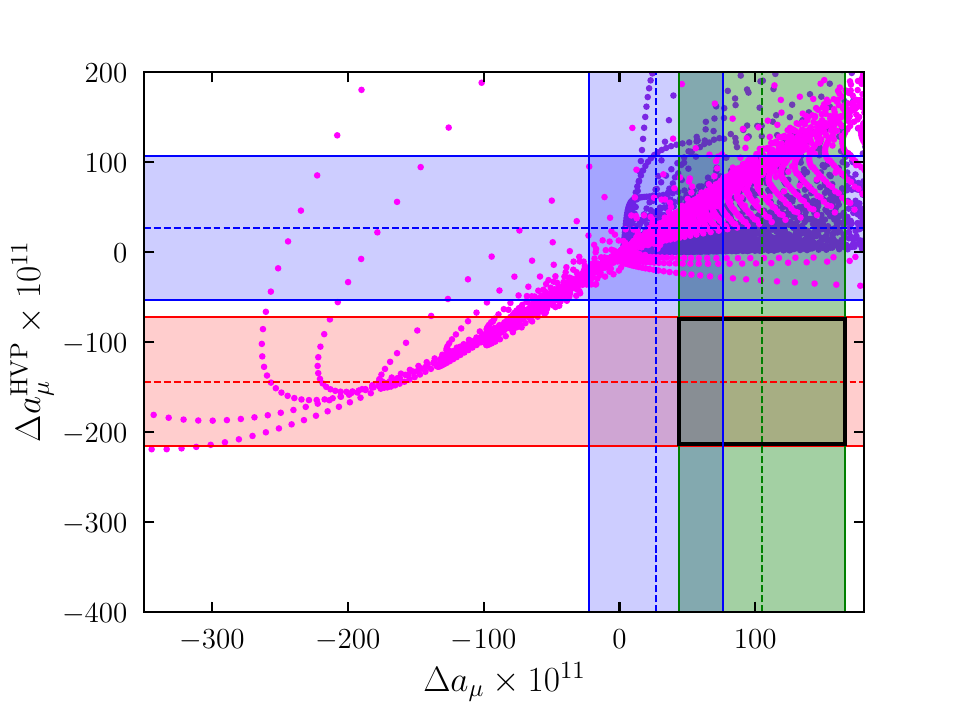}
    \caption{Plot of the two new physics contributions, $\Delta a_{\mu}$ (c.f. Eq.~\eqref{eq:amu-test}) and $\Delta a^{\rm{HVP}}_{\mu}$ (c.f. Eq.~\eqref{eq:hvp_test_def}), for $X=B$ (baryon number gauge boson) and $X=A'$ (dark photon). 
    The zeros of the $x$ and $y$ axis are set at $a_{\mu}^{\rm{lat}}$ and $(a^{\rm{HVP}}_{\mu})^{\rm{lat}}$ (from BMW 2020 \cite{Borsanyi:2020mff}). 
    The green band is the experimental world average of $a^{\rm exp}_{\mu}$~\cite{Muong-2:2023cdq}(with errors in quadrature with the $a^{\rm{lat}}_{\mu}$ error) while the red band represents $(\amu^{\rm HVP})^{\rm DD}_{\mu}$ from TI 2020 \cite{Aoyama:2020ynm}(errors in quadrature with the $(a^{\rm{HVP}}_{\mu})^{\rm{lat}}$ errors). The blue bands are $a_{\mu}$ and $a^{\rm{HVP}}_{\mu}$ from the CMD-3 experiment~\cite{Bryzgalov:2024uws,CMD-3:2023rfe}. 
    The purple and pink points depict the model prediction for the case of $A'$ and $B$ respectively. 
    To satisfy both the $a_{\mu}$ and HVP tests by a model, upto $\pm 1\sigma$, points must lie inside the black square.}
    \label{fig:rmu}
\end{figure}

Light and weakly coupled new physics has been extensively studied in the last few years. 
It is, however, in general challenging to directly search for new resonances if they dominantly couple to the hadronic sector of the SM and their masses are of the order of the QCD scale.
It is interesting to note however, that indirect precision measurements provide powerful probes of these types of models. 
Among them are the muon anomalous magnetic moment, $\amu$ and the photon hadronic vacuum polarization (HVP).
The latter is sensitive to new physics contribution to the cross-section for $e^+e^-\to{\rm hadrons}$, which also affects several other well-measured observables such as $\amu$, $\sin^2\theta_W$ and the running of $\alpha$. Two precision tests can be defined, namely the $\Delta\amu$ and the $\Delta\amu^{\rm HVP}$ test.
The first is a comparison between the experimental and theoretical values of the muon anomalous magnetic moment.
The second tests only the HVP ``part'' of $a_{ \mu }$ by comparing the result of the lattice calculation for it to the value obtained using the data-driven approach, i.e., based on a measured cross-section of $e^+ e^- \rightarrow$ hadrons. 

In this work, we have studied the implications of the above tests to a new flavor-universal vector, $X$, with mass around the GeV scale.
We find that in models where the hadronic coupling, $g_q$, is larger than the leptonic coupling, $g_\ell$, the contribution to  $\amu$ from the  $X$-photon mixing diagram can be as significant as the naive 1-loop diagram. 
In addition, the $X$ on-shell contribution to the $e^+e^-\to{\rm hadrons}$ cross-section should be subtracted in the data-driven approach for computing the photon HVP contribution to $a_{ \mu }$, resulting in reducing the sensitivity of $a_{ \mu }$ to 1-loop effect of $X$ by a factor of $g_\l^2/g_q^2$ as compared to the naive expectation. 
In this case, $\amu$ is strongly sensitive to the hadronic coupling of $X$, albeit 
$a_{ \mu }$ being a purely leptonic quantity.

We demonstrate the above by considering two specific cases, dark photon, $A'$, and baryon number gauge boson, $B$. 
As a summary plot, in Fig.~\ref{fig:rmu}, we reconcile our two benchmark models against both the tests. 
We see that based on the current scenario \textcolor{black}{i.e. using BMW 2020 as the lattice and Theory Intitative 2020 as the data-driven result}, both the tests cannot be satisfied simultaneously, leading to a complete exclusion of these models. \textcolor{black}{The data-driven HVP by Theory Initiative 2020 is obtained after averaging the $e^+e^- \to \rm hadrons$ cross-sections measured by BaBar \cite{BaBar:2012bdw,BaBar:2021cde}, KLOE \cite{KLOE-2:2017fda,KLOE:2010qei,KLOE:2012anl}, SND \cite{Achasov:2006vp,SND:2020nwa} and CMD-2 \cite{CMD-2:2006gxt,CMD-2:2005mvb,Aulchenko:2006dxz}. Another noteworthy result is the data-driven value obtained from the $e^+e^- \to \rm hadrons$ cross-section  measured by CMD-3 \cite{CMD-3:2023alj,CMD-3:2023rfe}, shown by the blue bands in Fig.~\ref{fig:rmu}. Due to a  discrepancy between the CMD-3 and KLOE measurements of the $e^+e^- \to \pi^+ \pi^-$ cross-section, there is an estimated $4\sigma$ tension in the corresponding HVPs. Thus, it is difficult to obtain a reliable average of the HVP from CMD-3 and Theory Initiative 2020. However, as seen in Fig.~\ref{fig:rmu}, the CMD-3 HVP taken in isolation is consistent with BMW 2020 (at 1$\sigma$ level), and does not lead to a complete exclusion of the two aforementioned models.} In addition, we pointed out that by using the on-tape BaBar and Belle data sets one can perform a direct search for $B$ in the $3\pi$ final state in events with ISR and probe the unexplored region of the parameter space. 

\section*{Acknowledgments}
We gratefully acknowledge Fabio Anulli, Vladimir Druzhinin, Phoebe Hamilton, Hassan Jawahery, Evgeny Kozyrev, Frank Porter,  Eugeny Solodov, and Yuki Sue for their invaluable insights and thoughtful feedback. We thank Zackaria Chacko, Tom Cohen, Anson Hook, Lorenzo Ricci, and Raman Sundrum for useful discussions. The work of KA, AB, and KP is supported by the National Science Foundation under grant number PHY2210361 and the Maryland Center for Fundamental Physics. 
The work of MJ is supported by the Deutsche Forschungsgemeinschaft under
Germany’s Excellence Strategy EXC 2121 ``Quantum Universe” — 390833306, as well as by the grant 491245950. 
The work of GP is supported by grants from NSF-BSF, ISF and Minerva. 
The work of YS is supported by grants from the NSF-BSF (grant No. 2021800), the ISF (grant No. 597/24). 
Finally, KA and GP are grateful to the organizers of the ``Physics of this Universe" workshop (Johns Hopkins University, 2022), where the seeds of this work were planted. 

\newpage
\appendix

\renewcommand{\theequation}{\thesection.\arabic{equation}}
\setcounter{section}{0}
\renewcommand{\thesection}{\Alph{section}}

\section{Some field theoretic proofs}\label{App:Field_theory}

In this section, we prove the various field theoretic results used in the main text. In Appendix~\ref{app:long-contributions} we show that the longitudinal parts of the propagator of a massive gauge boson (in the $a_{\mu}$ loop) and longitudinal parts of vacuum polarization functions, do not contribute to $a_{\mu}$. In Appendix~\ref{app:2-loop-renorm} we derive the renormalization scheme dependence of $a^X_{\mu}$, $a^{\gamma-X}_{\mu}$ and $a^{XX}_{\mu}$ followed by a proof of the renormalization scheme independence of the sum of these quantities in Appendix~\ref{app:renorm-independent}   
\subsection{Longitudinal contributions to $a_{\mu}$}\label{app:long-contributions}

\begin{figure}[h]
\centering
\begin{tikzpicture}

\draw[thick,electron] (0,0) to  [out=-45,in=135] (1.2,-1.2) ;
\draw[thick,electron] (1.2,-1.2) to  [out=-45,in=135] (1.8,-1.8) node[label=$p'$] {};
\draw[thick,antielectron] (0,0) to [out=-135,in=45] (-1.2,-1.2);
\draw[thick,antielectron] (-1.2,-1.2) to  [out=-135,in=45] (-1.8,-1.8) node[label=$p$] {};
\draw[thick,photon] (0,0) to [out=90,in=-90] (0,1.1); 
\draw[thick,photon] (-1.1,-1.1) to [out=0,in=-180] (1.1,-1.1) ;
\node[] at (1.2,-0.4) {$p'-k$};
\node[] at (-1.2,-0.4) {$p-k$};
\node[] at (0,-0.8) {$X$};
\node[] at (0,-1.4) {$k$};
\node[] at (-1.2,0.6) {$q=p-p'$};
\node[] at (0.3,0.6) {$\gamma$};
\end{tikzpicture}
\caption{1-loop $(g-2)_{\mu}$ with momenta labelled}
\label{fig:long-fig}
\end{figure}
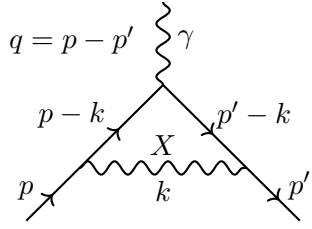
The propagator of a massive gauge boson has a piece proportional to $k^{\mu}k^{\nu}$. At the 2 loop order, we also need to include propagators connected by vacuum polarization (VP) blobs. The VP tensors relevant to our computations take the forms 
\begin{align}
    \Pi^{\mu\nu}_{\gamma \gamma}(k^2)&=(k^2-k^{\mu}k^{\nu})g^{\mu\nu}\Pi_{\gamma \gamma}(k^2)+k^{\mu}k^{\nu}\Pi^L_{\gamma \gamma}(k^2) \label{eqn:pi-gamgam}\,,\\
    \Pi^{\mu\nu}_{\gamma X}(k^2)&=(k^2-k^{\mu}k^{\nu})g^{\mu\nu}\Pi_{\gamma X}(k^2)+k^{\mu}k^{\nu}\Pi^L_{\gamma X}(k^2)\label{eqn:pi-gamx}\,,\\
    \Pi^{\mu\nu}_{XX}(k^2)&=((k^2-m_X^2)g^{\mu\nu}-k^{\mu}k^{\nu}{})\Pi_{XX}(k^2)+k^{\mu}k^{\nu}\Pi^L_{XX}(k^2)\label{eqn:pi-xx}~.
\end{align}
The contribution coming from the longitudinal part of the free $X$ propagator, i.e. the part of the propagator proportional to  $k^{\mu}k^{\nu}$ in the diagram is given by (see Fig.~\ref{fig:long-fig})
\begin{align}
    a_{\mu} &\propto \int_{}{} \frac{d^4k}{(2\pi)^4} \bar{u}(p')\frac{\gamma^{\beta}(\slashed{p}'-\slashed{k}+m_{\mu})\gamma^{\mu}(\slashed{p}-\slashed{k})+m_{\mu}) \gamma^{\alpha}k^{\alpha}k^{\beta}}{((p-k)^2-m^2_{\mu})((p'-k)^2-m^2_{\mu})(k^2-m_X^2)} u(p) \label{eqn:amu-long}\\
    &=\int_{}{} \frac{d^4k}{(2\pi)^4} \bar{u}(p')\frac{\slashed{k}(\slashed{p}'-\slashed{k}+m_{\mu})\gamma^{\mu}(\slashed{p}-\slashed{k})+m_{\mu})\slashed{k}}{((p-k)^2-m^2_{\mu})((p'-k)^2-m^2_{\mu})(k^2-m_X^2)} u(p)\nn\\
    &=\int_{}{} \frac{d^4k}{(2\pi)^4} \bar{u}(p')\frac{(2(p'\cdot k)-\slashed{p'}\slashed{k}'-{k}^2+m_{\mu}\slashed{k})\gamma^{\mu}(2(p\cdot k)-\slashed{k}\slashed{p}-{k}^2+m_{\mu}\slashed{k})}{((p-k)^2-m^2_{\mu})((p'-k)^2-m^2_{\mu})(k^2-m_X^2)} u(p). \nn
\end{align}
Using $\slashed{p}u(p)=m_{\mu} u(p)$ and $\bar{u}(p')\slashed{p}'=m_{\mu} \bar{u}(p')$, we get
\begin{equation}
    a_{\mu}\propto\int_{}{} \frac{d^4k}{(2\pi)^4} \bar{u}(p')\frac{(2(p'\cdot k)-k^2)\gamma^{\mu}(2(p\cdot k)-k^2)}{((p-k)^2-m^2_{\mu})((p'-k)^2-m^2_{\mu})(k^2-m_X^2)} u(p).
\end{equation}
Since this term is proportional only to $\bar{u}(p')\gamma_{\mu}u(p)$, it will not contribute to the magnetic form factor, which is proportional to $\bar{u}(p')i\sigma^{\mu\nu}(p-p')_{\nu}u(p)$. Note that if we want a VP insertion in the diagram then the integrand in \eqref{eqn:amu-long} now just has a VP tensor multiplied with it. This means that, for the VP pieces proportional to $k^{\mu}k^{\nu}$, our analysis falls through exactly as we showed above and we get no contribution to $a_{\mu}$.  Hence, any term proportional to $k^{\mu}k^{\nu}$ does not contribute to $a_{\mu}$, and can be safely neglected in our computations.

\subsection{2-loop calculations for $a_{\mu}$ and renormalization schemes}\label{app:2-loop-renorm}
In this section, we aim to obtain analytical expressions for computing 2 loop $a_{\mu}$ contributions (i.e. of the type in Fig.~\ref{Fig:feynmann_diagram_HVP}). This means we have to deal with one VP insertion for each contribution. Note that we only need to deal with the scalar parts of the VP functions i.e. $\Pi_{\gamma\gamma},\Pi_{\gamma X}$ or $\Pi_{XX}$, because the remaining pieces do not contribute to $a_{\mu}$ as shown in Appendix~\ref{app:long-contributions}. Since calculating the VP functions perturbatively is not possible, we turn to using the K\"all\'en–Lehmann (KL) spectral representation
\begin{align}
    \frac{{\Pi}(k^2)-{\Pi}(p_0^2)}{k^2-p_0^2}=-\int_{0}^{\infty} ds&\frac{\text{Im}~{\Pi}(s)}{\pi(k^2-s+i\varepsilon)(s-p_0^2)} =-\int_{0}^{\infty} ds\frac{\text{Im}~\bar{\Pi}(s)}{\pi(k^2-s+i\varepsilon)(s-p_0^2)} \label{eqn:kl-rep}\\
    &\bar{\Pi}(k^2)={\Pi}(k^2)-\text{Re}~\Pi(p_0^2) \,,\label{eqn:pi-renorm}
\end{align}
where $p_0$ is an arbitrary reference momentum, $\Pi=\Pi_{\gamma\gamma},\Pi_{\gamma X}$ or $\Pi_{XX}$ and the bar denotes that $\Pi(k^2)$ is renormalized at $p_0^2$ as shown above. To relate the imaginary part of $\Pi$ to a measured cross section, we use the optical theorem, which gives us the following relations for the three cross sections and the corresponding VP functions (where $\int_{}{}d\Pi$ is the phase space integral)
\begin{align}
    \sigma^{\gamma}(s)&=\sigma(e^+e^-\to \gamma^* \to {\rm had}) = \frac{e^4\,\text{Im}~\bar{\Pi}_{\gamma \gamma}(s)}{s}\,,  \label{eqn:sigma-gam}\\
    \sigma^{\gamma-X}(s)&=\int_{}{} d\Pi~2~\text{Re}\left(\mathcal{A}_{e^+e^- \rightarrow \gamma^* \rightarrow\text{had} }(s)~\mathcal{A}^{*}_{e^+e^- \rightarrow X^{*}\rightarrow\text{had} }(s) \right)  \nonumber\\
    &= 2g_{\ell}g_qe^2\text{Im}\left(\frac{\bar{\Pi}_{\gamma X}(s)}{s-m_X^2 + i\epsilon}\right)= 2g_{\ell}g_qe^2\text{Im}(\bar{\Pi}_{\gamma X}(s))\frac{s-m_X^2}{(s-m_X^2)^2 + \epsilon^2}\,, \label{eqn:sigma-gamx}\\
    \sigma^{X}(s)&=\sigma(e^+e^-\to X^*\to{\rm had}) (s)= g_{\ell}^2g_q^2e^2\,\text{Im}\left(\frac{\bar{\Pi}_{X X}(s)}{s-m_X^2 + i\epsilon}\right)\nonumber\\
    &=g_{\ell}^2g_q^2e^2\,\text{Im}~\bar{\Pi}_{XX}(s)\frac{s-m_X^2}{(s-m_X^2)^2 + \epsilon^2} ~.\label{eqn:sigma-xx}
\end{align}

Note that fixing a renormalization scheme for the three VP functions also fixes the scheme for the relevant couplings i.e. we have the couplings renormalized as $e=e(p_0^2), g_{\ell}=g_{\ell}(p_0^2), g_q=g_q(p_0^2)$. Keeping this in mind, we can now calculate the final $a_{\mu}$ contributions. First, we consider the SM/photon 2-loop contribution i.e
\begin{align}
    {a}^{\gamma\gamma}_{\mu}&=\int_{}{}\frac{d^4k}{(2\pi)^4} \frac{e^4(p_0^2) \text{Re}~\bar{\Pi}_{\gamma \gamma}(k^2)}{k^2} \mathcal{F}_{\mu}(k)\\
    &=\int_{}{}\frac{d^4k}{(2\pi)^4} \frac{e^4(p_0^2) \text{Re}~({\Pi}_{\gamma \gamma}(k^2)-{\Pi}_{\gamma \gamma}(p_0^2))}{k^2} \mathcal{F}_{\mu}(k)\,,
\end{align}
where $\mathcal{F}_{\mu} (k)$ contains all the pieces from the muon propagators and the external photon vertex in the feynmann diagram. We have used Eq.~\eqref{eqn:pi-gamgam} along with the results of Appendix~\ref{app:long-contributions} to get the first equality. Using Eq.~\eqref{eqn:kl-rep} for $\Pi=\Pi_{\gamma \gamma}$ and Eq.~\eqref{eqn:sigma-gam} leads to 
\begin{align}
    {a}^{\gamma\gamma}_{\mu}&=\int_{0}^{\infty}ds e^4(p_0^2)\frac{\text{Im}\bar{\Pi}_{\gamma \gamma}(s)}{\pi(s-p_0^2)(k^2-s)}\int_{}{}\frac{d^4k}{(2\pi)^4} \frac{k^2-p_0^2}{k^2} \mathcal{F}_{\mu}(k)\nn\\
    &=\int_{0}^{\infty}ds e^4(p_0^2)\frac{\text{Im}\bar{\Pi}_{\gamma \gamma}(s)}{\pi s} s \frac{d^4k}{(2\pi)^4} \left(\frac{1}{(s-p_0^2)k^2}+\frac{1}{s(k^2-s)}-\frac{1}{sk^2}\right) \mathcal{F}_{\mu}(k)\nn\\
&=\frac{1}{4\pi^3}\int_{0}^{\infty}ds\sigma^{\gamma}(s)\left(K(s)+\frac{p_0^2}{s-p_0^2}K(0)\right)\,,
\end{align}
where we have used the definiton of the kernel function $K(s)$ to get the last equality. Moving on to the BSM contributions, using Eq.~\eqref{eqn:pi-gamx} along with the results of Appendix~\ref{app:long-contributions}, we get
\begin{align}
    a^{\gamma -X}_{\mu}=\int_{}^{}\frac{d^4k}{(2\pi)^4} \frac{2e^2g_{q}g_{\ell}\bar{\Pi}_{\gamma X}(k^2)k^2}{k^2(k^2-m_X^2)} \mathcal{F}_{\mu}(k)=\int_{}^{}\frac{d^4k}{(2\pi)^4} \frac{2e^2g_{q}g_{\ell}\bar{\Pi}_{\gamma X}(k^2)}{(k^2-m_X^2)} \mathcal{F}_{\mu}(k) \,,
\end{align}
which upon using Eq.~\eqref{eqn:kl-rep} for $\Pi_{\gamma X}$ leads to
\begin{align}
    a^{\gamma-X}_{\mu}&=-\int_{}{}\frac{d^4k}{(2\pi)^4}\int_{0}^{\infty} ds\frac{2e^2g_\ell g_q\text{Im}(\bar{\Pi}_{\gamma X}(s))}{\pi(k^2-s)(s-p_0^2)}\frac{k^2-p_0^2}{(k^2-m_X^2)} \mathcal{F}_{\mu}(k)\nn\\
    &=\int_{0}^{\infty} ds\frac{2e^2g_{\ell} g_q\text{Im}(\bar{\Pi}_{\gamma X}(s))}{(s-m_X^2)(s-p_0^2)} \int_{}{}\frac{d^4k}{(2\pi)^4} \frac{1}{\pi}\left(\frac{s-p_0^2}{k^2-s}+\frac{p^2_0-m_X^2}{k^2-m_X^2}\right) \mathcal{F}_{\mu}(k)~. \nn
\end{align}
Finally, using Eq.~\eqref{eqn:sigma-gamx} in the off-shell limit i.e. $s\neq m_X^2$ leads to the simplification
\begin{align}
 a^{\gamma-X}_{\mu}&=\frac{1}{4\pi^3}\int_{0}^{\infty} ds\sigma^{\gamma-X}(s)\left(K(s)+\frac{p_0^2-m_X^2}{s-p_0^2}K(m_X^2)\right) \\
    & K_{\gamma-X}(s)=K(s)+\frac{p_0^2-m_X^2}{s-p_0^2}K(m_X^2) \,,\label{eqn:finalform-gamx} 
\end{align}
where $K_{\gamma-X}(s)$ can be viewed as the kernel for $a^{\gamma-X}_{\mu}$. Similarly, we can also compute the $XX$ contribution. Using Eq.~\eqref{eqn:pi-xx} along with the results of Appendix~\ref{app:long-contributions} leads to
\begin{equation}
a^{XX}_{\mu}=\int_{}^{}\frac{d^4k}{(2\pi)^4} \frac{g_{q}^2g_{\ell}^2\bar{\Pi}_{XX}(k^2-m_X^2)}{(k^2-m_X^2)^2} \mathcal{F}_{\mu}(k)=\int_{}^{}\frac{d^4k}{(2\pi)^4} \frac{g_{q}^2g_{\ell}^2\bar{\Pi}_{XX}}{(k^2-m_X^2)} \mathcal{F}_{\mu}(k)~.\label{eqn:step-1xx}
\end{equation}
Now using Eq.~\eqref{eqn:kl-rep} for $\Pi_{X X}$ gives
\begin{align}
    a^{XX}_{\mu}&=-\int_{}{}\frac{d^4k}{(2\pi)^4}\int_{0}^{\infty} ds\frac{g_{\ell}^2g_q^2\text{Im}(\bar{\Pi}_{XX}(s))}{\pi(k^2-s)(s-p_0^2)}\frac{k^2-p_0^2}{(k^2-m_X^2)} \mathcal{F}_{\mu}(k)\nn\\
    &= \frac{1}{4\pi^3}\int_{0}^{\infty} ds\frac{g_{\ell}^2g_q^2\text{Im}(\bar{\Pi}_{X X}(s))}{(s-m_X^2)(s-p_0^2)} \left((s-p_0^2)K(s)+(p_0^2-m_X^2)K(m_X^2)\right) \nn~.\label{eqn:step-2xx} 
\end{align}
Finally, using Eq.~\eqref{eqn:sigma-xx} in the off-shell limit
\begin{align}
 a^{XX}_{\mu}&=\frac{1}{4\pi^3}\int_{0}^{\infty} ds~\sigma^{\text{off-shell}~X}(s)\left(K(s)+\frac{p_0^2-m_X^2}{s-p_0^2}K(m_X^2)\right) \\
    & K_{XX}(s)=K(s)+\frac{p_0^2-m_X^2}{s-p_0^2}K(m_X^2)\,,\label{eqn:finalform-xx}
\end{align}
where $K_{XX}(s)$ can be viewed as the kernel for $a^{XX}_{\mu}$. Therefore, we see that the expressions of the 2 loop contributions to $a_{\mu}$ depend on the reference momentum $p_0$. Importantly, we see that
\begin{align}
    a^{\gamma \gamma}_{\mu}&=(a^{\rm HVP}_{\mu})_{e^+e^-\to {\rm had}}^{\gamma\gamma} ~~~~\text{for }p_0^2=0 \,,\\
    a^{\gamma -X}_{\mu}&=(a^{\rm HVP}_{\mu})_{e^+e^-\to {\rm had}}^{\gamma-X} ~~~~\text{for }p_0^2=m_X^2 \,,\\
    a^{XX}_{\mu}&=(a^{\rm HVP}_{\mu})_{e^+e^-\to {\rm had}}^{XX}~~~~\text{for }p_0^2=m_X^2\,,
\end{align}
i.e. the $a_{\mu}$ contributions exactly match the HVP contributions for particular scheme choices. These choices simplify things immensely since we just get $K_{\gamma-X}(s)=K_{XX}(s)=K(s)$ and so, these are the ones that we follow in this paper.  Naively, this might seem that we have a dependence on the renormalization scheme for physical observables. However, these contributions individually are not physical. One must keep in mind that the 1-loop contributions $a^\gamma_{\mu}$ and $a^X_{\mu}$, due to their dependence on $e^2$ and $g_\ell^2$, are also renormalization scheme dependent. As we show in the next section, it is the sum of the 1-loop and 2-loop contributions that is independent of the choice of $p_0$ and therefore, is physical.

\subsection{Renormalization scheme independence of $a_{\mu}$ }\label{app:renorm-independent}

First, we consider the SM contributions. We start by defining
\begin{align}
    &{a}_{\mu}^{\gamma\gamma}(p_0^2)=a^{\gamma\gamma}_{\mu}~~\text{for arbitrary $p_0^2$} \,,\\
    &a_{\mu}^{\gamma\gamma}(0)=a^{\gamma\gamma}_{\mu}~~\text{for $p_0^2=0$ scheme}~.
\end{align}
The running of the electric charge is given by
\begin{equation}
    \alpha(p_0^2)=\frac{\alpha(0)}{1-4\pi \alpha(0)\text{Re}\left({\Pi}_{\gamma \gamma}(p_0^2)-{\Pi}_{\gamma \gamma}(0)\right)}~.
\end{equation}
This gives us
\begin{equation*}
    {a}^{\gamma}_{\mu}(p_0^2)=\frac{\alpha(p_0^2)}{2\pi}=\frac{\alpha}{2\pi}+2\alpha^2(0)\text{Re}\left({\Pi}_{\gamma \gamma}(p_0^2)-{\Pi}_{\gamma \gamma}(0)\right) \,,
\end{equation*}
\begin{align}
    {a}^{\gamma\gamma}_{\mu}(p_0^2)&=\int_{}{}\frac{d^4k}{(2\pi)^4} \frac{e^4(p_0^2) \text{Re}~({\Pi}_{\gamma \gamma}(k^2)-{\Pi}_{\gamma \gamma}(p_0^2))}{k^2} \mathcal{F}_{\mu}(k)\nn\\
    &=\int_{}{}\frac{d^4k}{(2\pi)^4} \frac{e^4(p_0^2) \text{Re}~({\Pi}_{\gamma \gamma}(k^2)-{\Pi}_{\gamma \gamma}(0))}{k^2} \mathcal{F}_{\mu}(k)\nn\\
&+\text{Re}\left({\Pi}_{\gamma \gamma}(0^2)-{\Pi}_{\gamma \gamma}(p_0^2)\right)\int_{}{}\frac{d^4k}{(2\pi)^4} e^4(p_0^2) \frac{1}{k^2} \mathcal{F}_{\mu}(k)\nn\\
    &={a}^{\gamma\gamma}_{\mu}(0)-2\alpha^2(0)\text{Re}\left({\Pi}_{\gamma \gamma}(p_0^2)-{\Pi}_{\gamma \gamma}(0)\right) +\mathcal{O}(\Pi_{\gamma \gamma}^2)\,,
\end{align}
where $\mathcal{O}(\Pi_{\gamma \gamma}^2)$ is the higher order contribution involving 2 VP blobs. Combining everything
\begin{align}
    {a}^{\gamma}_{\mu}(p_0^2)+{a}^{\gamma\gamma}_{\mu}(p_0^2)={a}^{\gamma}_{\mu}(0)+{a}^{\gamma\gamma}_{\mu}(0)~.
\end{align}
Hence, the SM 1-loop + 2-loop $a_{\mu}$ contributions indeed turn out to be renormalization scheme independent. For the BSM contributions, we only consider the running of $g_{\ell}$ to keep the proof brief. Since the running of $g_{\ell}$ proceeds through $\gamma-X$ mixing at the leading loop order, we only need to focus on the $\gamma-X$ and 1-loop $X$ contributions to $a_{\mu}$. The running of $g_\ell$ through $\gamma-X$ mixing is given by:

\begin{equation}
    g_\ell(m_X^2)=g_\ell(p_0^2)+e^2g_q\text{Re}\left(\Pi_{\gamma X}(m_X^2)-\Pi_{\gamma X}(p_0^2)\right)+ \mathcal{O}(\Pi^2)~.
    \label{gl-run}
\end{equation}
We make the notations below for the arguments that follow:
\begin{align}
    &{a}_{\mu}^{X}(p_0^2)=a^{\text{$X$}}_{\mu}~~\text{for arbitrary $p_0^2$}\,,\\
    &{a}_{\mu}^{X}(m_X^2)=a^{\text{$X$}}_{\mu}~~\text{for $p_0^2=m_X^2$}\,,\\
    &{a}_{\mu}^{\gamma-X}(p_0^2)=a^{\gamma-X}_{\mu}~~\text{for arbitrary $p_0^2$}\,,\\
    &a_{\mu}^{\gamma-X}(m_X^2)=a^{\gamma-X}_{\mu}~~\text{for $p_0^2=m_X^2$}\,,\\
    &{a}_{\mu}^{\gamma-X-\gamma}(p_0^2)=a^{\gamma-X-\gamma}_{\mu}~~\text{for arbitrary $p_0^2$}\,,\\
    &a_{\mu}^{\gamma-X-\gamma}(m_X^2)=a^{\gamma-X-\gamma}_{\mu}~~\text{for $p_0^2=m_X^2$}\,,
\end{align}
where the $\gamma-X-\gamma$ contribution comes from the diagram shown in Fig.~\ref{fig:gam-b-gam}. We have:

\begin{figure}[t]
\centering
\begin{tikzpicture}
\begin{scope}[xshift=10cm]
\draw[thick,electron] (5,0) to  [out=-45,in=135] (6.8,-1.8) ;
\draw[thick,electron] (6.8,-1.8) to  [out=-45,in=135] (8.8,-3.8) node[label=$\mu$] {};
\draw[thick,antielectron] (5,0) to [out=-135,in=45] (3.2,-1.8);
\draw[thick,antielectron] (3.8,-1.2) to  [out=-135,in=45] (1.2,-3.8) node[label=$\mu$] {};
\draw[thick,photon] (5,0) to [out=90,in=-90] (5,1.1); 
\draw[thick,photon] (2.4,-2.6) to [out=0,in=-180] (7.6,-2.6) ;
\node[draw,circle,fill=gray,inner sep=1.5pt] (amp1) at (6,-2.6) {Blob};
\node[draw,circle,fill=gray,inner sep=1.5pt] (amp1) at (4,-2.6) {Blob};
\node[] at (5.3,0.6) {$\gamma$};
\node[] at (3,-3.2) {$\gamma$};
\node[] at (5,-3.2) {$X$};
\node[] at (7,-3.2) {$\gamma$};
\end{scope}
\end{tikzpicture}
\caption{$\gamma-X-\gamma$ contribution to $g-2$}
\label{fig:gam-b-gam}
\end{figure}
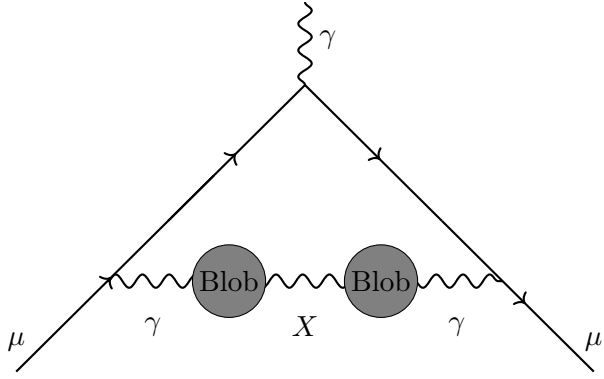

\begin{align}
        {a}_{\mu}^{\gamma-X}(p_0^2)&  =\int_{}{}\frac{d^4k}{(2\pi)^4} \frac{2e^2g_{\ell}(p_0^2)g_q \text{Re}\left({\Pi}_{\gamma X}(k^2)-{\Pi}_{\gamma X}(p_0^2)\right)}{(k^2-m_X^2)} \mathcal{F}_{\mu}(k)\label{eqn:step-1}\\
        &=\int_{}{}\frac{d^4k}{(2\pi)^4} \frac{2e^2g_{\ell}(p_0^2)g_q \text{Re}\left({\Pi}_{\gamma X}(k^2)-{\Pi}_{\gamma X}(m_X^2)\right)}{(k^2-m_X^2)} \mathcal{F}_{\mu}(k)\nonumber\\
        &+2e^2g_{\ell}(p_0^2)g_q \text{Re}\left({\Pi}_{\gamma X}(m_X^2)-{\Pi}_{\gamma X}(p_0^2)\right)4\pi K(m_X^2) \label{eqn:step-2}
\end{align}
\begin{align}
        &=\int_{}{}\frac{d^4k}{(2\pi)^4} \frac{2e^2g_{\ell}(m_X^2)g_q \text{Re}\left({\Pi}_{\gamma X}(k^2)-{\Pi}_{\gamma X}(m_X^2)\right)}{(k^2-m_X^2)} \mathcal{F}_{\mu}(k)\nonumber \\
        &-\int_{}{}\frac{d^4k}{(2\pi)^4} \frac{2e^4g_q^2 \left(\text{Re}~({\Pi}_{\gamma X}(k^2)-{\Pi}_{\gamma X}(m_X^2))\right)\left(\text{Re}~({\Pi}_{\gamma X}(m_X^2)-{\Pi}_{\gamma X}(p_0^2))\right)}{(k^2-m_X^2)} \mathcal{F}_{\mu}(k)\label{eqn:step-3}\nn\\
        &~~~~~~~~~~~~~~~ +2e^2g_{\ell}(p_0^2)g_q \text{Re}\left({\Pi}_{\gamma X}(m_X^2)-{\Pi}_{\gamma X}(p_0^2)\right)4\pi K(m_X^2) 
\end{align}
\begin{align}
        &=a^{\gamma-X}_{\mu}(m_X^2)-\int_{}{}\frac{d^4k}{(2\pi)^4} \frac{2e^4g_q^2 \left(\text{Re}~({\Pi}_{\gamma X}(k^2)-{\Pi}_{\gamma X}(m_X^2))\right)\left(\text{Re}~({\Pi}_{\gamma X}(m_X^2)-{\Pi}_{\gamma X}(p_0^2))\right)}{(k^2-m_X^2)} \mathcal{F}_{\mu}(k)\nn\\
        &~~~~~~~~~~~~~~~ +2e^2g_{\ell}(p_0^2)g_q \text{Re}\left({\Pi}_{\gamma X}(m_X^2)-{\Pi}_{\gamma X}(p_0^2)\right)4\pi K(m_X^2) ~.
\end{align}
Note that, we use Eq.~\eqref{eq:1_loop_X} to go from Eq.~\eqref{eqn:step-1} to Eq.~\eqref{eqn:step-2} and Eq.~\eqref{gl-run} to obtain Eq.~\eqref{eqn:step-3} from Eq.~\eqref{eqn:step-2}.   
This gives the total $a_{\mu}$ contribution upto 1 VP insertion:
\begin{align}
    &{a}_{\mu}^{X}(p_0^2)+{a}_{\mu}^{\gamma-X}(p_0^2)=\frac{1}{4\pi^2} K(m_X^2) \left(g_{\ell}^2(p_0^2)+2e^2g_{\ell}(p_0^2)g_q {\rm Re}\left({\Pi}_{\gamma X}(m_X^2)-{\Pi}_{\gamma X}(p_0^2)\right) \right)
    +a_{\mu}^{\gamma-X}(m_X^2)\nonumber\\
    &-\int_{}{}\frac{d^4k}{(2\pi)^4} \frac{2e^4g_q^2 \left(\text{Re}~({\Pi}_{\gamma X}(k^2)-{\Pi}_{\gamma X}(m_X^2))\right)\left(\text{Re}~({\Pi}_{\gamma X}(m_X^2)-{\Pi}_{\gamma X}(p_0^2))\right)}{(k^2-m_X^2)} \mathcal{F}_{\mu}(k)\label{eqn:step-4}\\
    &=\frac{1}{4\pi^2} K(m_X^2) \left( g_{\ell}^2(m_X^2)+a_{\mu}^{\gamma-X}(m_X^2)-\int_{}{}\frac{d^4k}{(2\pi)^4} \frac{e^4g_q^2 \left(\text{Re}~({\Pi}_{\gamma X}(m_X^2)-{\Pi}_{\gamma X}(p_0^2))\right)^2}{(k^2-m_X^2)} \right)\mathcal{F}_{\mu}(k)\nn\\
    &-\int_{}{}\frac{d^4k}{(2\pi)^4} \frac{2e^4g_q^2 \left(\text{Re}~({\Pi}_{\gamma x}(k^2)-{\Pi}_{\gamma X}(m_X^2))\right)\left(\text{Re}~({\Pi}_{\gamma X}(m_X^2)-{\Pi}_{\gamma X}(p_0^2))\right)}{(k^2-m_X^2)} \mathcal{F}_{\mu}(k)\nn\\
    &={a}_{\mu}^{X}(m_X^2)+{a}_{\mu}^{\gamma-X}(m_X^2)-\int_{}{}\frac{d^4k}{(2\pi)^4} \frac{e^4g_q^2 \left(\text{Re}~({\Pi}_{\gamma X}(m_X^2)-{\Pi}_{\gamma X}(p_0^2))\right)^2}{(k^2-m_X^2)} \mathcal{F}_{\mu}(k) \nonumber\\
    &-\int_{}{}\frac{d^4k}{(2\pi)^4} \frac{2e^4g_q^2 \left(\text{Re}~({\Pi}_{\gamma X}(k^2)-{\Pi}_{\gamma X}(m_X^2))\right)\left(\text{Re}~({\Pi}_{\gamma X}(m_X^2)-{\Pi}_{\gamma X}(p_0^2))\right)}{(k^2-m_X^2)} \mathcal{F}_{\mu}(k)~. \label{eqn:step-5}
\end{align}
where we use Eq.~\eqref{eq:1_loop_X} to massage the first term of Eq.~\eqref{eqn:step-4} to the first term of Eq.~\eqref{eqn:step-5}.
Now we also include the 2 VP insertions contribution given by:
\begin{align}
{a}_{\mu}^{\gamma- X -\gamma}(p_0^2)&=\int_{}{}\frac{d^4k}{(2\pi)^4} \frac{e^4g_q^2 \left(\text{Re}\left({\Pi}_{\gamma X}(k^2)-{\Pi}_{\gamma X}(0)\right)\right)^2}{(k^2-m_X^2)} \mathcal{F}_{\mu}(k)\nn\\
    &=\int_{}{}\frac{d^4k}{(2\pi)^4} \frac{e^4g_q^2 \left(\text{Re}~({\Pi}_{\gamma X}(k^2)-{\Pi}_{\gamma X}(m_X^2))\right)^2}{(k^2-m_X^2)} \mathcal{F}_{\mu}(k)\nn\\
    &+\int_{}{}\frac{d^4k}{(2\pi)^4} \frac{e^4g_q^2 \left(\text{Re}~({\Pi}_{\gamma X}(m_X^2)-{\Pi}_{\gamma X}(0))\right)^2}{(k^2-m_X^2)} \mathcal{F}_{\mu}(k)\nonumber\\
    &+\int_{}{}\frac{d^4k}{(2\pi)^4} \frac{2e^4g_q^2 \left(\text{Re}~({\Pi}_{\gamma X}(k^2)-{\Pi}_{\gamma X}(m_X^2))\right)\left(\text{Re}~({\Pi}_{\gamma X}(m_X^2)-{\Pi}_{\gamma X}(0))\right)}{(k^2-m_X^2)} \mathcal{F}_{\mu}(k)\nonumber
    \end{align}
\begin{align}
    &=a_{\mu}^{\gamma- X -\gamma}(m_X^2)+\int_{}{}\frac{d^4k}{(2\pi)^4} \frac{e^4g_q^2 \left(\text{Re}~({\Pi}_{\gamma X}(m_X^2)-{\Pi}_{\gamma X}(0))\right)^2}{(k^2-m_X^2)} \mathcal{F}_{\mu}(k)\nonumber\\
   &
    +\int_{}{}\frac{d^4k}{(2\pi)^4} \frac{2e^4g_q^2 \left(\text{Re}~({\Pi}_{\gamma X}(k^2)-{\Pi}_{\gamma X}(m_X^2))\right)\left(\text{Re}~({\Pi}_{\gamma X}(m_X^2)-{\Pi}_{\gamma X}(0))\right)}{(k^2-m_X^2)} \mathcal{F}_{\mu}(k)~.
\end{align}
This finally leads to:
\begin{equation}
    {a}_{\mu}^{X}(p_0^2)+{a}_{\mu}^{\gamma-X}(p_0^2)+ {a}_{\mu}^{\gamma- X -\gamma}(p_0^2)={a}_{\mu}^{X}(m_X^2)+{a}_{\mu}^{\gamma-X}(m_X^2)+ {a}_{\mu}^{\gamma- X -\gamma}(m_X^2)~.
\end{equation}
Hence, we have shown the sum of the 1-loop and 2-loop BSM contributions is also scheme independent (we have shown it up to 2 VP insertions but one can continue to check this for higher orders).

\section{Further explorations of the tests and parameter space}

\subsection{Contour plots for the $a_{\mu}$ and HVP tests} \label{app:contour_plots}
\begin{figure}[b]
    \centering
    \begin{subfigure}{0.45\textwidth}
    \includegraphics[width=\textwidth]{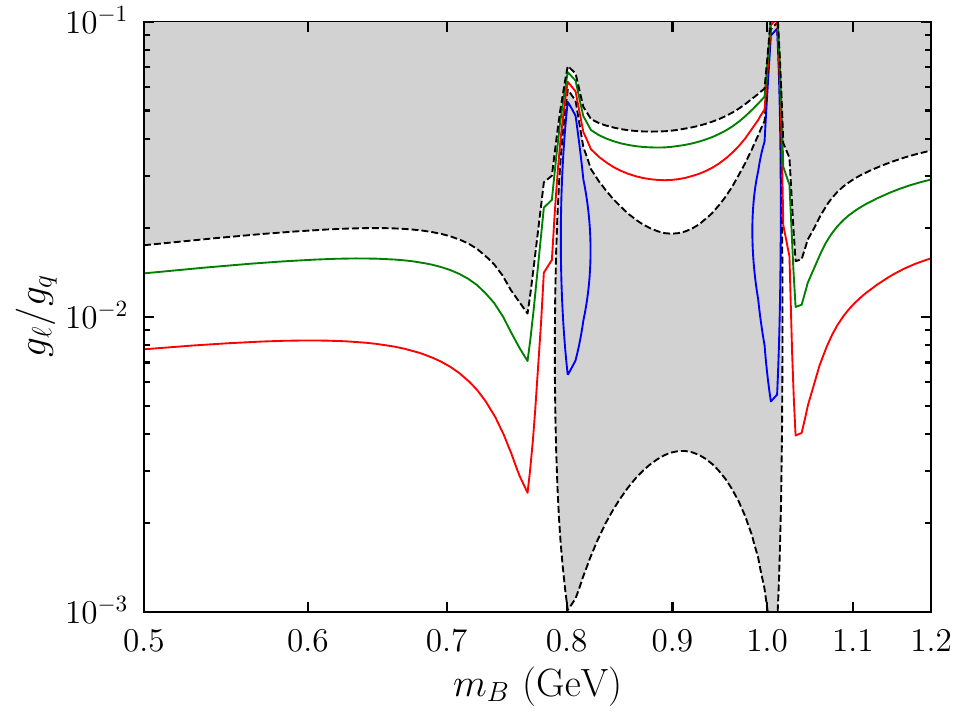}
    \caption{Lattice $a_{\mu}$ test}
    \label{fig:contour-del-amu-lat}
    \end{subfigure}~
    \begin{subfigure}{0.45\textwidth}
    \includegraphics[width=\textwidth]{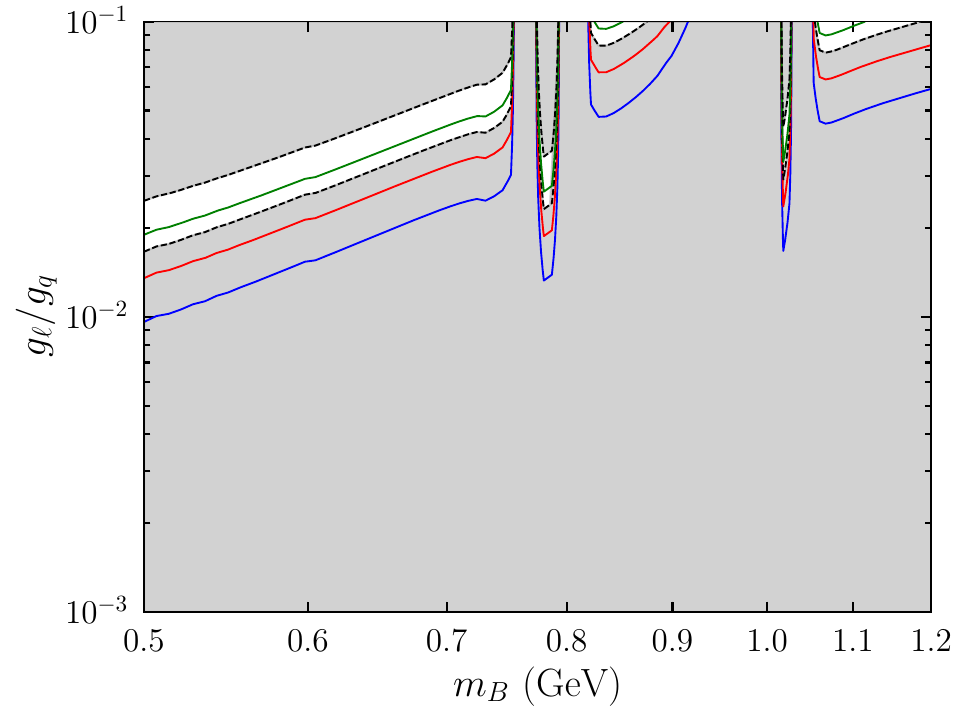}
    \caption{Data-driven $a_{\mu}$ test}
    \label{fig:contour-del-amu-dd}
    \end{subfigure}

    \caption{Contour plots for $\Delta a_{\mu}$ ,in units of $10^{-11}$, in the parameter space of $g_{\l}/g_q$ vs. $m_B$,  for the $B$ gauge boson model. Here we have fixed $g_q=0.15$. ({\it Left:}) We consider the lattice $\amu$ test (c.f. Eq.~\eqref{eqn:amu-lattice-def}). The blue, red and green contour lines are for $\Delta a_{\mu}=-100,50,150$. ({\it Right:}) We consider the data-driven $\amu$ test (c.f. Eq.~\eqref{eqn:amutest:data}). The blue, red and green contour lines are for $\Delta a_{\mu}=50,100,200$. In both figures, the black dashed lines represent the contours for $\Delta a_{\mu}= \text{current difference }\pm 2 \sigma$, and hence the grey regions are currently excluded by the corresponding $a_{\mu}$ tests. }
    \label{fig:contour-del-amu}
    
\end{figure}
\begin{figure}[t]
    \centering
    \begin{subfigure}{0.45\textwidth}
        \includegraphics[width=\textwidth]{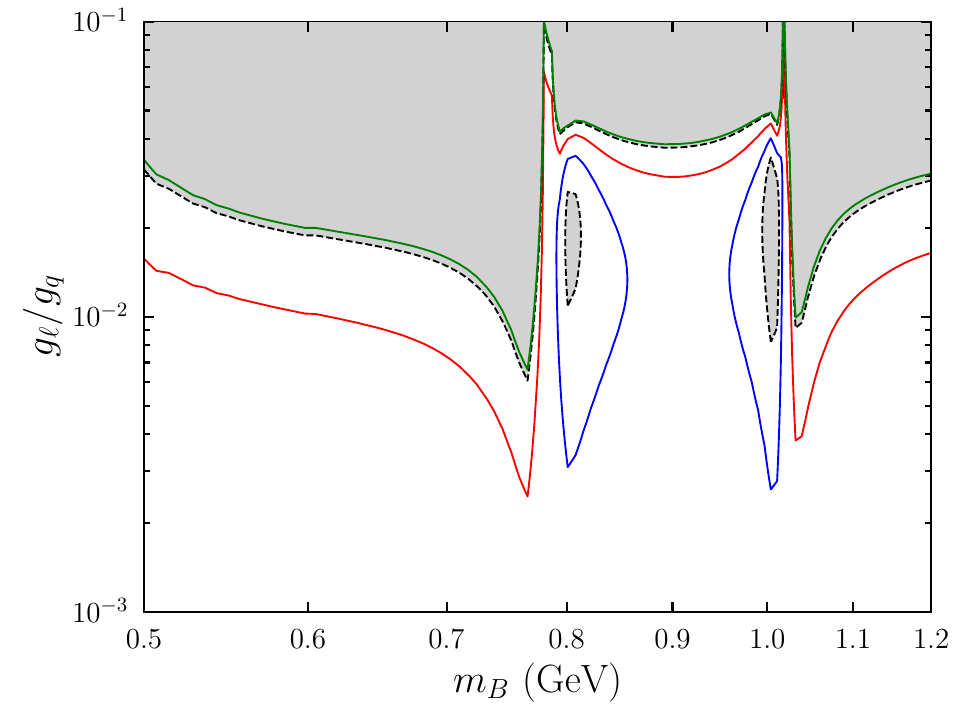}
        \caption{$(\amu^{\rm HVP})^{\text{DD}}=(\amu^{\rm HVP})^{\rm{lat}}~$}
        \label{fig:contour-del-hvp-lat}
    \end{subfigure}~
    \begin{subfigure}{0.45\textwidth}
        \includegraphics[width=\textwidth]{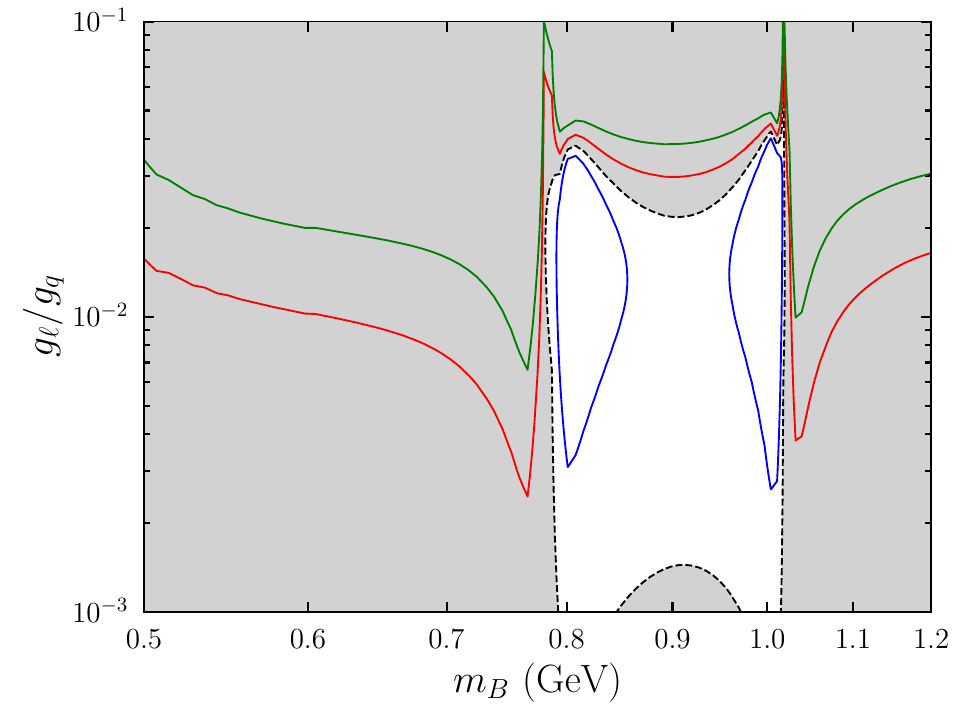}
        \caption{$(\amu^{\rm HVP})^{\text{DD}}=(\amu^{\rm HVP})^{\text{TI}}~$}
        \label{fig:contour-del-hvp-ti}
    \end{subfigure}
    \caption{Contour plots of $\Delta a^{\text{HVP}}_{\mu}$ (see Eq.~\eqref{eq:hvp_test_def}), in units of $10^{-11}$, in the parameter space of $g_{\l}/g_q$ vs. $m_B$,  for the $B$ gauge boson model. Here we have fixed $g_q=0.15$. 
    ({\it Left:}) We consider a futuristic conservative scenario of $(a^{ \rm  HVP }_{ \mu })^{\text{DD}}  =  (a^{ \rm  HVP }_{ \mu })^{\rm{lat}}.({\it Right:})$ We consider the current status of $(a^{ \rm  HVP }_{ \mu })^{\text{DD}}  =  (a^{ \rm  HVP }_{ \mu })^{\rm{lat}} -144$ without the $B$ contribution.  
    The blue, red and green contour lines are for $\Delta \amu^{\text{HVP}}=-50,50,150$  while the black dashed lines represent the contours for $\Delta a_{\mu}= \pm 2 \sigma$, and hence, the gray regions are excluded by the HVP test. 
    }
    \label{fig:contour-del-hvp}
\end{figure}

In this section, we address the value added by the $a_{\mu}$ and HVP test individually, before we combine them with existing constraints in the next section. In Fig.~\ref{fig:contour-del-amu-lat} we focus on the lattice $a_{\mu}$ test by plotting the contours for $\Delta a_{\mu}=a^{B}_{\mu}+a^{\gamma-B}_{\mu}$ in the plane of $g_{\ell}/g_{q}$ and $m_B$. The dashed black lines show the contours for $\Delta a_{\mu}= a^{\text{exp}}_{\mu}-a^{\rm{lat}}_{\mu}\pm 2 \sigma$, and the solid colored lines show contours for some fixed values of $\Delta a_{\mu}$. 
 In Fig.~\ref{fig:contour-del-amu-dd} we focus on the data-driven $a_{\mu}$ test by plotting contours for $\Delta a_{\mu}=a^{B}_{\mu}-(a^{\rm{HVP}}_{\mu})^{B}_{e^+e^-\to \rm{had}}$ in the plane of $g_{\ell}/g_{q}$ and $m_B$. In Fig.~\ref{fig:contour-del-hvp}, we focus on the HVP by plotting contours for $\Delta a^{\rm{HVP}}_{\mu}=a^{BB}_{\mu}+a^{\gamma-B}_{\mu}$. Fig.~\ref{fig:contour-del-hvp-lat} is for the futuristic case where $\Delta a_{\mu}= (a^{\text{ HVP}}_{\mu})^{\text{DD}}=(a^{\text{ HVP}}_{\mu})^{\rm{lat}}_{\mu}$ while Fig.~\ref{fig:contour-del-hvp-ti} is for the case where $(a^{\text{ HVP}}_{\mu})^{\text{DD}}=(a^{\text{ HVP}}_{\mu})^{\text{TI}}$ . The dashed black lines show the contours for $\Delta a_{\mu}= (a^{\text{ HVP}}_{\mu})^{\text{DD}}-(a^{\text{ HVP}}_{\mu})^{\rm{lat}}\pm 2 \sigma$, and the solid colored lines again show contours for some fixed values of $\Delta a^{\text{HVP}}_{\mu}$. The gray regions, in both the plots, are excluded by the respective tests. Since, the experimental,lattice and data-driven numbers (the differences between them and their errors) can change but the various contours remain fixed, these plots give an idea about how the contributions from $B$ can be excluded/included if these numbers shift. We already see that we rule out a lot of parameter space just by considering these two tests. In the most extreme case i.e. when we use Fig.~\ref{fig:contour-del-hvp-ti}, we rule out almost all of the parameter space through a single test.

\subsection{Plots for different $g_\ell/g_q$ for the $B$ boson}\label{app:additional_figues_gq_mb}

\begin{figure}[h]
\centering
    \begin{subfigure}{0.45\textwidth}
    \captionsetup{justification=centering}
    \includegraphics[width=\textwidth]{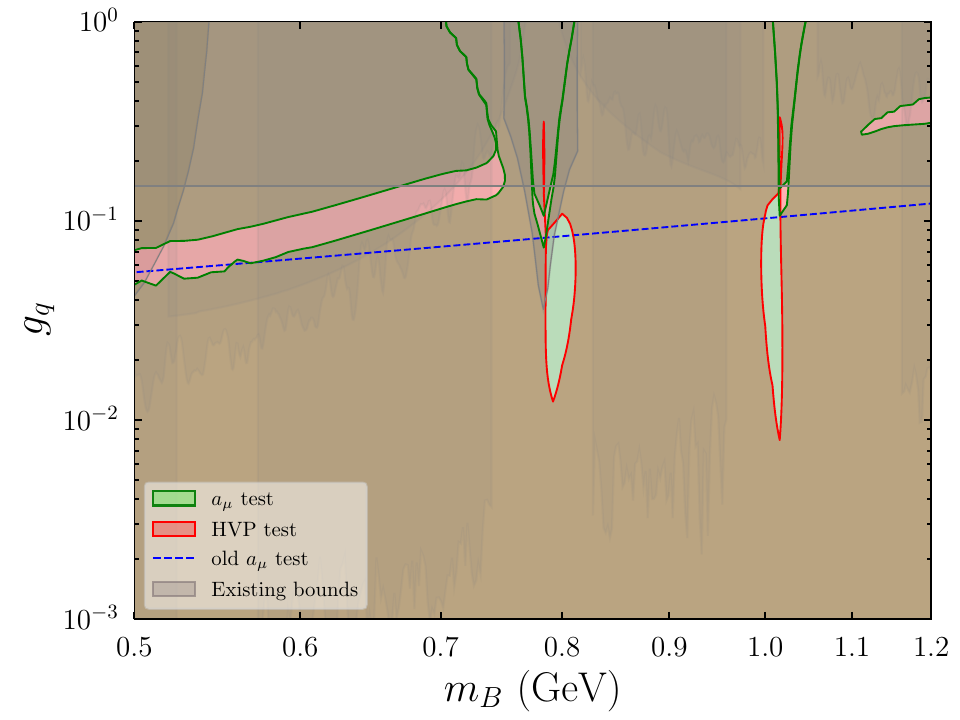}
    \caption{$(a^{ \rm  HVP }_{ \mu })^{\text{DD}}  =  (a^{ \rm  HVP }_{ \mu })^{\text{TI}}$, $g_\l/g_q=0.05$}
    \label{fig:05_dd_ti}
    \end{subfigure}~
    \begin{subfigure}{0.45\textwidth}
    \captionsetup{justification=centering}\includegraphics[width=\textwidth]{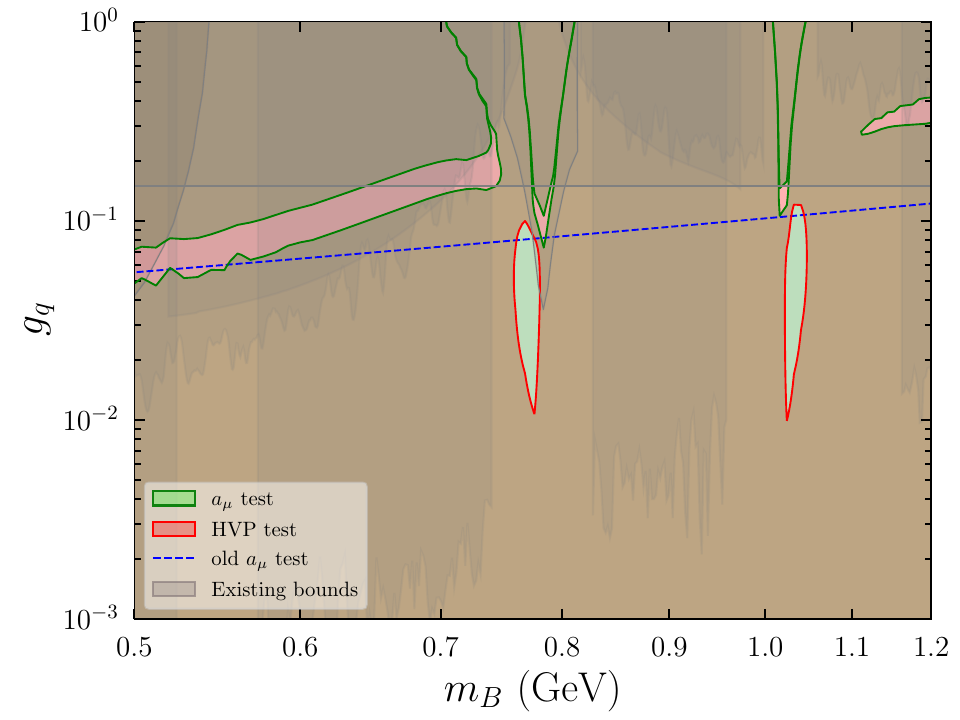}
    \caption{$(a^{ \rm  HVP }_{ \mu })^{\text{DD}} = (a^{ \rm  HVP }_{ \mu })^{\text{TI}}$, $g_\l/g_q=-0.05$}
    \label{fig:05_neg_dd_ti}
    \end{subfigure}
    \caption{ Parameter space of the quark coupling $g_q$ as the function of the mass of \zprime mass ($m_B$) for $g_\l/g_q=0.05$ (Fig.~\ref{fig:05_dd_ti}) and $g_\l/g_q=-0.05$ (Fig.~\ref{fig:05_neg_dd_ti}). 
    In this plot, we consider the data-driven $a_{\mu}$ test (c.f. Eq.~\eqref{eqn:amutest:data}). Both for the HVP test (c.f. Eq.~\eqref{eq:hvp_test_def}) and the $a_{\mu}$, we take TI 2020 as the data driven value. The colour scheme is the same as the main text.}
    \label{fig:add_fig_05_ti}
\end{figure}
    \begin{figure}[h]
\centering

    \begin{subfigure}
    {0.45\textwidth}
    \captionsetup{justification=centering}\includegraphics[width=\textwidth]{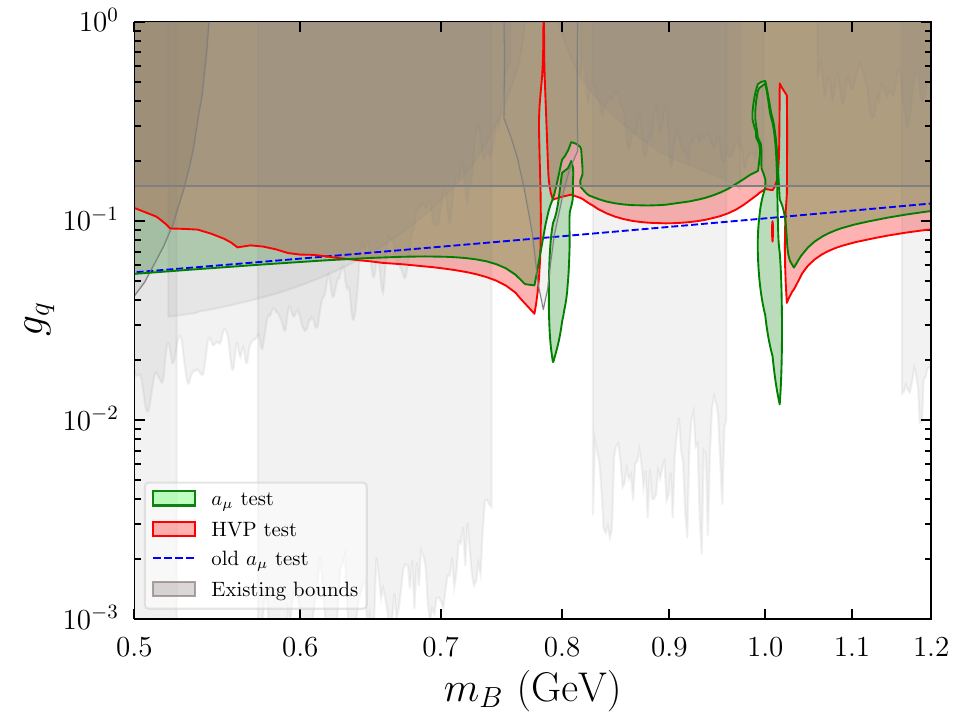}
    \caption{$(a^{ \rm  HVP }_{ \mu })^{\text{DD}}= (a^{ \rm  HVP }_{ \mu })^{\rm{lat}}$, $g_\l/g_q=0.05$}
    \label{fig:05_dd_lat}
    \end{subfigure}
    \begin{subfigure}{0.45\textwidth}
    \captionsetup{justification=centering}\includegraphics[width=\textwidth]{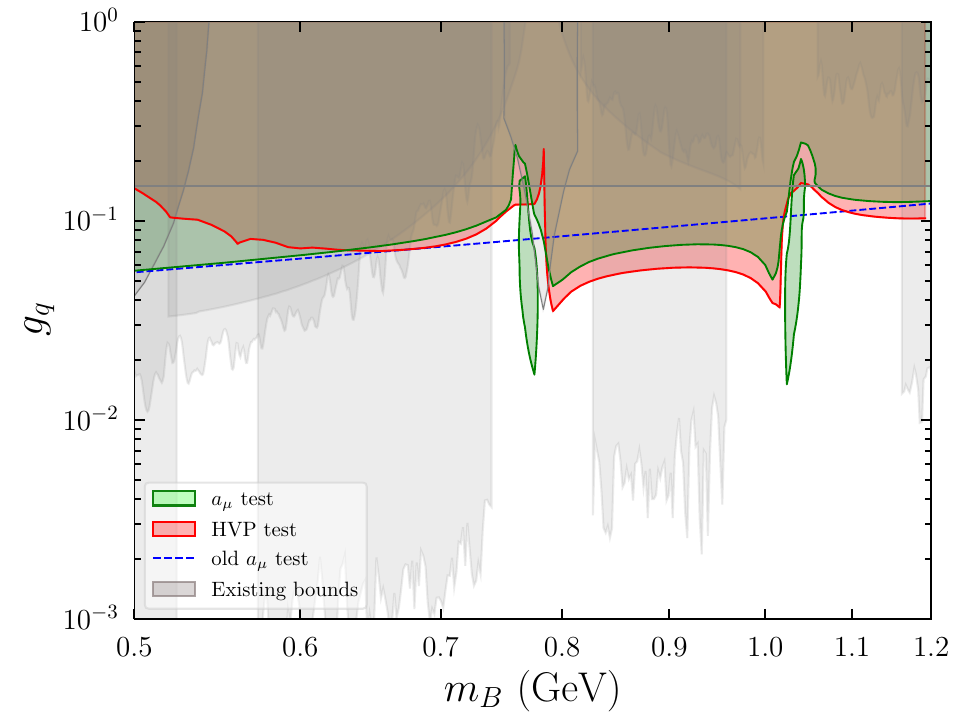}
    \caption{$(a^{ \rm  HVP }_{ \mu })^{\text{DD}}= (a^{ \rm  HVP }_{ \mu })^{\rm{lat}}$, $g_\l/g_q=-0.05$}
    \label{fig:05_neg_dd_lat}
    \end{subfigure}

    \caption{Parameter space of the quark coupling $g_q$ as the function of the mass of \zprime mass ($m_B$) for  $g_\l/g_q=0.05$ (Fig.~\ref{fig:05_dd_lat}) and $g_\l/g_q=-0.05$ (Fig.~\ref{fig:05_neg_dd_lat}). 
    We perform the lattice $a_{\mu}$ test (c.f. Eq.~\eqref{eqn:amu-lattice-def}) by taking the BMW result as the SM contribution. 
    For the HVP test (c.f. Eq.~\eqref{eq:hvp_test_def}), we consider a futuristic conservative possibility of the central values of the data driven HVP and the lattice result to match with each other within their respective
    error bars. The colour scheme is the same as the main text.} 

\label{fig:add_fig_05_lat}
\end{figure}

\begin{figure}[t]

\centering
    \begin{subfigure}{0.45\textwidth}
    \captionsetup{justification=centering}
    \includegraphics[width=\textwidth]{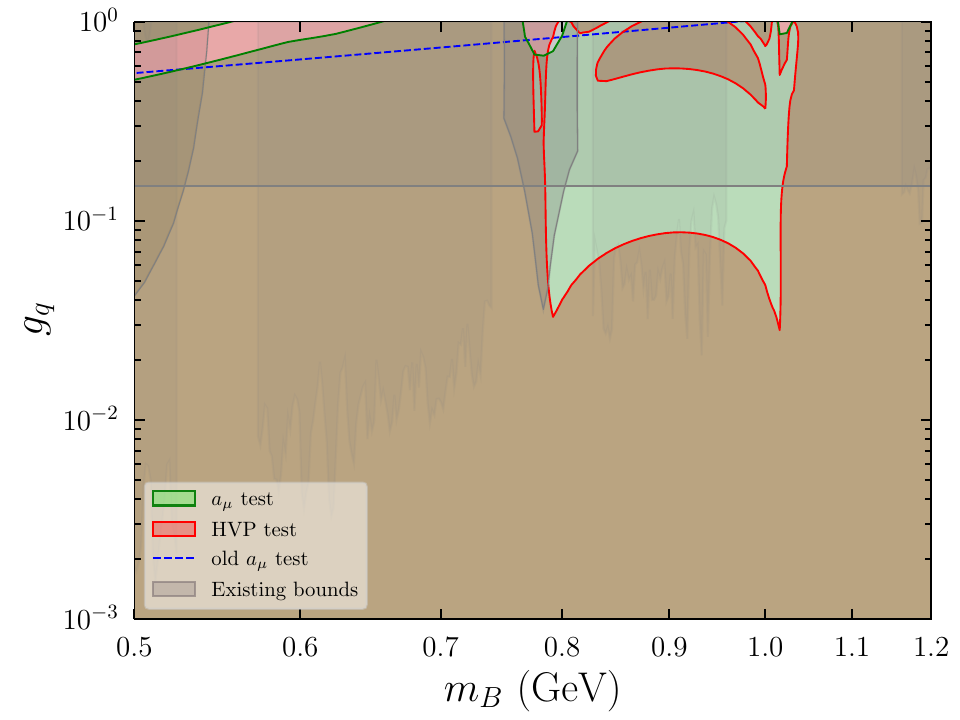}
    \caption{$(a^{ \rm  HVP }_{ \mu })^{\text{DD}}  =  (a^{ \rm  HVP }_{ \mu })^{\text{TI}}$, $g_\l/g_q=0.005$}
    \label{fig:005_dd_ti}
    \end{subfigure}~
    \begin{subfigure}{0.45\textwidth}
    \captionsetup{justification=centering}\includegraphics[width=\textwidth]{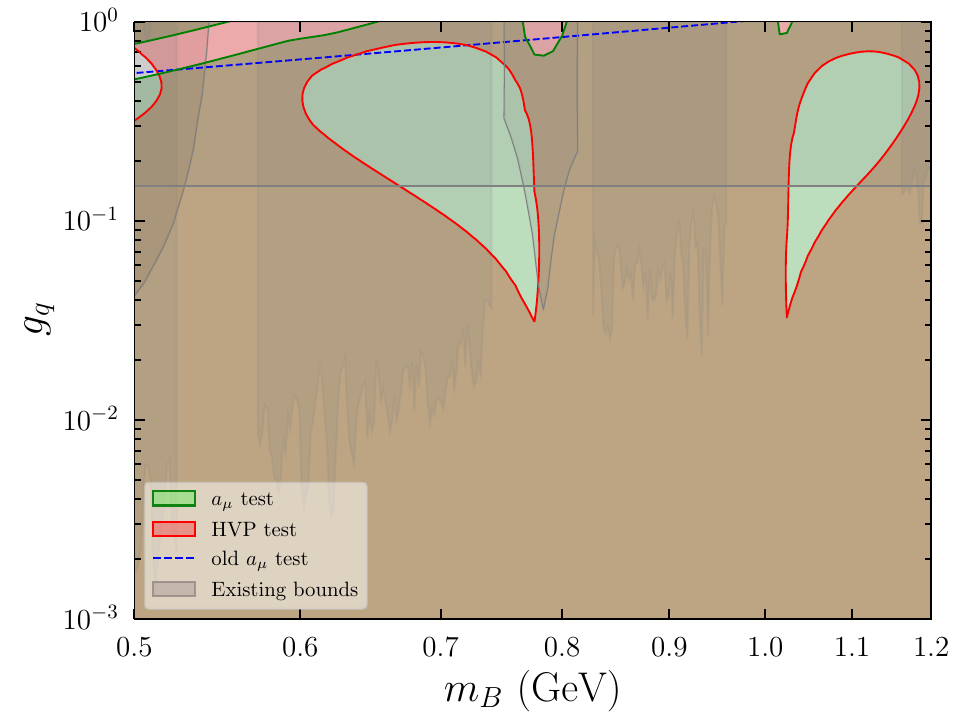}
    \caption{$(a^{ \rm  HVP }_{ \mu })^{\text{DD}} = (a^{ \rm  HVP }_{ \mu })^{\text{TI}}$, $g_\l/g_q=-0.005$}
    \label{fig:005_neg_dd_ti}
    \end{subfigure}
    \caption{  
Parameter space of the quark coupling $g_q$ as the function of the mass of \zprime mass ($m_B$) for $g_\l/g_q=0.005$ (Fig.~\ref{fig:005_dd_ti}) and $g_\l/g_q=-0.005$ (Fig.~\ref{fig:005_neg_dd_ti}). Restmatches Fig.~\ref{fig:add_fig_05_ti}.}
    \label{fig:add_fig_005_ti}
    \end{figure}
    \begin{figure}[h]
    \centering
    
    \begin{subfigure}{0.45\textwidth}
    \captionsetup{justification=centering}
    \includegraphics[width=\textwidth]{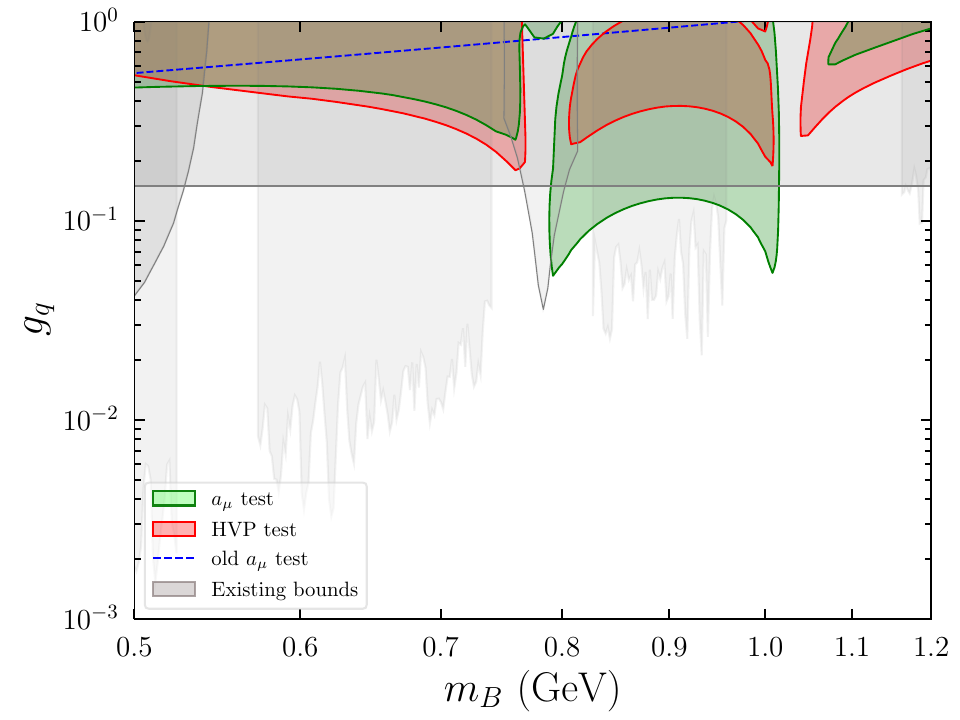}
    \caption{$(a^{ \rm  HVP }_{ \mu })^{\text{DD}}= (a^{ \rm  HVP }_{ \mu })^{\rm{lat}}$, $g_\l/g_q=0.005$}
    \label{fig:005_dd_lat}
    \end{subfigure}~
    \begin{subfigure}{0.45\textwidth}
    \captionsetup{justification=centering}\includegraphics[width=\textwidth]{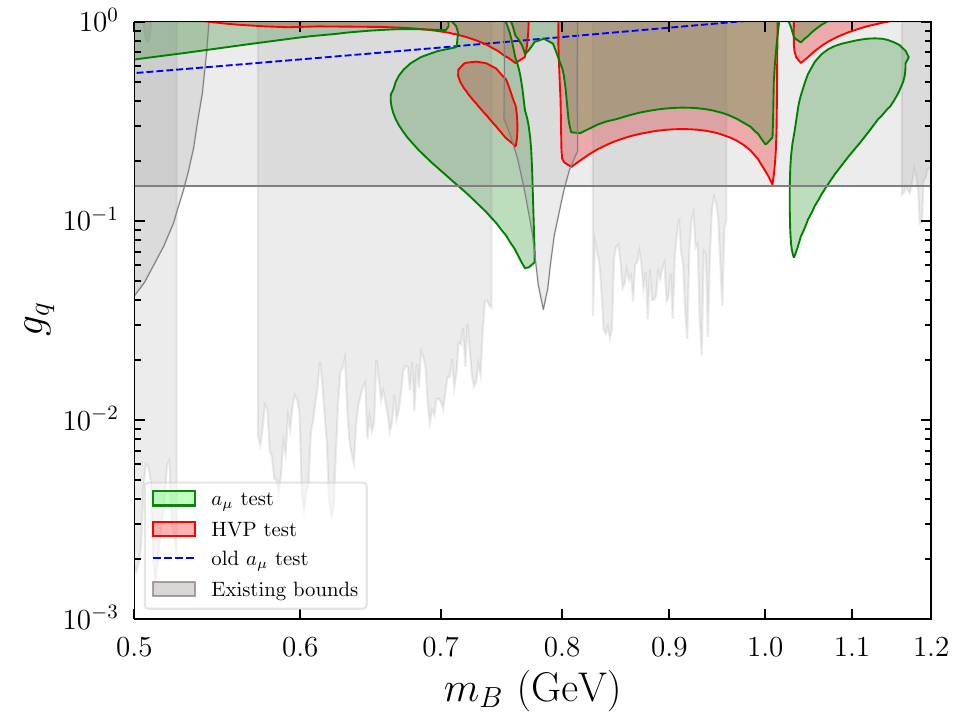}
    \caption{$(a^{ \rm  HVP }_{ \mu })^{\text{DD}}= (a^{ \rm  HVP }_{ \mu })^{\rm{lat}}$, $g_\l/g_q=-0.005$}
    \label{fig:005_neg_dd_lat}
    \end{subfigure}

    \caption{Parameter space of the quark coupling $g_q$ as the function of the mass of \zprime mass ($m_B$) for $g_\l/g_q=0.005$ (Fig.~\ref{fig:005_dd_ti}) and $g_\l/g_q=-0.005$ (Fig.~\ref{fig:005_neg_dd_ti}). Rest matches Fig.~\ref{fig:add_fig_05_lat} .} 

\label{fig:add_fig_005_lat}
\end{figure}

In this section, we provide plots for the baryon number gauge boson, i.e. $g_q$ vs. $m_B$ for some additional values of $g_q/g_\l$ (which we denote by $r$ here). We start by considering the current scenario i.e. taking the TI 2020 number for the data-driven HVP. Using this number, we perform the HVP test along with the data-driven $\amu$ test. We do it for two values of $r$, namely $r=0.05$ and $r=0.005$ in Fig. \ref{fig:05_dd_ti} and \ref{fig:005_dd_ti} respectively. We see that for both the $\amu$ and HVP tests, we require New Physics(NP) at the $\pm 2\sigma$ level. Due to the resonant $B$ term being positive and large (see Eq.~\eqref{eq:amu_estimates_onshellx2}) $\Delta\amu^{\text{HVP}}$ being negative in the HVP test, most of the points in the parameter space are excluded. Only points close to the hadronic resonances manage to make $\amu^{\gamma-B}$ large and negative enough to satisfy the HVP test. With the data-driven $\amu$ test, we still exclude most of the parameter space due to the cancellations between the $(\amu^{\rm HVP})^{B}_{e^+e^-\to \rm{had}}$ and $\amu^B$ in Eq.~\eqref{eqn:amutest:data}. Since both these quantities depend on $g_l^2$, the numbers for $\pm 2\sigma$ bounds on $g_q$ from the data-driven $\amu$ test become larger with increasing $r$ (compare Fig. \ref{fig:gq_mb_SM_dd}, \ref{fig:05_dd_ti} and \ref{fig:005_dd_ti}) and thus the bounds are weaker(since existing constraints already require $g_q \leq0.15$). \\
To examine the effects of switching the sign of $\amu^{\gamma-B}$, we consider the same $r$ values but of opposite signs in Fig.~\ref{fig:05_neg_dd_ti} and \ref{fig:005_neg_dd_ti}. Since the $\amu$ test only depends on $g_l^2$, the sign flip does not affect it. Due to $\amu^{\gamma-B}$ changing signs, we see that there are no regions satisfying the HVP tests for both signs for a particular $r$. Another signature of the sign fip is that the regions satisfying the HVP tests are on the opposite side of the hadronic resonances  (compared to the positive $r$ cases ).\\
In  Fig.~\ref{fig:05_dd_lat} and \ref{fig:005_dd_lat}, we perform the lattice $a_{\mu}$ test (c.f. Eq.~\eqref{eqn:amu-lattice-def}) for $g_\l/g_q=0.05$ and $0.005$ respectively. 
As mentioned in Sec.~\ref{subsec:baryon_number}, for the HVP test, we consider the futuristic assumption, $(a_\mu^{\text{HVP}})^{\text{DD}}=(a_\mu^{\text{HVP}})^{\rm{lat}}$.
For the choice of the lattice $a_{\mu}$ test, no NP is required to satisfy either of the tests within the $2\sigma$ value. 
making $g_\l,g_q\to 0$ consistent with that choice. 
For this choice, both the tests depend on the $\amu^{\gamma-B}$ which could be of either sign, and can be large close to the $\omega$ and/or $\phi$ mass. 
Near these masses, a large but negative $\amu^{\gamma-B}$  can offset the positive $(\amu^{\rm HVP})^{B}_{e^+e^-\to \rm{had}}$ and $\amu^{B}$ contributions. This is the reason we see closed red curves and the closed green curves, in the $m_B\sim[0.8,1]\GeV$ region, are excluded because those obtain values below the $2\sigma$ region for the respective tests. 
Other regions show the parameter space which is above the $2\sigma$ value, and thus are excluded. For a positive $\amu^{\gamma-B} $, we see that for both the HVP and lattice $a_{\mu}$ test, we need to satisfy a condition which is roughly: $c_1g_qg_\l+c_2g_\l^2<2\sigma$, where $c_1,c_2 >0$. Solving this equation shows that  the $+2\sigma$ bound on $r=0.005$ is the weakest and that on $r=0.05$ is the strongest. \\
 Just like before, we consider the negative $r$ values for these tests as well. These are shown in Fig.~\ref{fig:05_neg_dd_lat} and \ref{fig:005_neg_dd_lat}. This time, the sign flip affects both the tests. We see that the $-2\sigma$ excluded regions for both tests are now just outside the [0.8,1] GeV region, since the $\amu^{\gamma-B}$ is large there, and it flips sign. The HVP test requires $\amu^{\gamma-X}+(\amu^{\rm HVP})^{X}_{e^+e^-\to \rm{had}}$ to lie in $[-2\sigma,2\sigma]$. Thus, the $+2\sigma$ bound from of the HVP test becomes more stringent in the [0.8,1] GeV region, because of $\amu^{\gamma-B}$ being positive (on top of $(\amu^{\rm HVP})^{B}_{e^+e^-\to \rm{had}}$ always being positive).

\subsection{Plots of $g_q$ vs. $g_\ell$ for $B$ boson with a different $m_B$ }\label{app:additional_figures_gq_gl}
In Section \ref{subsec:baryon_number}, we charted the parameter space of $g_q$ and $g_\ell$ for $B$, using a fixed mass $m_B=0.9$ GeV. Here, in Fig. \ref{fig:gl_gq_0900}, we present the plots for another mass, $m_B=1.085$ GeV. In Fig. \ref{fig:gl_gq_dd_sm_09} and \ref{fig:gl_gq_dd_sm_neg_09} we use the TI 2020 number for the data-driven HVP and perform the HVP (c.f. \eqref{eq:hvp_test_def}) and data-driven $\amu$ (c.f. Eq.~\eqref{eqn:amutest:data}) tests. 
Similar to the case of Section \ref{subsec:baryon_number}, we see that not many points satisfy the data-driven $\amu$ test because of the cancellation between $\amu^{B}$ and $(\amu^{\rm HVP})^{B}_{e^+e^-\to \rm{had}}$. We also see that because $\Delta \amu^{\rm{HVP}} $ is very negative, we do not satisfy the HVP test as well. For this case, the HVP test is better than the $\amu$ test.
In Fig. \ref{fig:gl_gq_dd_lat_09} and \ref{fig:gl_gq_dd_lat_neg_09} we take the futuristic case where the lattice and data-driven HVP results are equal and perform the HVP (c.f. \eqref{eq:hvp_test_def}) and lattice $\amu$ (c.f. Eq.~\eqref{eqn:amu-lattice-def}) tests. Here, we do not rule out arbitrarily small couplings. The HVP test fares better than the lattice $\amu$ test place for $g_\l \times g_q >0 $, whereas the lattice $\amu$ test does slightly better for $g_\l \times g_q <0$.  
\begin{figure}[t]
\centering

    \begin{subfigure}{0.45\textwidth}
    \captionsetup{justification=centering}
    \includegraphics[width=\textwidth]{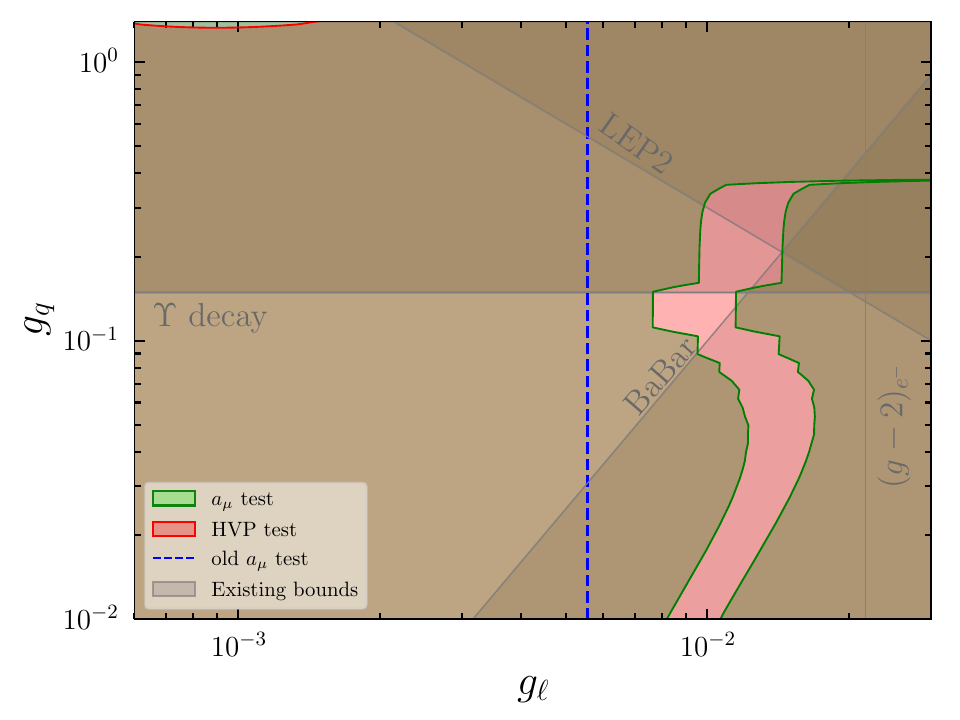}
    \caption{$(a^{ \rm  HVP }_{ \mu })^{\text{DD}}  =  (a^{ \rm  HVP }_{ \mu })^{\text{TI}}$, $g_\l\times g_q>0$}
    \label{fig:gl_gq_dd_sm_09}
    \end{subfigure}~
    \begin{subfigure}{0.45\textwidth}
    \captionsetup{justification=centering}
    \includegraphics[width=\textwidth]{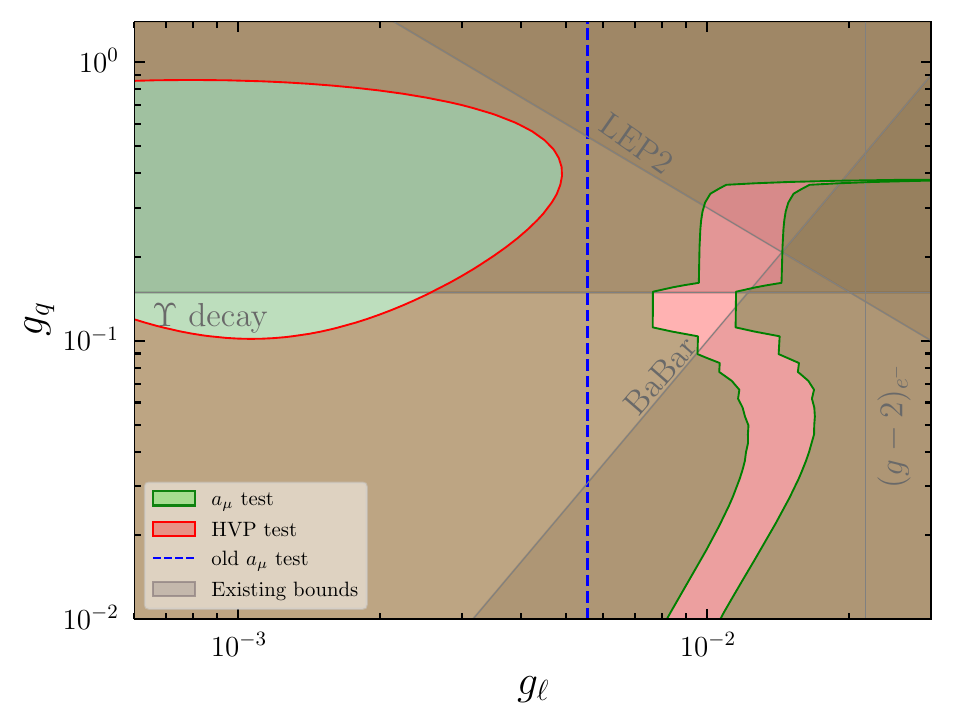}
    \caption{$(a^{ \rm  HVP }_{ \mu })^{\text{DD}} = (a^{ \rm  HVP }_{ \mu })^{\text{TI}}$, $g_\l\times g_q<0$}
    \label{fig:gl_gq_dd_sm_neg_09}
    \end{subfigure}\\
    \begin{subfigure}{0.45\textwidth}
    \captionsetup{justification=centering}
    \includegraphics[width=\textwidth]{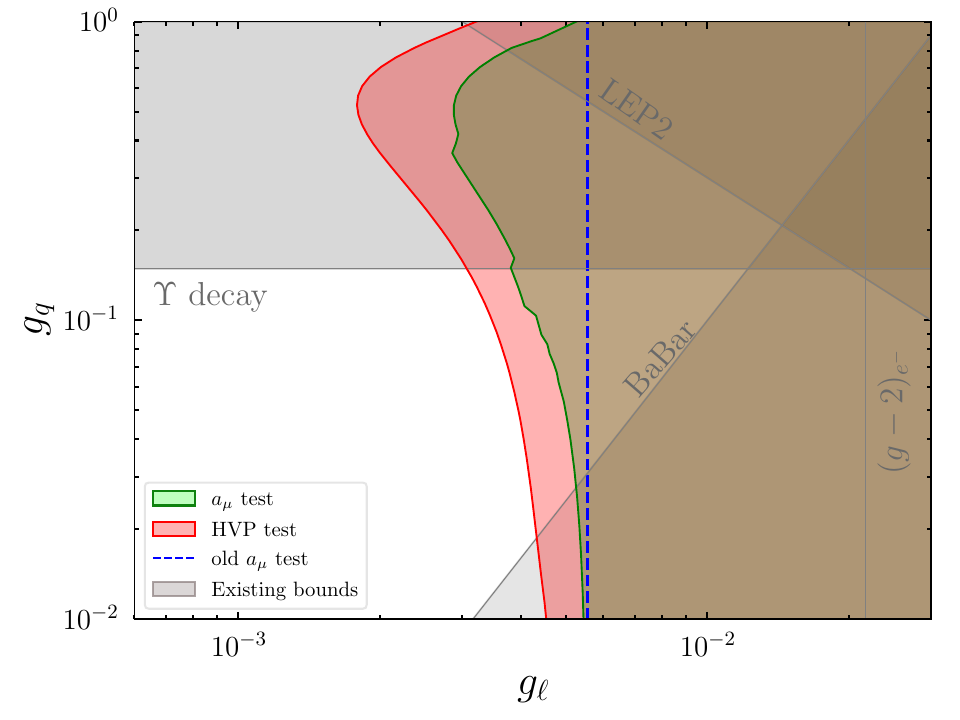}
    \caption{$(a^{ \rm  HVP }_{ \mu })^{\text{DD}}= (a^{ \rm  HVP }_{ \mu })^{\rm{lat}}$, $g_\l\times g_q>0$}
    \label{fig:gl_gq_dd_lat_09}
    \end{subfigure}~
    \begin{subfigure}{0.45\textwidth}
    \captionsetup{justification=centering}
    \includegraphics[width=\textwidth]
    {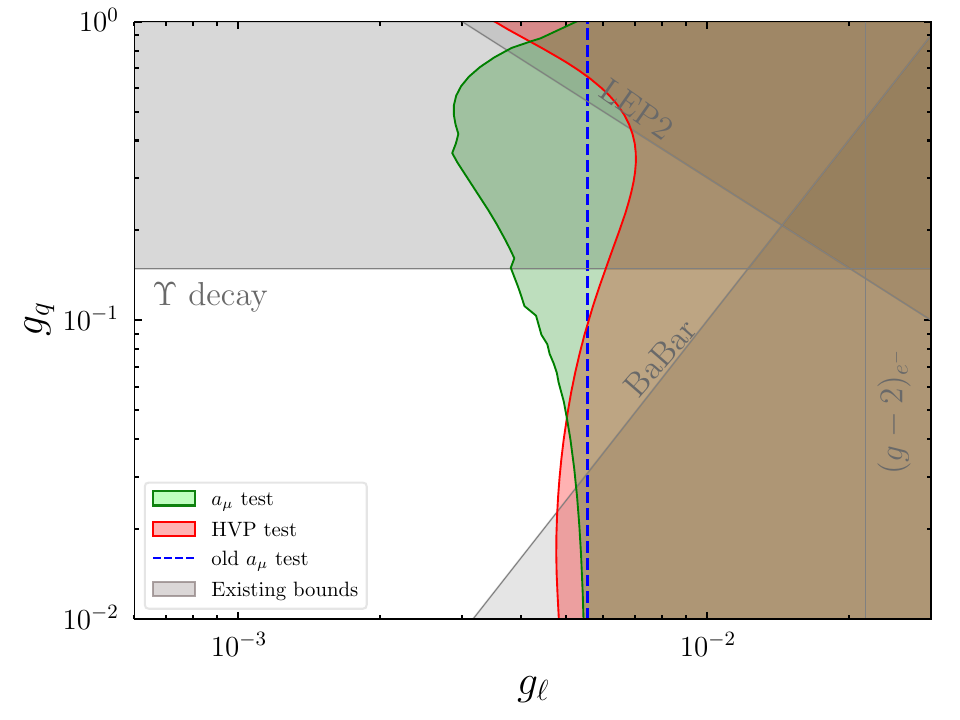}
    \caption{$(a^{ \rm  HVP }_{ \mu })^{\text{DD}}= (a^{ \rm  HVP }_{ \mu })^{\rm{lat}}$, $g_\l\times g_q<0$}
    \label{fig:gl_gq_dd_lat_neg_09}
    \end{subfigure}
    
    \caption{Plot of $g_q$ vs $g_\l$ for a fixed of $m_B=1.085$ GeV both for positive and negative ratio. 
    For the plots with $g_\l\times g_q<0$, we always consider $g_q>0$ and plot $|g_\l|$. 
    In Fig.~\ref{fig:gl_gq_dd_sm} and~\ref{fig:gl_gq_dd_sm_neg}, we perform the data-driven $a_{\mu}$ test (c.f. \eqref{eqn:amutest:data}) and take the TI 2020 value for the HVP test, whereas in Fig.~\ref{fig:gl_gq_dd_lat} and ~\ref{fig:gl_gq_dd_lat_neg}, we consider the lattice $\amu$ test (c.f. \eqref{eq:amu-test}) and take the data-driven HVP to be equal to the lattice result for the HVP test. The colour scheme is the same as the main text.
}
\label{fig:gl_gq_0900}
\end{figure}
\subsection{Plots for various data-driven vs lattice HVP scenarios}\label{app:additional_figures_dd_lat}
In Section \ref{subsec:baryon_number}, we charted the parameter space of both dark photon and $B$ boson for the futuristic case where the $(\amu^{ \rm HVP})^{\rm{DD}}=(\amu^{ \rm HVP})^{\rm{lat}}$. We then performed the lattice $\amu$ and HVP tests. To highlight the complementary nature of the two tests, we now present the results for two cases: Case A: $(\amu^{ \rm HVP})^{\rm{DD}}=(\amu^{ \rm HVP})^{\rm{lat}}+ 1\sigma$ and Case B: $(\amu^{ \rm HVP})^{\rm{DD}}=(\amu^{ \rm HVP})^{\rm{lat}}- 1\sigma$. These cases do not impact the $\amu$ test and only the bounds from the HVP tests will change. We note that, to satisfy the HVP test within $\pm 2 \sigma$, $\amu^{\gamma-X}+(\amu^{\rm HVP})^{X}_{e^+e^-\to \rm{had}}$ has to lie in the range $[-\sigma,3\sigma]$ and $[-3\sigma,\sigma]$ for Case A and B respectively. Since $(\amu^{\rm HVP})^{X}_{e^+e^-\to \rm{had}}$ is always positive, it is more difficult to satisfy the HVP test on the $+2\sigma$ side for Case B. Thus, we expect the $+2\sigma$ bounds to be stronger for this condition. However, we could have $a_{\mu}^{\gamma-B}$ become negative and large (which happens for $m_B$ near hadronic resonances). For such situations, from the allowed ranges shown above, satisfying the HVP test on the $-2\sigma$ is easier if we take Case B, and thus results in weaker $-2\sigma$ bounds. To further illustrate these observations, we consider the two models separately:\\
\\
{\bf\em Dark photon:} For the dark photon, we present the results in Fig. \ref{fig:add_figure_dp}. We see that for Case A the HVP test bounds are completely subsumed by the lattice $\amu$ test bounds, while for Case B, the HVP bounds are roughly equal to the $\amu$ bounds.\\
\\
{\bf \em $B$ boson:} For the $B$ boson, we present the results in Fig.~\ref{fig:add_fig_05_plus}-\ref{fig:add_fig_005_minus}. Again, for Case A, we see that the HVP test is not much better than the lattice $\amu$ test for $m_B \gtrsim 0.7$ GeV. For lower masses, such as $m_B \simeq 0.5$ GeV, the $\amu$ bounds are stronger by $\mathcal{O}(2)$. For Case B, we obtain stronger $+2\sigma$ HVP bounds as argued before. Compared to the $+2\sigma$ HVP bounds of Case A and the $\amu$ bounds, the $+2\sigma$ HVP bounds for Case B can be stronger by $\mathcal{O}(2)$. The $-2\sigma$ HVP bounds for Case A are much stronger than those for Case B (where we even do not have any bounds in some cases). However, they are weaker than the $-2\sigma$ $\amu$ bounds.

\begin{figure}[h]
    \centering
\begin{subfigure}{0.45\textwidth}
\includegraphics[width=\textwidth]{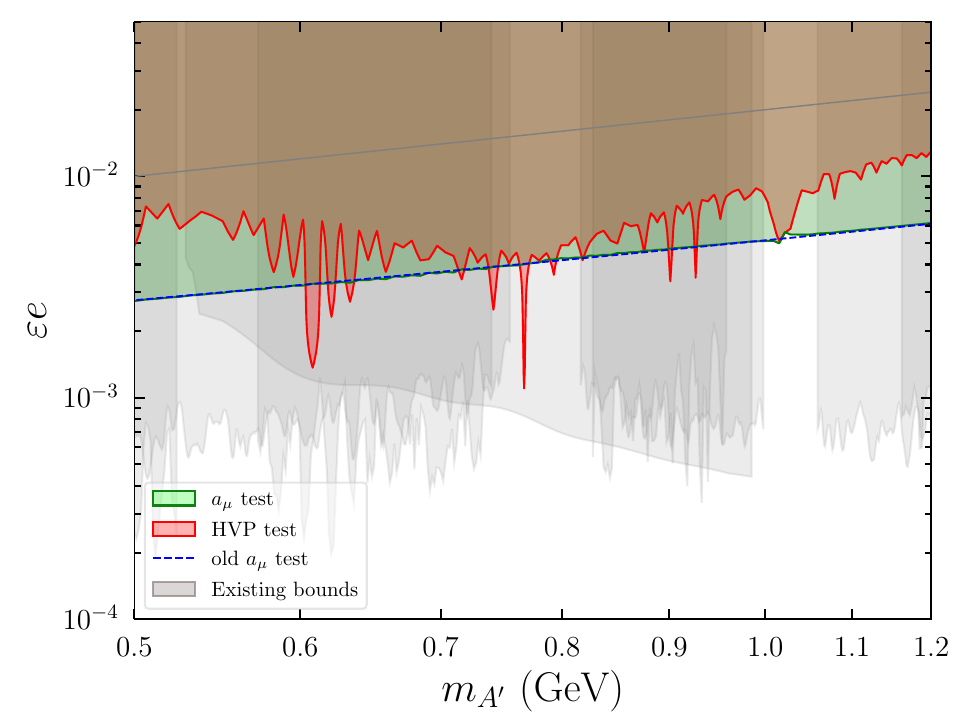}
    \caption{$(a^{ \rm  HVP }_{ \mu })^{\text{DD}}  =  (a^{ \rm  HVP }_{ \mu })^{\text{lat}} +1\,\sigma$}
        \label{fig:dp_plus1}
    \end{subfigure}~
    \begin{subfigure}{0.45\textwidth}
    \includegraphics[width=\textwidth]{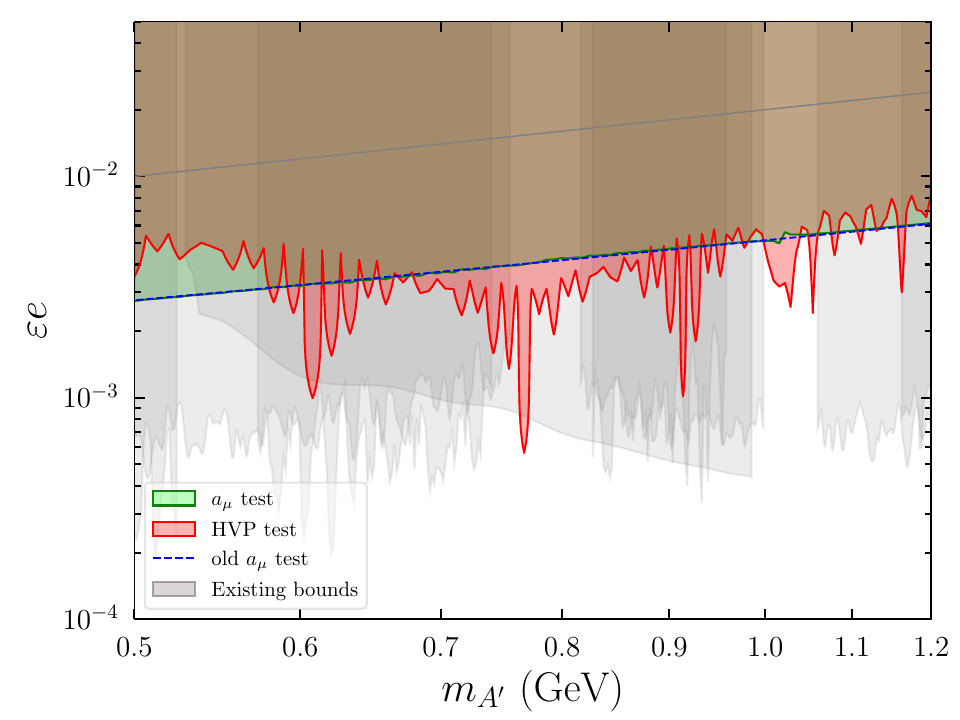}
    \caption{$(a^{ \rm  HVP }_{ \mu })^{\text{DD}}  =  (a^{ \rm  HVP }_{ \mu })^{\text{lat}} -1\,\sigma$}
    \label{fig:dp_minus1}
    \end{subfigure}
    \caption{Parameter space of the dark photon coupling $\varepsilon e$ as a function of its mass $m_{A'}$ for the lattice $\amu$ test (c.f. Eq.~\eqref{eqn:amu-lattice-def}). We use BMW result for the $\amu$ test~\cite{Boccaletti:2024guq}. 
    For the HVP test, we  consider the possibility that $(a^{\rm HVP}_{\mu})^{\rm DD}=(\amu^{\rm HVP})^{\rm lat}+1\sigma$ (Fig.~\ref{fig:dp_plus1}) or $(a^{\rm HVP}_{\mu})^{\rm DD}=(\amu^{\rm HVP})^{\rm lat}-1\sigma$ (Fig.~\ref{fig:dp_minus1}).  
    The colour scheme is the same as the main text. }
    \label{fig:add_figure_dp}
\end{figure}

\begin{figure}[h]
\centering
    \begin{subfigure}
    {0.45\textwidth}
    \captionsetup{justification=centering}
\includegraphics[width=\textwidth]{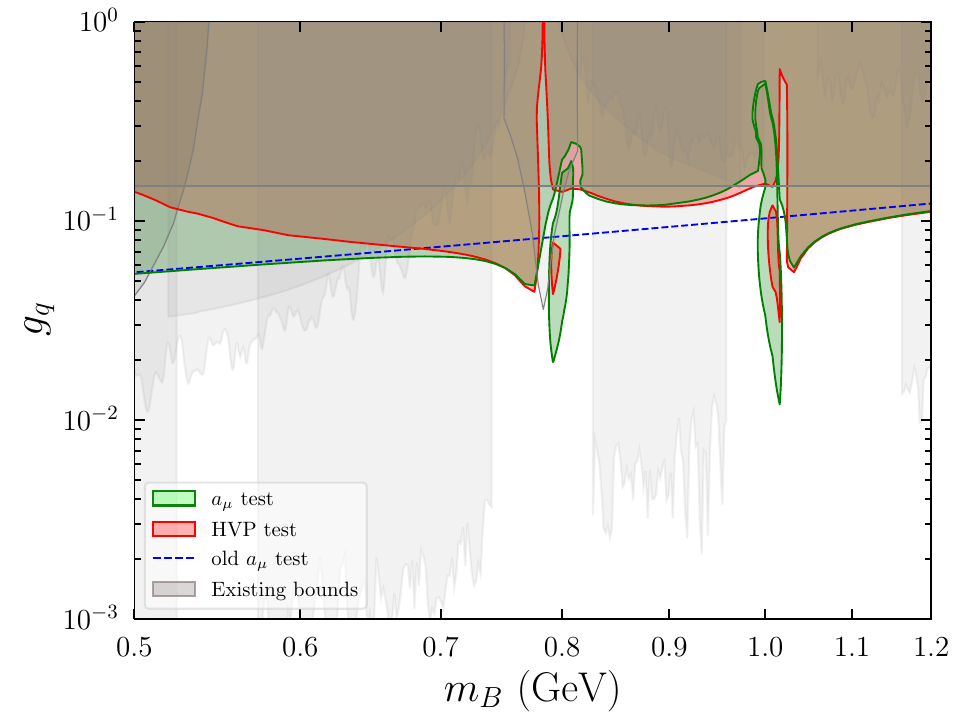}
    \caption{$(a^{ \rm  HVP }_{ \mu })^{\text{DD}}  =  (a^{ \rm  HVP }_{ \mu })^{\text{lat}}+1\sigma$ \\$g_\l/g_q=0.05$}
    \label{fig:05_dd_lat_plus1}
    \end{subfigure}~
    \begin{subfigure}
    {0.45\textwidth}
    \captionsetup{justification=centering}\includegraphics[width=\textwidth]{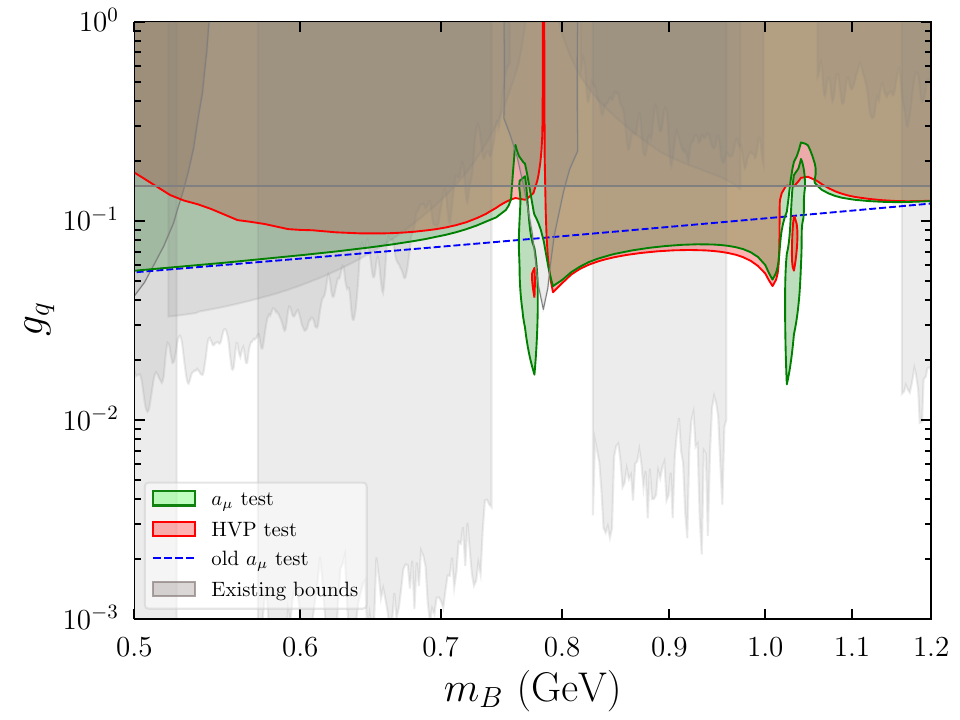}
    \caption{$(a^{ \rm  HVP }_{ \mu })^{\text{DD}} = (a^{ \rm  HVP }_{ \mu })^{\text{lat}}+1\sigma$ \\~~~~ $g_\l/g_q=-0.05$}
    \label{fig:05_neg_dd_lat_plus1}
    \end{subfigure}
    \caption{Parameter space of the quark coupling $g_q$ as the function of the mass of \zprime mass ($m_B$) for  $g_\l/g_q=0.05$ (Fig.~\ref{fig:05_dd_lat_plus1}) and $g_\l/g_q=-0.05$ (Fig.~\ref{fig:05_neg_dd_lat_plus1}). 
    We perform the lattice $a_{\mu}$ test (c.f. Eq.~\eqref{eqn:amu-lattice-def}) by taking the BMW result as the SM contribution. 
    For the HVP test, we  consider the possibility that $(a^{\rm HVP}_{\mu})^{\rm DD}=(\amu^{\rm HVP})^{\rm lat}+1\sigma$.
    The colour scheme is the same as the main text.}
    \label{fig:add_fig_05_plus}
\end{figure}
\begin{figure}[h]
\centering
\begin{subfigure}{0.45\textwidth}
    \captionsetup{justification=centering}
    \includegraphics[width=\textwidth]{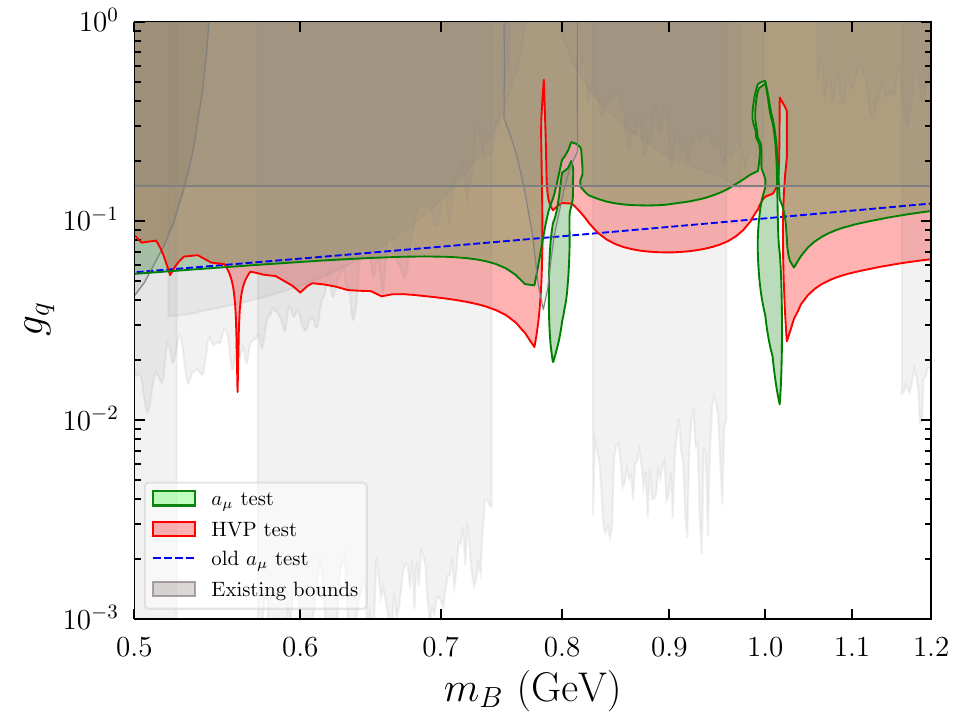}
    \caption{$(a^{ \rm  HVP }_{ \mu })^{\text{DD}}= (a^{ \rm  HVP }_{ \mu })^{\rm{lat}}-1\sigma$\\ $g_\l/g_q=0.05$}
    \label{fig:05_dd_lat_minus1}
    \end{subfigure}~
    \begin{subfigure}{0.45\textwidth}
    \captionsetup{justification=centering}
    \includegraphics[width=\textwidth]{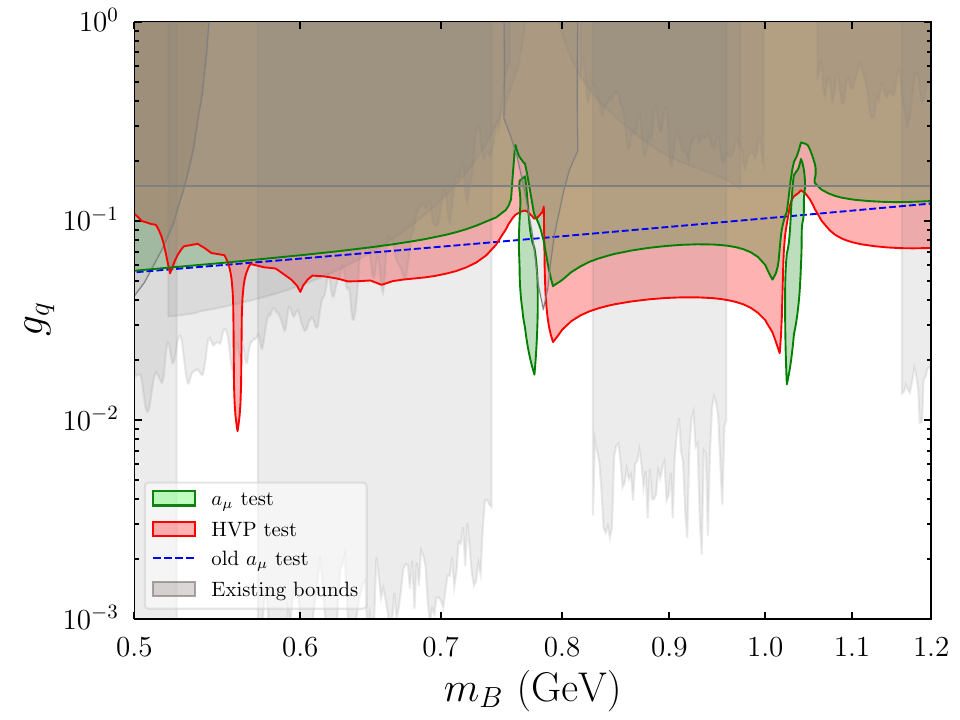}
    \caption{$(a^{ \rm  HVP }_{ \mu })^{\text{DD}}= (a^{ \rm  HVP }_{ \mu })^{\rm{lat}}-1\sigma$\\ $g_\l/g_q=-0.05$}
    \label{fig:05_neg_dd_lat_minus1}
    \end{subfigure}

    \caption{Parameter space of the quark coupling $g_q$ as the function of the mass of \zprime mass ($m_B$) for  $g_\l/g_q=0.05$ (Fig.~\ref{fig:05_dd_lat_minus1}) and $g_\l/g_q=-0.05$ (Fig.~\ref{fig:05_neg_dd_lat_minus1}). 
    We perform the lattice $a_{\mu}$ test (c.f. Eq.~\eqref{eqn:amu-lattice-def}) by taking the BMW result as the SM contribution. 
    For the HVP test, we  consider the possibility that $(a^{\rm HVP}_{\mu})^{\rm DD}=(\amu^{\rm HVP})^{\rm lat}-1\sigma$ .
    Rest matches Fig. \ref{fig:add_fig_05_plus}.} 

\label{fig:add_fig_05_minus}
\end{figure}

\begin{figure}[t]
\centering
    \begin{subfigure}{0.45\textwidth}
    \captionsetup{justification=centering}
    \includegraphics[width=\textwidth]{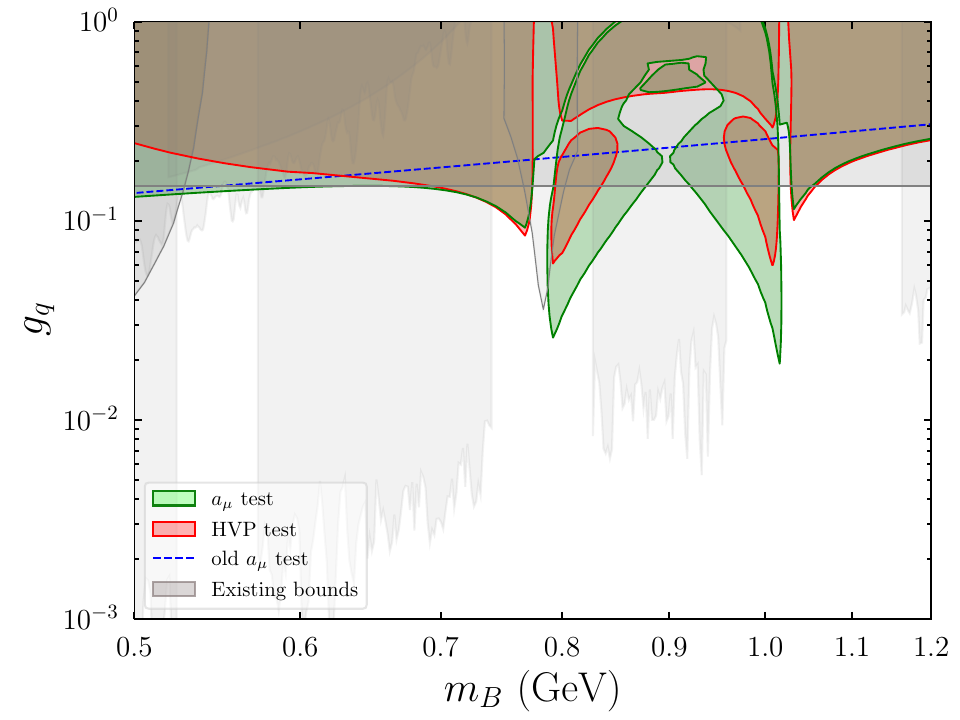}
    \caption{$(a^{ \rm  HVP }_{ \mu })^{\text{DD}}  =  (a^{ \rm  HVP }_{ \mu })^{\text{lat}}+1\sigma$\\ $g_\l/g_q=0.02$}
    \label{fig:02_dd_lat_plus1}
    \end{subfigure}~
    \begin{subfigure}{0.45\textwidth}
    \captionsetup{justification=centering}
    \includegraphics[width=\textwidth]{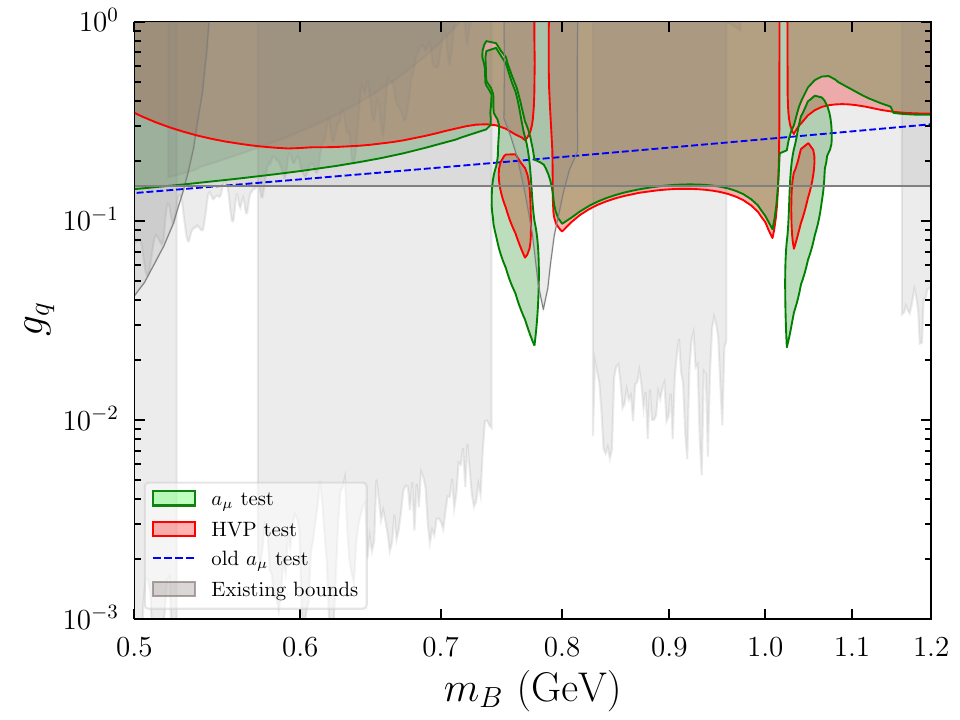}
    \caption{$(a^{ \rm  HVP }_{ \mu })^{\text{DD}} = (a^{ \rm  HVP }_{ \mu })^{\text{lat}}+1\sigma$\\ $g_\l/g_q=-0.02$}
    \label{fig:02_neg_dd_lat_plus1}
    \end{subfigure}
    
    \caption{Parameter space of the quark coupling $g_q$ as the function of the mass of \zprime mass ($m_B$) for  $g_\l/g_q=0.02$ (Fig.~\ref{fig:02_dd_lat_plus1}) and $g_\l/g_q=-0.02$ (Fig.~\ref{fig:02_neg_dd_lat_plus1}). 
    We perform the lattice $a_{\mu}$ test (c.f. Eq.~\eqref{eqn:amu-lattice-def}) by taking the BMW result as the SM contribution. 
    For the HVP test, we  consider the possibility that $(a^{\rm HVP}_{\mu})^{\rm DD}=(\amu^{\rm HVP})^{\rm lat}+1\sigma$ .
    Rest matches Fig. \ref{fig:add_fig_05_plus}.}
    \label{fig:add_fig_02_plus}
    \end{figure}
    \begin{figure}[t]
    \centering\begin{subfigure}{0.45\textwidth}
    \captionsetup{justification=centering}
    \includegraphics[width=\textwidth]{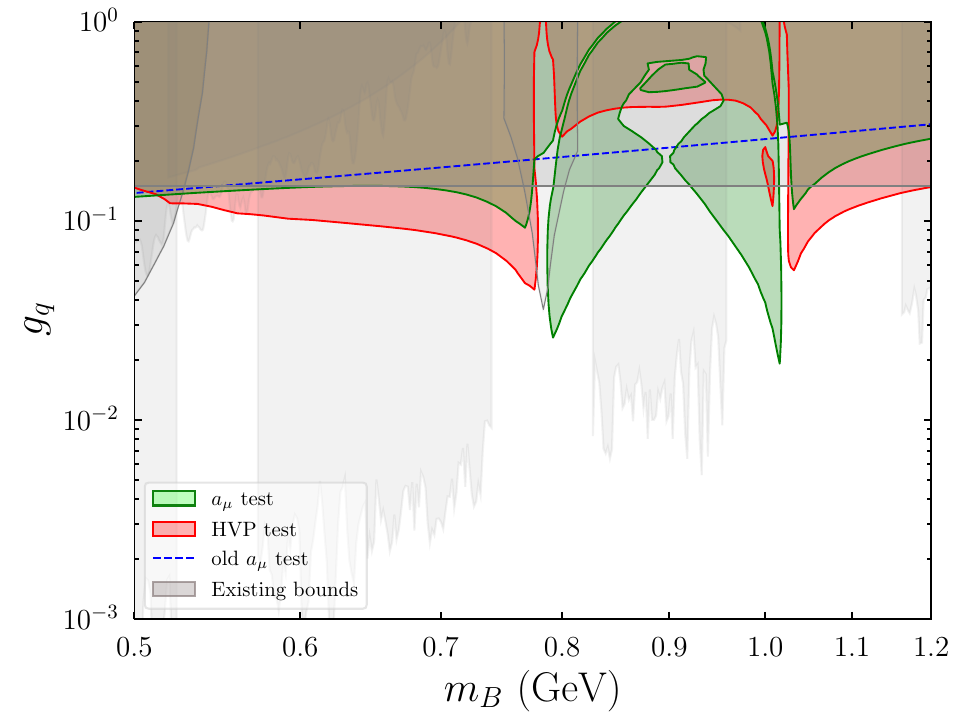}
    \caption{$(a^{ \rm  HVP }_{ \mu })^{\text{DD}}= (a^{ \rm  HVP }_{ \mu })^{\rm{lat}}-1\sigma$\\ $g_\l/g_q=0.02$}
    \label{fig:02_dd_lat_minus1}
    \end{subfigure}~
    \begin{subfigure}{0.45\textwidth}
    \captionsetup{justification=centering}
    \includegraphics[width=\textwidth]{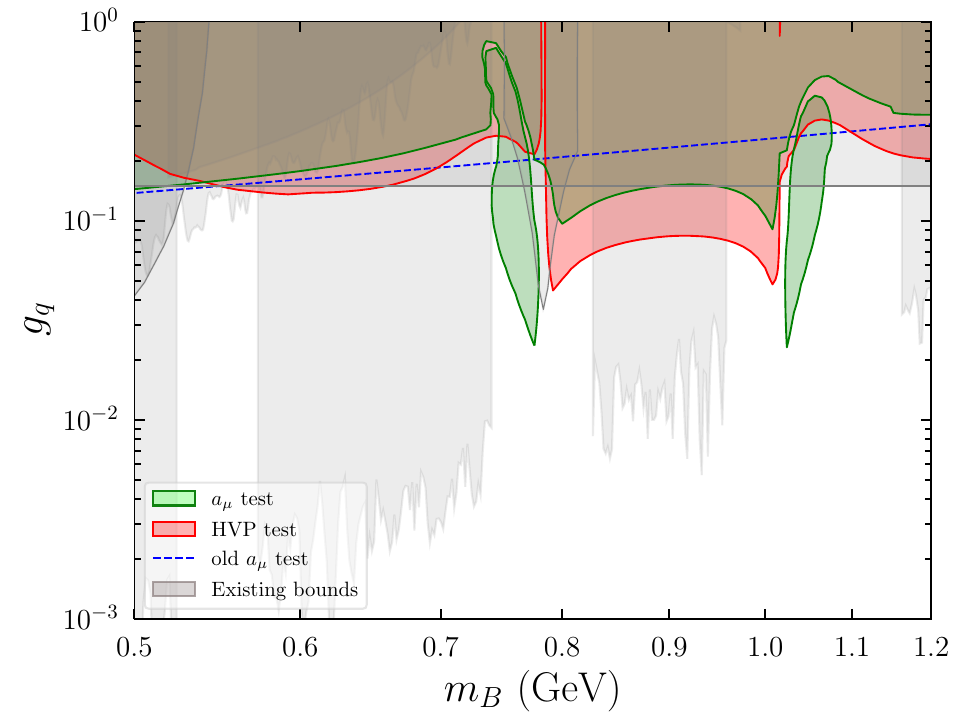}
    \caption{$(a^{ \rm  HVP }_{ \mu })^{\text{DD}}= (a^{ \rm  HVP }_{ \mu })^{\rm{lat}}-1\sigma$\\ $g_\l/g_q=-0.02$}
    \label{fig:02_neg_dd_lat_minus1}
    \end{subfigure}

    \caption{Parameter space of the quark coupling $g_q$ as the function of the mass of \zprime mass ($m_B$) for  $g_\l/g_q=0.02$ (Fig.~\ref{fig:02_dd_lat_minus1}) and $g_\l/g_q=-0.02$ (Fig.~\ref{fig:02_neg_dd_lat_minus1}). 
    We perform the lattice $a_{\mu}$ test (c.f. Eq.~\eqref{eqn:amu-lattice-def}) by taking the BMW result as the SM contribution. 
    For the HVP test, we  consider the possibility that $(a^{\rm HVP}_{\mu})^{\rm DD}=(\amu^{\rm HVP})^{\rm lat}-1\sigma$ .
    Rest matches Fig. \ref{fig:add_fig_05_plus}.}
\label{fig:add_fig_02_minus}
\end{figure}
\begin{figure}[t]
\centering
    \begin{subfigure}{0.45\textwidth}
    \captionsetup{justification=centering}\includegraphics[width=\textwidth]{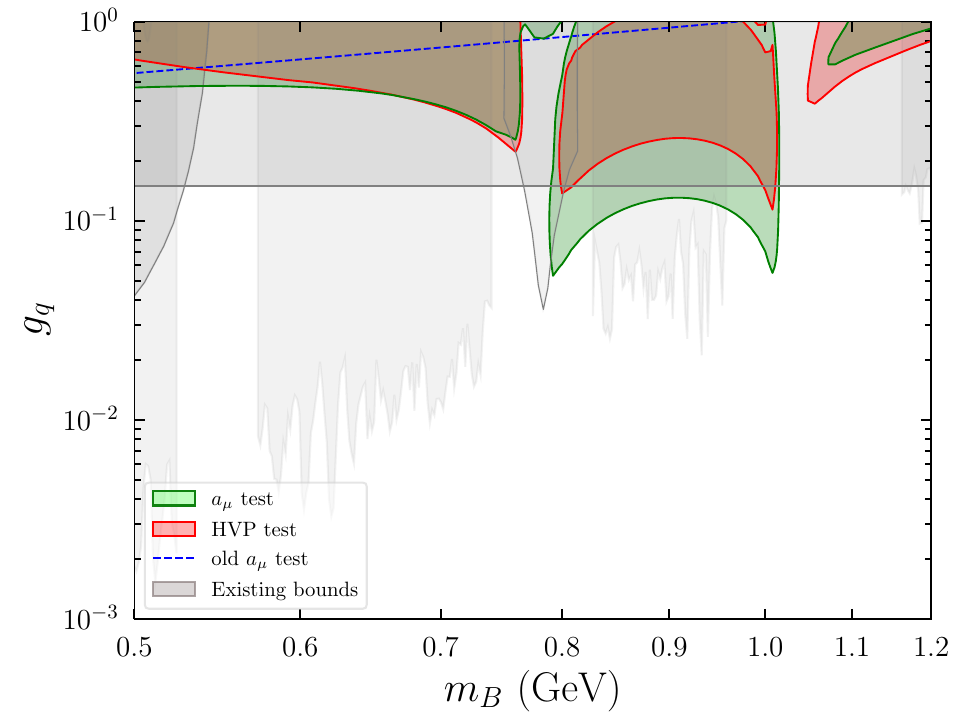}
    \caption{$(a^{ \rm  HVP }_{ \mu })^{\text{DD}}  =  (a^{ \rm  HVP }_{ \mu })^{\text{lat}}+1\sigma$\\ $g_\l/g_q=0.005$}
    \label{fig:005_dd_lat_plus1}
    \end{subfigure}~
    \begin{subfigure}{0.45\textwidth}
    \captionsetup{justification=centering}\includegraphics[width=\textwidth]{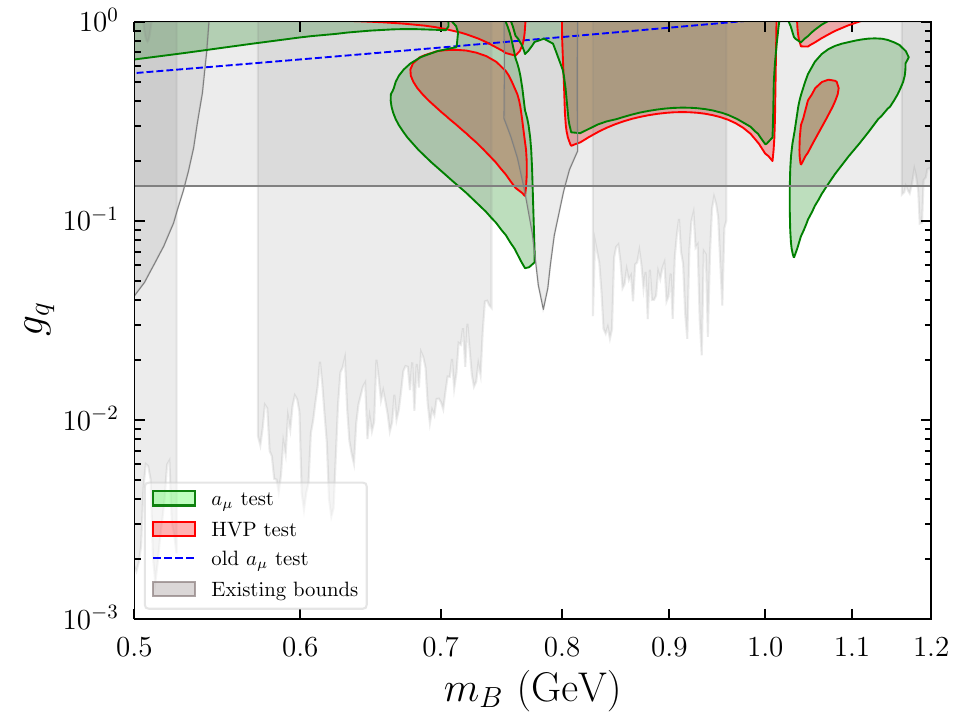}
    \caption{$(a^{ \rm  HVP }_{ \mu })^{\text{DD}} = (a^{ \rm  HVP }_{ \mu })^{\text{lat}}+1\sigma$\\ $g_\l/g_q=-0.005$}
    \label{fig:005_neg_dd_lat_plus1}
    \end{subfigure}
    
    \caption{Parameter space of the quark coupling $g_q$ as the function of the mass of \zprime mass ($m_B$) for  $g_\l/g_q=0.005$ (Fig.~\ref{fig:005_dd_lat_plus1}) and $g_\l/g_q=-0.005$ (Fig.~\ref{fig:005_neg_dd_lat_plus1}). 
    We perform the lattice $a_{\mu}$ test (c.f. Eq.~\eqref{eqn:amu-lattice-def}) by taking the BMW result as the SM contribution. 
    For the HVP test, we  consider the possibility that $(a^{\rm HVP}_{\mu})^{\rm DD}=(\amu^{\rm HVP})^{\rm lat}+1\sigma$ .
    Rest matches Fig. \ref{fig:add_fig_05_plus}.}
    \label{fig:add_fig_005_plus}
    \end{figure}
    
\begin{figure}
    \centering\begin{subfigure}{0.45\textwidth}
    \captionsetup{justification=centering}\includegraphics[width=\textwidth]{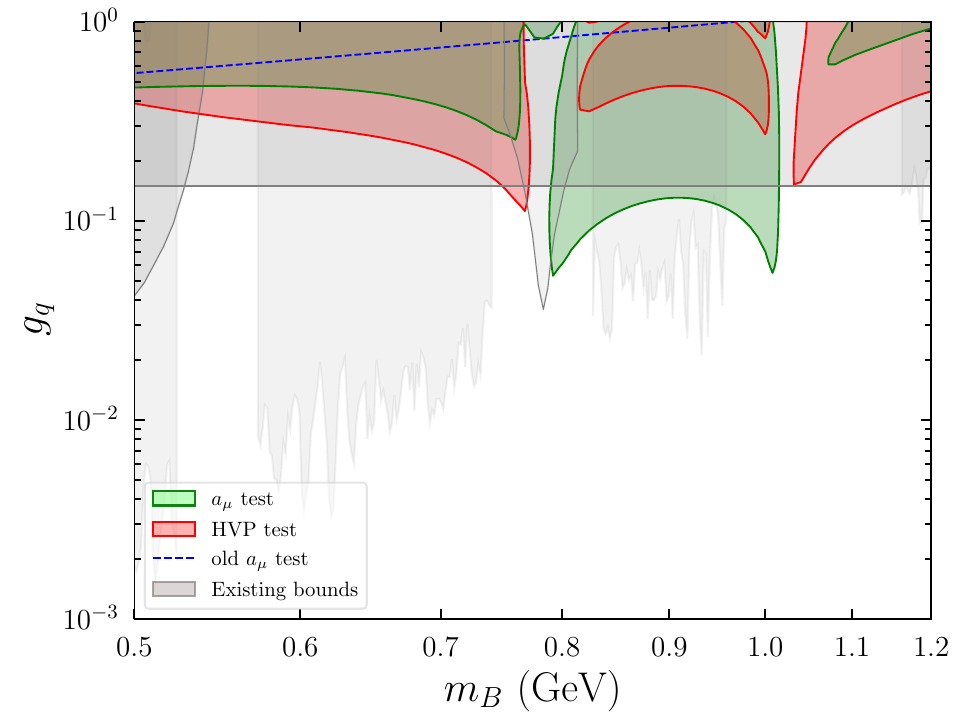}
    \caption{$(a^{ \rm  HVP }_{ \mu })^{\text{DD}}= (a^{ \rm  HVP }_{ \mu })^{\rm{lat}}-1\sigma$\\ $g_\l/g_q=0.005$}
    \label{fig:005_dd_lat_minus1}
    \end{subfigure}~
    \begin{subfigure}{0.45\textwidth}
    \captionsetup{justification=centering}\includegraphics[width=\textwidth]{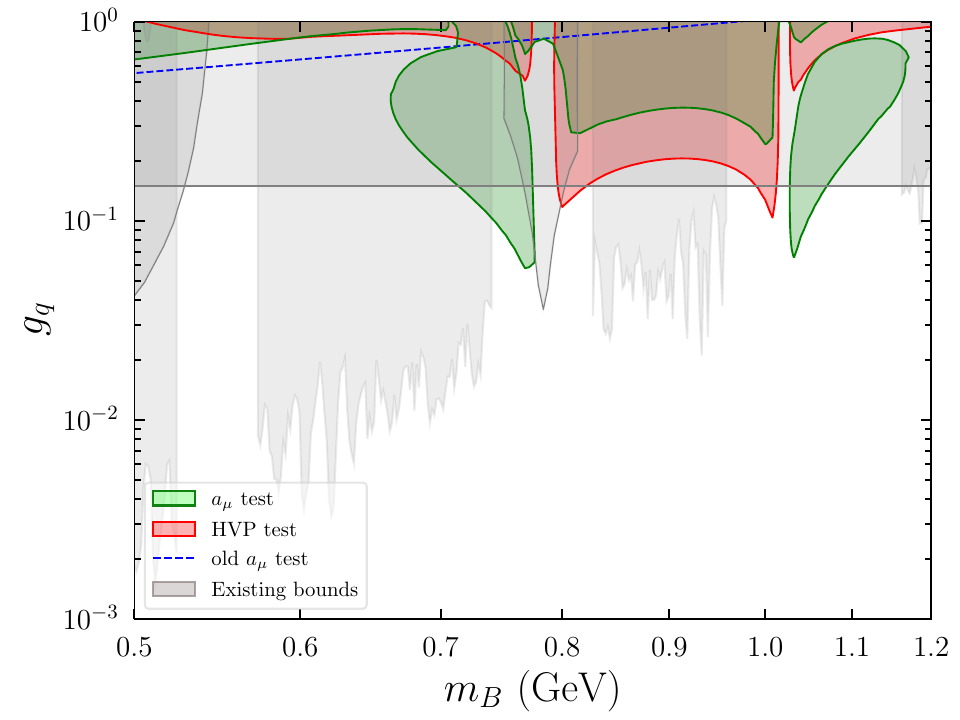}
    \caption{$(a^{ \rm  HVP }_{ \mu })^{\text{DD}}= (a^{ \rm  HVP }_{ \mu })^{\rm{lat}}-1\sigma$\\ $g_\l/g_q=-0.005$}
    \label{fig:005_neg_dd_lat_minus1}
    \end{subfigure}

    \caption{Parameter space of the quark coupling $g_q$ as the function of the mass of \zprime mass ($m_B$) for  $g_\l/g_q=0.005$ (Fig.~\ref{fig:005_dd_lat_minus1}) and $g_\l/g_q=-0.005$ (Fig.~\ref{fig:005_neg_dd_lat_minus1}). 
    We perform the lattice $a_{\mu}$ test (c.f. Eq.~\eqref{eqn:amu-lattice-def}) by taking the BMW result as the SM contribution. 
    For the HVP test, we  consider the possibility that $(a^{\rm HVP}_{\mu})^{\rm DD}=(\amu^{\rm HVP})^{\rm lat}-1\sigma$ .
    Rest matches Fig. \ref{fig:add_fig_05_plus}.}
\label{fig:add_fig_005_minus}
\end{figure}
\clearpage
\section{Window quantities}
\label{app:window}
\begin{figure}[h]
    \centering
\includegraphics[width=0.55\linewidth]{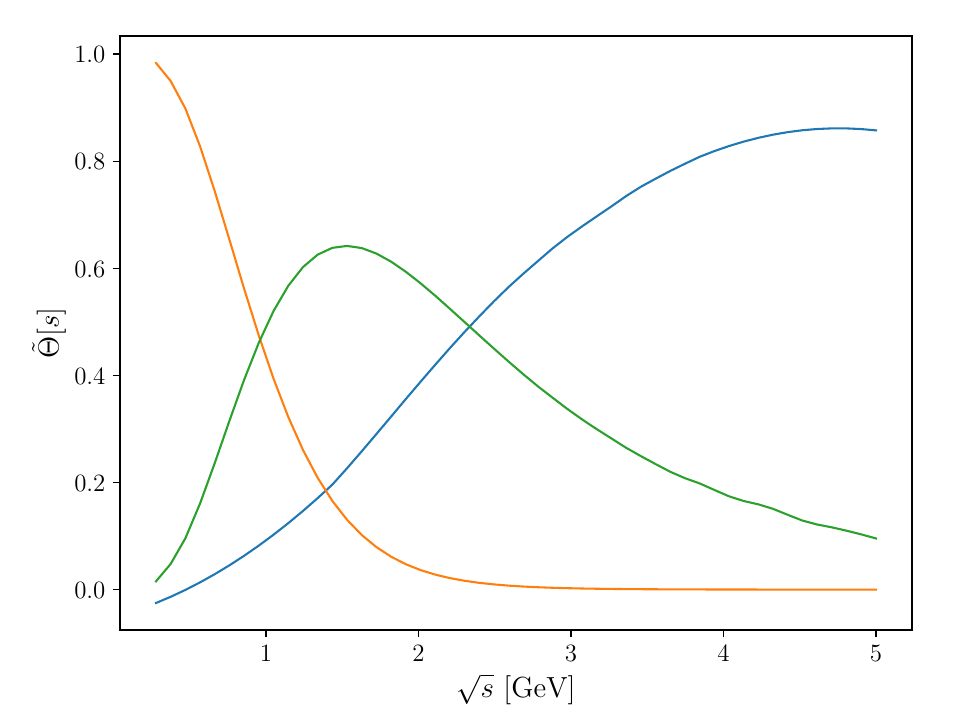}
    
    \caption{The weight functions of the various windows as a function of center-of-mass energy are shown; in blue (short-distance), green (intermediate-distance), and orange (long-distance).
}
\label{fig:theta_s}
\end{figure}

\begin{table}[h]
\centering
\begin{tabular}{|c|c|c|c|}
\hline
 & $a_\mu^{\rm SD}\times 10^{10}$ & $a_\mu^{\rm W}\times 10^{10}$  & $a_\mu^{\rm LD}\times 10^{10}$\\
\hline
ETMC ~\cite{Alexandrou:2022amy}  & 69.27(34)   & 235.0(1.1) & --\\
\hline
BMW 24~\cite{Boccaletti:2024guq} & --  & 235.94(70)    & --   \\
\hline
CLS/Mainz~\cite{Ce:2022kxy,Kuberski:2024bcj,Djukanovic:2024cmq}&68.85(45) &237.30(1.46)&423.2(5.4) \\
\hline
RBC/UKQCD~\cite{RBC:2024fic,RBC:2023pvn} &--&235.56(82)&411.4(4.9) (iso, LQC)\\
\hline
Fermilab/HPQCD/MILC~\cite{Bazavov:2024xsk}&69.01(21) & 236.57(96) &--\\
\hline
\hline
 Colangelo etc. '22 ~\cite{Colangelo:2022vok} &68.4(5) & 229.4(1.4)   & 395.1(2.4)  \\
 \hline
Davier etc. '23~\cite{Davier:2023fpl}&--& \makecell[c]{230.8(1.5) (BaBar) \\  227.3(1.3) (KLOE)\\ 234.2(1.6) (CMD-3)}&--\\
\hline
\makecell[c]{Benton '24~\cite{Benton:2024kwp}\\(LQC)}&\makecell[c]{46.96(48) (KNT19)\\47.63(50) (CMD-3)} & \makecell[c]{199.0(1.1) (KNT19)\\205.8(1.6) (CMD-3)} &\makecell[c]{389.9(1.7) (KNT19)\\ 406.5(3.2) (CMD-3)}\\
\hline
\end{tabular}
\caption{Lattice (up) and data-driven (down) results of window observables.}
\label{tab:win}
\end{table}
In this section, we provide the data for the window quantities discussed in Section \ref{sec:fine_grained_HVP}. The profile of the weight functions, $\tilde{\Theta}(s)$ as the function of the center of mass of energy $s$, is shown in Fig.~\ref{fig:theta_s} \cite{Colangelo:2022vok}.
Many collaborations have calculated $\amu^{\rm HVP}$ in the three windows, and we compile the results from some of them in Table~\ref{tab:win}. 
So far the intermediate distance window has the most results, while the results at the long-distance window appeared only very recently in~\cite{RBC:2024fic} (iso-spin
symmetric light-quark connected (lqc) contribution) and ~\cite{Djukanovic:2024cmq}. These results manifest good agreements with each other. Also shown in the table are recent results from the data-driven method~\cite{Colangelo:2022vok,Davier:2023fpl}. We also include the recent data-driven results for the LQC-only contribution~\cite{Benton:2024kwp}, to be compared with the results in~\cite{RBC:2024fic}. We find that for the short-distance window, these two methods are consistent with each other. 
Large deviation shows up for the intermediate and long-distance regime when BaBar/KLOE results are used in the data-driven method, while results from CMD-3 agree with the lattice within the current error bars.

\bibliographystyle{JHEP}
\bibliography{g2}

\end{document}